\def\kms{km~s$^{-1}$}
\def\Msun{M$_\odot$}
\def\Rsun{R$_\odot$}
\def\SB9{$S\!_{B^9}$}

\documentclass[
    final            % use final for the camera ready runs
  ,numberedheadings % uncomment this option for numbered sections
  ]
  {aipproc}

\layoutstyle{6x9}

\begin{document}

\title{Detection methods of binary stars with low- and intermediate-mass components}


\author{A. Jorissen}{
  address={Institut d'Astronomie et d'Astrophysique, Universit\'e libre de Bruxelles, CP 226, Boulevard du Triomphe, B-1050 Bruxelles, Belgium}
}

\author{A. Frankowski\thanks{Presently at Department of Physics, Technion, Haifa 32000, Isra\"el}}{
  address={Institut d'Astronomie et d'Astrophysique, Universit\'e libre de Bruxelles, CP 226, Boulevard du Triomphe, B-1050 Bruxelles, Belgium}
}

\begin{abstract}
This paper reviews methods which can be used to detect binaries
involving low- and intermediate-mass
stars, with special emphasis on evolved systems.
Besides the traditional
methods involving radial-velocity or photometric monitoring, the paper
discusses as well less known methods involving astrometry or maser
\mbox{(non-)detection}. An extensive list of internet resources (mostly
catalogues/databases of orbits and individual measurements) for the study of
binary stars is provided at the end of the paper.    
\end{abstract}

\maketitle

%%%%%%%%%%%%%%%%%%%%%%%%%%%%%%%%%%%%%%%%%%%%
%% MAINMATTER
%%%%%%%%%%%%%%%%%%%%%%%%%%%%%%%%%%%%%%%%%%%%

\section{Scope of the paper}

Binary stars are home to so many different physical processes that an exhaustive review of them would fill a thick textbook. It is therefore necessary to delineate right away the scope of the present text, which focuses on the 
methods to detect binaries. Besides the traditional
methods involving radial-velocity or photometric monitoring, the discussion covers as well less known methods involving
astrometry or maser (non-)detection. For a more traditional review
focused on the evolutionary aspects (including a detailed description of the  evolution of orbital elements), we refer the reader to the recent book by \citet{Eggleton-2006}, or to the older Saas-Fee course \citep{Shore-1994}. 
            
      Because of the authors' own biases, this review on binary detection methods 
must be understood in the general scientific context of  
systems involving low- and intermediate-mass ('L\&IM') stars, i.e.,
stars which end their lifes as white dwarfs (WDs).
For this reason,
some of the  detection methods discussed in the present review only apply 
to  L\&IM binary systems; conversely, methods that are specific to massive
binaries (Wolf-Rayet binaries, High-mass X-ray binaries, binaries
involving black holes or pulsars) will not be addressed here.

This review is organized as follows. 
The specificities of L\&IM stars are briefly discussed in Sect.~\ref{Sect:LIM}. Basic concepts about binaries (orbital elements and Roche lobe) are summarized in Sect.~\ref{Sect:basics}. 
Possible ways to detect binary systems (sometimes specific to L\&IM binaries) are reviewed in Sects.~\ref{Sect:spectroscopy} to \ref{Sect:diverse}, covering spectroscopic methods (Sect.~\ref{Sect:spectroscopy}), photometric methods (Sect.~\ref{Sect:photometry}), astrometric methods (Sect.~\ref{Sect:astrometry}) and miscellaneous
other methods  (Sect.~\ref{Sect:diverse}). We stress the difficulties associated with these  methods. The reader is referred to
a recent lecture by Halbwachs ({\footnotesize \url{http://www.astro.lu.se/ELSA/pages/PublicDocuments/Halbwachs.pdf}}) for a good introduction on the technical aspects of the main methods, and to the book \textit{Observing and Measuring Visual Double Stars} \citep{Argyle-2004}
for methods relating to visual binaries, which are not addressed here. A summary of what may be known about masses for the different kinds of binaries is presented in Sect.~\ref{Sect:summary}, and finally Sect.~\ref{Sect:resources} provides an extensive list of internet resources (mostly
catalogues/databases of orbits and individual measurements) for the study of
binary stars.      

\section{The families of L\&IM binaries}
\label{Sect:LIM}

 \begin{figure}
  \includegraphics[height=.6\textheight]{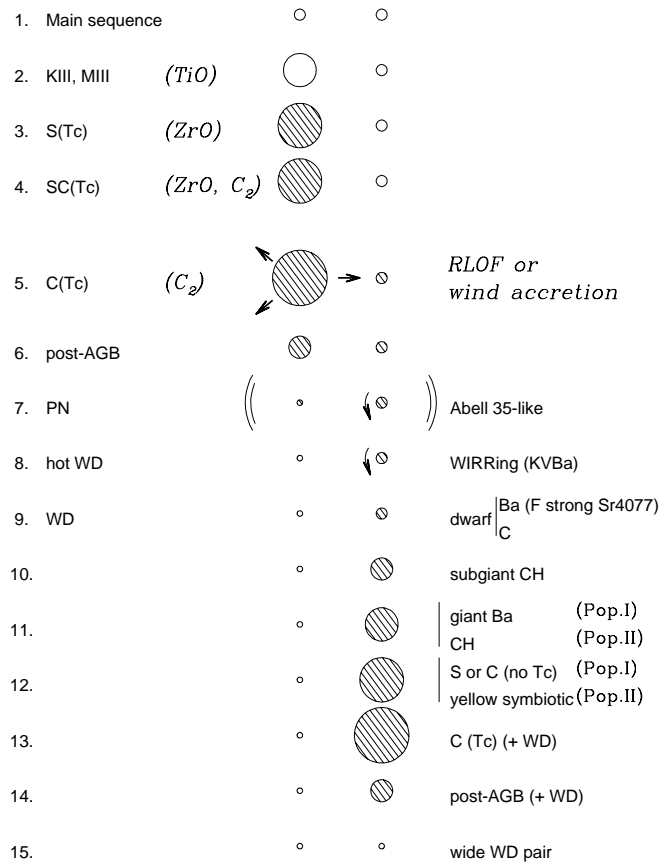}
  \caption{\label{Fig:evolution_Ba}
The evolution of a
system consisting initially of two L\&IM
main-sequence stars. The left column corresponds to the normal
evolutionary sequence of single stars, while the right column
represents the various classes of stars with chemical peculiarities 
specifically produced by mass transfer across the binary system. Hatched circles denote 
stars with atmospheres enriched in carbon or heavy elements. 
(From \citep{Jorissen-03})}
\end{figure}

 \begin{figure}
  \includegraphics[height=.4\textheight]{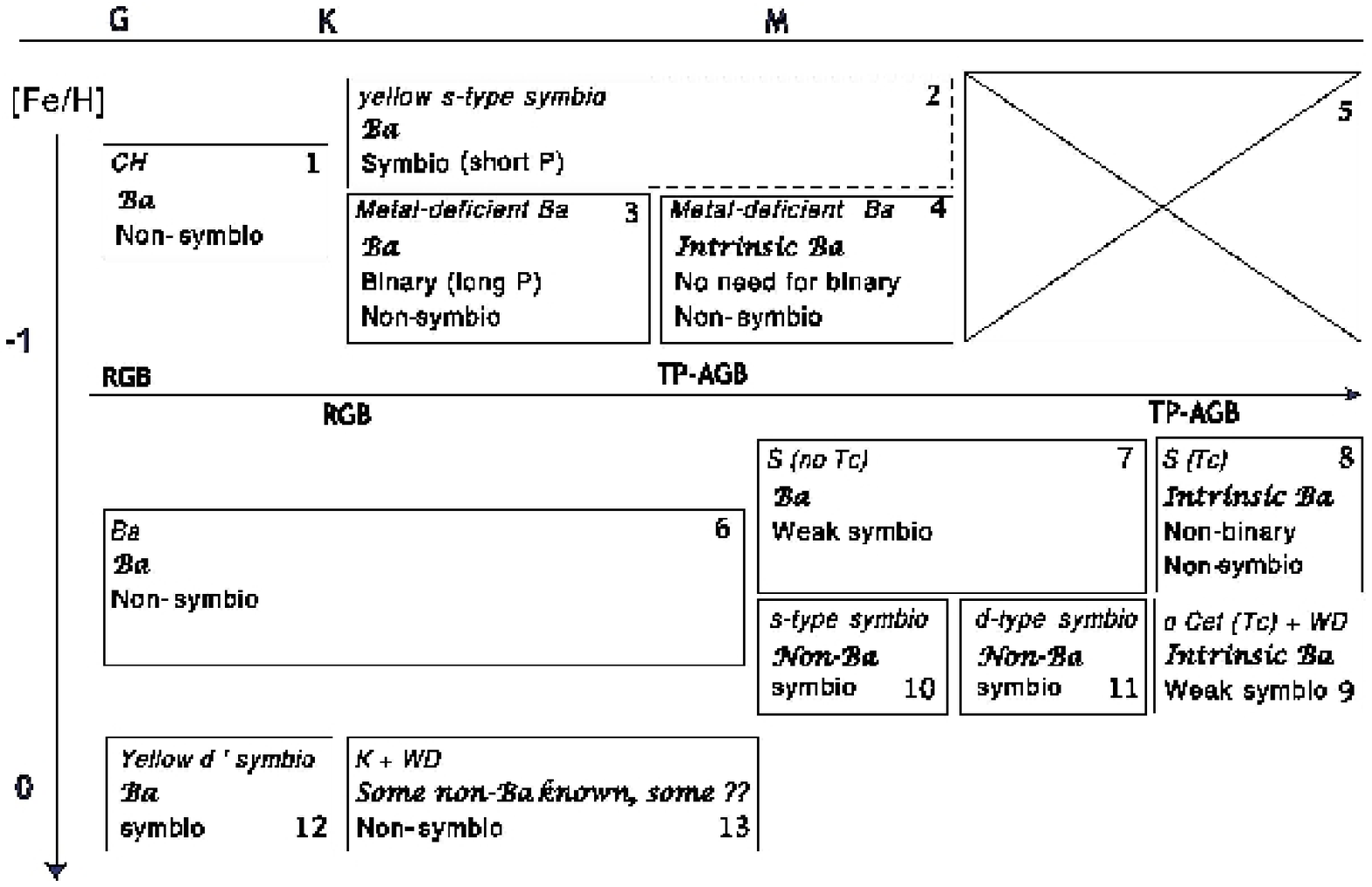}
  \caption{\label{Fig:mdBa}
The various kinds of symbiotic stars (SyS) and of peculiar red giants, in a plane
spectral-type vs. metallicity. Within each box, the first line (slanted font) lists the
stellar family, the second line (calligraphic font) indicates whether or not stars from that
family are enriched in s-process elements (either from internal
  nucleosynthesis -- ''Intrinsic Ba'' -- or from mass transfer across
  a binary system -- ''Ba'' standing for
Extrinsic Ba). The last lines (regular font) provide the binary properties
of that family: binary (short or long periods) or non-binary, SyS or non-SyS. Each
box has been assigned a number, for easy reference in the text. The horizontal line with an
arrow in the middle of the figure is to indicate that the
  thermally-pulsing AGB phase (where the s-process nucleosynthesis
  takes place) involves
different spectral types at low- and near-solar metallicity.
(From \citep{Frankowski-2007a})}
\end{figure}

What makes L\&IM stars so interesting is that 
in the course of their evolution they go through the asymptotic giant
branch (AGB) phase of evolution.
AGB stars have two
very important properties: they are home to  a rich internal
nucleosynthesis and exhibit strong mass loss. 
Besides controlling the AGB evolutionary timescale, 
the strong mass loss has important side effects when it occurs in a binary system,
like the development of symbiotic activity, of associated X-ray
emission, of maser emission (possibly suppressed by the perturbation
induced by the companion), etc... 
In many 'after-AGB' systems\footnote{The term after-AGB binaries is
used here to refer to binary 
systems where
at least one component has gone through the AGB.
After-AGB systems should not be confused with the more
restricted class of post-AGB systems, denoting the short 
transition phase between AGB and planetary nebula stages of (single or binary) 
stellar evolution.}, 
the mass transfer from 
an AGB star has left its mark on the companion, enhancing its abundances with 
the products of the AGB nucleosynthesis, most remarkably C, F, and
elements heavier than iron produced by the s-process 
of nucleosynthesis \citep[see][for a recent
review]{Lattanzio-Wood-2003}. An exemplary case of after-AGB 
systems are barium stars: G-K type giants remarkable for their overabundances 
of Ba \citep{McClure-84b}. Related families include the Abell-35
subclass of planetary nebulae 
\citep{Bond-1993}, barium dwarfs \citep[including the so-called
WIRRing stars;][]{Jeffries-Stevens-96}, subgiant and giant CH
stars \citep{McClure-1984:b,McClure-1997:a}, 
extrinsic S stars, as opposed to intrinsic S stars, which exhibit
spectral lines of the element technetium, a product of s-process
nucleosynthesis which has no stable
isotopes \citep{Jorissen-Mayor-92} and d'-type yellow
symbiotics \citep{Schmid-Nussbaumer-93}. But not all of the after-AGBs need to be s-process rich. The post-AGB 
binaries are an interesting case, as they are all by definition after-AGBs: some of 
them do exhibit s-process enhancement while others do
not \citep{VanWinckel-03,VanWinckel-2007}. Neither do symbiotic stars
(SyS) involving M giants and massive WD 
companions ($M_{WD} > 0.5$~$M_\odot$) exhibit s-process enhancements \citep{Jorissen-03a}. 
Finally, some of the cataclysmic variables with
massive WDs 
should also belong to the after-AGB family. An extensive list of
families of binary stars with WD companions is presented in the
dedicated reviews of \citet{Jorissen-03} and \citet{Parthasarathy-2007}. 
How the different families mentioned above fit in a coherent
evolutionary scheme is sketched in Fig.~\ref{Fig:evolution_Ba}.

While Fig.~\ref{Fig:evolution_Ba} provides a classification of L\&IM
binaries in terms of a temporal sequence, one may also try to order
them in terms of their physical properties, like metallicity, spectral
type and symbiotic activity. Symbiotic
activity is expected when a compact star is heated by matter
falling onto it from a mass-losing (often giant) companion. As it will
be discussed in more details later on, the hallmark of SyS
is a hot spectral continuum superimposed on cool
spectral features, but symbiotic activity also means     
outbursts, X-ray emission, high-excitation emission
lines, and nebular lines.

Fig.~\ref{Fig:mdBa} is an attempt to classify L\&IM binaries in the
plane metallicity -- spectral-type, and to correlate this location
with the presence or absence of symbiotic activity
\citep[for reviews,
see][]{Frankowski-2007a,Jorissen-03a,Jorissen-Zacs-2005}. 

Metallicity, plotted along the vertical axis, has a strong impact on
(i) the spectral appearance, which controls 
taxonomy (CH giants for
instance -- box~1 -- are the low-metallicity analogs of the barium stars
-- box~6); (ii) the efficiency of heavy-element synthesis, being more
efficient at low metallicities
\citep{Clayton-1988}, and (iii)  the location of evolutionary tracks in
the Hertzsprung-Russell diagram (hence the correspondence between
spectral type and 
evolutionary status, like the onset of thermally-pulsing AGB, where
the s-process operates, will depend on
metallicity).
Fig.~\ref{Fig:mdBa} therefore considers three different metallicity
ranges: (i) [Fe/H]$<-1$, corresponding to the halo population; (ii)
$-1 \le $[Fe/H]$ \le 0$, or disk metallicity; (iii) [Fe/H]$\ge 0$,
solar and super-solar metallicities
found in the young thin disk.

The horizontal axis in Fig.~\ref{Fig:mdBa} displays spectral type.
At a given metallicity, spectral type is a proxy for evolutionary 
status: the giant components in L\&IM binaries may either be located
on the first red giant branch
(RGB), in the core He-burning phase (which is hardly distinguishable from
the
lower RGB/AGB;
CH giants probably belong to that phase),
He-shell burning early AGB (E-AGB),
or
on the thermally-pulsing AGB (TP-AGB) phase, where the s-process operates. 

Symbiotic activity is expected in the middle of
this spectral sequence, because (i) at the left end, stars (like CH) are not
luminous enough to experience a mass loss sufficient to power
symbiotic activity; (ii) at the right end, the stars with the barium
syndrome need not be binaries. Indeed, 
in TP-AGB stars,
heavy-elements are synthesized in the stellar interior and dredged-up
to the surface, so that
``intrinsic Ba'' (or S) stars occupy the rightmost boxes -- 4 and 8 -- of
Fig.~\ref{Fig:mdBa}, and hence need not exhibit
any symbiotic activity since they are single stars (For examples of stars belonging to box~4, see \citep{Jorissen-Zacs-2005} and \citep{Drake-2008}).  
Such evolved giants which {\em are} nevertheless members of
binary systems (like Mira Ceti -- box~9) 
will of course exhibit symbiotic activity.
It is noteworthy that late M giants are inexistent in a
halo population (hence the crossed box~5), because evolutionary tracks
are bluer as compared to higher metallicities. Examples of such very
evolved (relatively warm) 
stars in a halo population (box~4) include CS~30322-023
\citep{Masseron-2006:a} and V~Ari \citep{VanEck-03}.

\section{Important preliminary notions}
\label{Sect:basics}

 \subsection{Orbital elements}
 \label{Sect:elements}
 
\begin{figure}
  \includegraphics[height=.4\textheight]{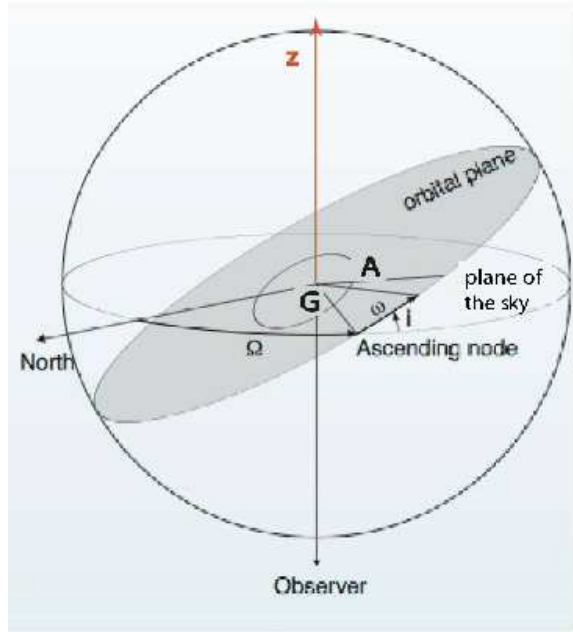}
  \caption{\label{Fig:orbital_elements}
  The orbital plane and orbit of component A around the centre of mass G. $z$ is the coordinate along the line of sight, pointing away from the observer. The plane of the sky is tangent to the celestial sphere and perpendicular to the line of sight. }
\end{figure}

The 7 elements used to describe an orbit are the semi-major axis $a$, the eccentricity $e$, the orbital period $P$, the time of passage (at periastron for a non-circular orbit)  $T_0$, the orbital inclination on the plane of the sky $i$, the longitude of periastron (for non-circular orbits) $\omega$, and finally the position angle of the ascending node $\Omega$. The angles $i, \omega$ and $\Omega$ are identified in Fig.~\ref{Fig:orbital_elements}. 

These 7 elements are also called {\it Campbell elements}, as opposed to the Thiele-Innes elements described below. For edge-on orbits, $i = 90^\circ$. The semi-major axis may either refer to the relative orbit of the two components (usually denoted $a$, in the case of visual binaries), to the orbit of one  component with respect to the centre of mass of the system (usually denoted $a_A$, $a_B$, in the case of spectroscopic or astrometric binaries), or to the orbit of the photocentre  with respect to the centre of mass of the system (usually denoted $a_0$, in the case of astrometric binaries). For astrometric and visual binaries, the semi-major axis is an angular quantity (sometimes denoted $a''$), so that the conversion to a linear quantity requires the knowledge of the parallax $\varpi$: $a({\rm AU}) = a''/\varpi$.

For computational reasons, it is often convenient to replace four of the
Campbell elements, $a, i, \omega, \Omega$, by the so called Thiele-Innes
elements (or constants).
The apparent motion of a  binary component in the plane of the sky (i.e., the plane locally tangent
to the celestial sphere) is described by the cartesian coordinates $(x,y)$ (with $x$ pointing towards the North)
\citep{Binnendijk-1960,Heintz-1978}:
\begin{equation}\label{Eq:xy}
\begin{array}{l}
x=AX + FY\\
y=BX + GY
\end{array}
\end{equation} 
with 
\begin{eqnarray*} 
X &=& \cos E -e \label{Eq:CXCY}\\ 
Y &=& \sqrt{1 - e^2} \sin E,
\end{eqnarray*}
where $(X,Y)$ are the coordinates in the true
orbit and $E$ is the eccentric anomaly related to the mean anomaly $M$ by 
Kepler's equation
\begin{equation}
M(t) = 2\pi \frac{t - T_0}{P} = E - e \sin E.
\end{equation}
The Thiele-Innes constants $A, B, F, G$ are related to the remaining orbital elements by
\begin{eqnarray}
A & = & a(+\cos\omega \cos\Omega - \sin\omega \sin\Omega \cos i) \\
B & = & a(+\cos\omega \sin\Omega + \sin\omega \cos\Omega \cos i) \\
F & = & a(-\sin\omega \cos\Omega - \cos\omega \sin\Omega \cos i) \\
G & = & a(-\sin\omega \sin\Omega + \cos\omega \cos\Omega \cos i). 
\end{eqnarray}

\subsection{Roche lobe}

The concept of the Roche lobe is central while dealing with binary stars.
Let us consider a circular binary and a test particle corotating with
this binary, such as a mass element on the surface of a star rotating
synchronously with the orbital realiasvolution. The Roche lobe corresponds to
the critical equipotential (in the corotating reference frame) surface
around the two binary-system components which contains the inner Lagrangian point. At this point
the net force  acting on the corotating test mass vanishes
(the net force is the vector addition of the two gravitational forces
and of the centrifugal force arising in a non-inertial reference frame
rotating uniformly with the binary).
If the star swells so as to fill its Roche lobe, matter will flow through the inner Lagrangian point onto the companion.
The radius of a sphere with the same radius as the Roche lobe around star A may be expressed by \citep{Eggleton-83}:
\begin{equation}
\label{Eq:RR2}
R_{R,A}/a = \frac{0.49\; q^{-2/3}}{0.6\; q^{-2/3} + \ln( 1 + q^{-1/3})},
\end{equation}
where $q = M_B/M_A$ is the mass ratio. 
In the case of a star 
losing mass (like an AGB star), the extra-force responsible for the 
ejection of the wind must be included in the effective potential. It 
distorts the shape of the equipotentials and reduces the size of the 
Roche lobe around the mass-losing star \citep{Schuerman-72}.
To take this effect into account, a generalisation of Eggleton's 
formula is necessary \citep{Dermine-Jorissen-2008}. 
This reasoning is strictly applicable only to circular
systems, but is often employed as an useful approximation also in the
case of non-zero eccentricity.
A more detailed discussion of the concept of Roche lobe is given by, e.g.,
\citet{Shore-1994} or \citet{Jorissen-03}.

 \section{Spectroscopy}
 \label{Sect:spectroscopy}
 
 \subsection{The method}

Spectroscopic detection of binaries relies on measuring the Doppler shifts
of stellar spectral lines as the binary components orbit their centre of
mass. Soon after the discovery of the first spectroscopic binaries in late
nineteen century \citep{Vogel-1890,Pickering-1890} this became the preferred
detection method as it is robust and through relatively simple means it
gives access to important physical parameters.
Time-dependent spectroscopy allows direct determination of $P, T_0, e,$ and
$\omega$ -- period is the easiest to constrain, the other parameters require
more detailed knowledge of the shape of the radial-velocity curve.
Radial velocities give no handle on $\Omega$, but convey entangled
information on $a$ and $i$. Joined with Kepler's third law, this leads
to partial, but very much sought after, knowledge concerning the component
masses (see Sect.~\ref{Sect:fM} and Table~\ref{Tab:masses} in Sect.~\ref{Sect:summary}).

The expression for the radial-velocity of star A is 
\begin{equation}
\label{Eq:Vr}
V_{r} ({\rm A}) = V_{r} ({\rm G})\; + \; {\rm d}z({\rm A}) / {\rm d}t
= V_{r} \left({\rm G})\;+ \; K_A \left( e \cos \omega + \cos (\omega + \nu(t)\right) \right),
\end{equation}
where $V_{r} (\mathrm{G})$ is the radial velocity of the centre of mass of the system,
$z$ is the spatial coordinate along the line of sight (Fig.~\ref{Fig:orbital_elements}), $\nu$ is the true anomaly (angle between periastron and the true position of the star on its orbit),
and \citep[e.g.,][]{Smart-1977} 
\begin{eqnarray}
\label{Eq:K}
K_A &=&\frac{2 \pi} {P} \frac{a_A \sin i}{(1 - e^2)^{1/2}} \nonumber \\
     &=&212.9 \; \left[ \frac{M_A ({\rm M}_\odot)}{P ({\rm d})} \right]^{1/3} \frac{q}{(1+q)^{2/3}} \;\; \frac{\sin i}{(1-e^2)^{1/2}} \;\; {\rm km\;s}^{-1} \\
     &=&212.9 \; \left[ \frac{M_B ({\rm M}_\odot)}{P ({\rm d})} \right]^{1/3} \left(\frac{q}{1+q}\right)^{2/3} \frac{\sin i}{(1-e^2)^{1/2}} \;\; {\rm km\;s}^{-1} \nonumber
\end{eqnarray}
is the semi-amplitude of the radial-velocity curve of component A, with $q = M_B/M_A$.
To fix the ideas, the semi-amplitude $K_A$ associated with companions of
various masses and orbital periods is given in Table~\ref{Tab:K}, for a
3~M$_\odot$ primary star in a circular orbit with $i = 90^\circ$. 

\begin{table}
\caption{\label{Tab:K}
The semi-amplitude $K_A$ (in \kms) associated with companions of various masses and orbital periods, for a 3~M$_\odot$ primary star in a circular orbit with $i = 90^\circ$. 
}
\begin{tabular}{lrrrrllllll}
\hline
$M_B$ (M$_\odot$) \hspace{4pt} $P=$ & 3 d & 30~d & 1~yr & 3 yr\\
\hline
3 &  134 & 62 & 27 & 19 \\
1 &  59  & 27 & 12 & 8   \\
0.6 & 38 & 17 & 8  & 5  \\
0.08 & 6 & 3  & 1  & 0.8 \\
\hline\\
\end{tabular}
\end{table}

\subsection{
One-dimensional observation: the uncertainty introduced by the inclination}

The first difficulty 
%%%AJ 
facing spectroscopic-binary detection
%%%
comes from the fact that the radial velocity is a one-dimensional measurement (along the line of sight), which implies that the knowledge of the orbit can only be partial. In particular, what is known is the projected orbit on the plane of the sky, implying a degeneracy between $i$ and $a_A$, since only the combination 
$a_A \sin i$ may be extracted from $K_A$.  The same expression holds for component B if that component is visible ('SB2 systems' standing for spectroscopic binaries with two observable spectra). 

If the components of a binary are of approximately equal luminosities, the spectrum will appear as 'composite' (as further discussed in Sect.~\ref{Sect:composite}), and a radial-velocity curve (Eq.~\ref{Eq:Vr}) will be available for both components. Since $a_A M_A = a_B M_B$ from the definition of the centre of mass, the mass ratio  $q = M_B/M_A$ may be derived from  $K_A/K_B$ (see Eq.~\ref{Eq:K}).
Then, $K_A+K_B$ joined with Kepler's third law
\begin{equation}
a^3/P^2  = G\; (M_A + M_B) / 4\pi^2,
\end{equation}
where $a = a_A + a_B$ is the semi-major axis of the {\em relative} orbit, gives access to $(M_A + M_B) \sin^3 i$ (since only  $a \sin i$ -- not $a$ -- may be extracted from $K_A + K_B$).
Combining the mass ratio and sum
(multiplied by $\sin^3 i$)
then yields $M_A \sin^3 i$ and $M_B \sin^3 i$ for the case of SB2 binaries.
A summary of what may be known about the masses for spectroscopic
binaries is provided by Table~\ref{Tab:masses} in Sect.~\ref{Sect:summary}.

If there is an independent way to derive the orbital inclination $i$ (the system is visual, astrometric or eclipsing; in the latter case $i$ is close to 90$^\circ$), then (and only then) may the individual masses be derived (see Table~\ref{Tab:masses}). 
A textbook case \citep[among many others; see the review in][]{Andersen-1991} is provided by the S star HD 35155 : this star is a spectroscopic binary with only one spectrum visible in the optical region  \citep{Udry-98a}, but an {\it International Ultraviolet Explorer} (IUE) spectrum reveals ultraviolet emission lines tied to the companion \citep{Ake-91,Jorissen-92c}, which thus gives access to the mass ratio. Eclipses have been observed in IUE spectra and optical photometry \citep{Jorissen-92c,Jorissen-96}, which implies an inclination close to 90$^\circ$. With all these data at hand, the masses inferred for the giant and its companion are 1.3 -- 1.8~\Msun\ and 0.45 -- 0.6~\Msun, respectively \citep{Jorissen-92c}.

If one is only interested in the {\em distribution} of the masses (rather than in the masses of individual objects) for a sample of SB2 binaries with their orbital planes inclined randomly on the sky, then statistical techniques may be used, as further discussed under the next item.

\subsection{
Only one spectrum is observable: the mass function}
\label{Sect:fM}

It must be stressed that the semi-major axis  $a = a_A + a_B$ of the relative orbit cannot be derived if  only one  spectrum is observable (for instance in the case 
of a faint companion,
 on the lower main sequence or a WD). Hence Kepler's third law, which involves $a$,  is not applicable.
Instead, for these binaries with only one observable spectrum (say A), only
a quantity called the mass function and denoted $f(M)$, can be derived:
\begin{eqnarray}
f(M) \;\; = \;\; \frac{(M_B \sin i)^3}{(M_A + M_B)^2}
&=& \frac{\;K_A^3\;P\;}{2 \pi\,G} \;\; (1-e^2)^{3/2}\\
&=& 1.036\times10^{-7} \;\, K_A^3({\rm km\;s}^{-1}) \; P({\rm d})\; (1-e^2)^{3/2}
\;{\rm M}_{\odot}
\end{eqnarray}

Still, there is a way to get better constraints on the masses if specific conditions are met, namely $M_B << M_A$ (exoplanet companion) with $M_A$ known from the mass-luminosity relationship  for main-sequence stars (if applicable). In those circumstances, \citet{Jorissen-2001}\footnote{Alternative methods have been proposed by \citet{Mazeh-1992:b},
\citet{Zucker-Mazeh-2001} and \citet{Tabachnik-Tremaine-2002}}
have shown that, for a sample of stars, it is possible to extract the
{\it distribution} of $M_B$ from the observed distribution $\Phi(Y)$ of
$Y(M_B) \equiv M_B \sin i $. This follows from
\begin{displaymath}
\textrm{\phantom{blabla (since $M_B << M_A$).}}
f(M) \approx \frac{M_B^3\;\sin^3 i}{M_A^2}
\textrm{\phantom{blabla} (since $M_B << M_A$).}
\end{displaymath}
As $M_A$ is known, $Y(M_B)$ may be derived from $Y(M_B) = f(M)^{1/3} M_A^{2/3}$.
% Prev. ver. was:
%$f(M_A,M_B) = \frac{M_B^3\;\sin^3 i}{(M_A+M_B)^2} \sim \frac{M_B^3\;\sin^3 i}{M_A^2} $
%since $M_B << M_A$. Hence $Y(M_B)$ may be derived from $Y(M_B) = f(M_A,M_B)^{1/3} M_A^{2/3}$, since $M_A$ is known.
%%%

In the more general case of a stellar rather than planetary companion, and with $M_A$ known (as above), the distribution of
$Q \equiv q\;(1+q)^{-2/3}$,
where $q = M_B/M_A$, may be obtained rather than that of $M_B$
\citep{Cerf-Boffin-1994}. Indeed, we have
$f(M) = M_A \; q^3\;(1+q)^{-2}  \sin^3 i =  M_A \; Q^3 \sin^3 i $.
Hence, $Y(Q) \equiv Q \sin i = f(M)^{1/3} M_A^{-1/3} $ is available from the observations.
% Prev. ver. was:
%$Q \equiv \frac{q}{(1+q)^{2/3}}$
%(where $q = M_B/M_A$) rather than $M_B$ may be obtained \citep{Cerf-Boffin-1994}, since 
%$f(M) = \frac{M_B^3\;\sin^3 i}{(M_A+M_B)^2} =  M_A \frac{q^3}{(1+q)^2}  \sin^3 i \equiv  M_A Q^3 \sin^3 i $.
%Hence, $Y(Q) = f(M)^{1/3} M_A^{-1/3} \equiv Q \sin i $ is available from the observations.
%%%

Then, the sought distributions $\Psi(M_B)$ or $\Psi(Q)$ obey
the relation:\\
\[
\Phi(Y) = \int_0^\infty \Psi(M_B)\; \Pi(Y | M_B)\; {\rm d}M_B
\quad {\rm or} \quad
\Phi(Y) = \int_0^\infty \Psi(Q)\; \Pi(Y |Q)\; {\rm d}Q.
\]
The kernel $\Pi(Y | M_B)$ corresponds to the conditional probability
of observing the value $Y$ given $M_B$. Under the assumption that the
orbits are oriented at random in space, the inclination angle $i$
distributes as $\sin i: \Pi(i)\; {\rm d}i = \sin i\; {\rm d}i$ and
the expression for the kernel is obtained
from:
% Skipped:
%from the relations:
%\[
%\Pi(\sin i)\; {\rm d}\sin i = \Pi(i)\; {\rm d}i \begin{footnotesize}\end{footnotesize}
%\]
%\[
%\Pi(M_B \sin i)\; {\rm d}(M_B \sin i) = \Pi(i)\; {\rm d}i 
%\]
\[
\Pi(M_B \sin i) M_B \cos i \;{\rm d}i = \sin i\; {\rm d}i.
\]
% Skipped:
%\[
%\Pi(M_B \sin i)  \;{\rm d}i = \frac{\sin i} {M_B \cos i}  \;{\rm d}i
%\]
%
%Finally, the kernel writes:
%\begin{equation}
%\Pi(Y | M_B) =  \frac{\sin i_0}  {M_B  \cos  i _0}, 
%\end{equation}
%where $i_0$ satisfies the relations $M_B \sin i_0 - Y = 0$ and $0 \le
%i_0 \le 90$. \\

Eliminating the inclination
% Skipped:
%$i_0$
%%%
in the above relation yields
\begin{equation}
\label{Eq:Pi}
\Pi(Y | M_B) =  \frac{Y}{M_B} \frac{1}{(M_B^2 - Y^2)^{1/2}}\quad\quad\mbox{\rm with}\quad Y \le M_B,
\end{equation}
and
\begin{equation}
\Phi(Y)   = Y \int_Y^\infty \Psi(M_B) \; \frac{1}{M_B (M_B^2 - Y^2)^{1/2}}
\; {\rm d}M_B .
\label{Eq:integral2}
\end{equation}
This integral equation must be solved for
$\Psi(M_B)$. It can be reduced to Abel's integral equation by the
substitutions \citep{Chandrasekhar-1950}
\begin{equation}
Y^2 = 1/\eta \hspace{6pt}{\rm and} \hspace{6pt} M_B^2 = 1/t.
\end{equation}
With these substitutions, Eq.(\ref{Eq:integral2}) becomes
\begin{equation}
\label{Eq:Abel}
\phi(\eta) = \int_0^{\eta} \frac{\psi(t)}{(\eta - t)^{1/2}} \; {\rm
  d}t,
\end{equation}
where
\begin{equation}
\phi(\eta) \equiv \Phi(\frac{1}{\sqrt{\eta}}) \hspace{6pt} {\rm and}
\hspace{6pt} \psi(t) \equiv \frac{1}{2\;\sqrt{t}}
\Psi(\frac{1}{\sqrt{t}}).
\end{equation}

The formal solution of Abel's equation
(Eq.~\ref{Eq:Abel}) is given by
\begin{equation}
\label{Eq:Abelsolution}
\psi(t) = \frac{1}{\pi} \int_0^t
\frac{\partial\phi}{\partial \eta}\frac{1}{(t - \eta)^{1/2}} \; {\rm
  d} \eta + \frac{1}{\pi} \frac{\phi(0)}{\sqrt{t}},
\end{equation}
where $\phi(0) = {\rm lim}_{Y \rightarrow \infty} \Phi(Y) = 0$.

It is difficult to implement numerically, since it
requires the differentiation of the observed frequency distribution
$\Phi(Y)$. Unless the observations are of high precision, it is well
known that this process can lead to misleading results.
Two approaches are possible to overcome
that difficulty:
\begin{itemize}
\item The observed frequency distribution is smoothed
in an optimal way  before being used in
Eq.~\ref{Eq:Abelsolution} \citep{Silverman-1986}. The solution $\Psi(t)$ is then computed
numerically using standard differentiation and integration schemes.
\item The Lucy-Richardson  inversion algorithm is applied to Abel's
  integral equation \citep{Cerf-Boffin-1994,Lucy-1974,Richardson-72}.
% a few words on what Lucy-Richarson algorithm is?
\end{itemize}
Both methods have been used by \citet{Jorissen-2001} to derive the
distribution of exoplanet masses.

\begin{figure}
 \includegraphics[height=.5\textheight]{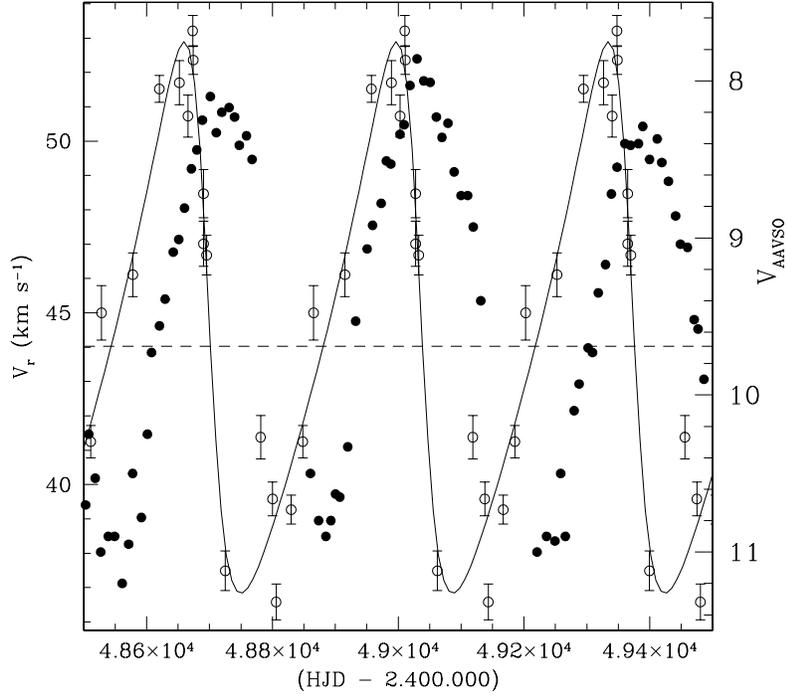}
\caption{\label{Fig:RCMi}
 AAVSO light curve (filled dots) for the Mira 
variable R CMi during three successive cycles, 
compared to its radial-velocity curve 
(solid line and open dots, from \cite{Udry-98a}). 
Radial-velocity measurements were not necessarily obtained at the
indicated dates, as they were
all folded onto the radial-velocity solution. Note that the light- and
velocity-curves are phase-shifted:
maximum light occurs when 
the velocity curve crosses the centre-of-mass velocity (represented by the horizontal dashed line).  
(From \citep{Jorissen-03})
}
\end{figure}

\subsection{Intrinsic velocity variations}
\label{Sect:intrinsic}

\subsubsection{Pseudo-orbital variations and Roche lobe radius}

\begin{figure}
 \includegraphics[height=.6\textheight,angle=-90]{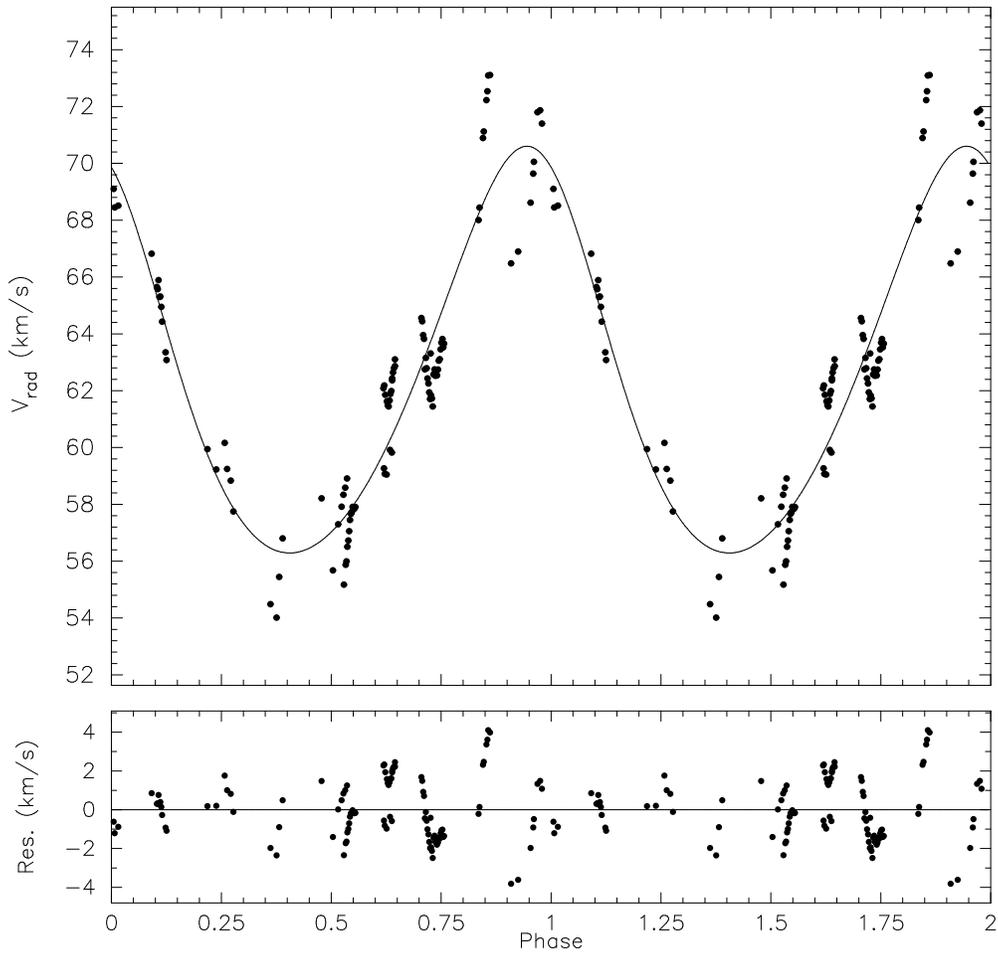}
\caption{\label{Fig:RVTau}
The radial-velocity curve of the RV~Tau variable IRAS 08544-4431: the variations with a large amplitude are caused by the orbital motion, whereas the small-amplitude variations are caused by the RV~Tau-like pulsations with period 90~d, as shown on the bottom panel, displaying the residuals with respect to the orbital solution.
 (From \citep{Maas-2003:a})
}
\end{figure}

The pulsation of the atmosphere of Mira, Cepheid or RV Tau variables
causes intrinsic velocity variations
and makes binary detection using radial velocities difficult or even impossible. Often the radial-velocity variations
associated with the pulsations are with the same period as
the photometric variations
%
% removed the specific \pi/2 reference for the phase-shift:
% RVs and brightness in some semiregulars (Lebzelter et al. 2000, A&A 361,
% 167, Lebzelter et al. 2005, A&A 431, 623) seem to be shifted half a period
% rather than 0.25. Also, when different curve shapes are involved, it is
% hard to talk about a constant phase shift (this is also the case for R CMi)
(although they are phase-shifted),
so that the intrinsic nature of these radial-velocity variations is very
clear.
A very spectacular example is provided by the comparison of the light curve of the 337~d Mira variable R CMi with its radial-velocity curve, as shown on Fig.~\ref{Fig:RCMi}.  The radial-velocity semi-amplitude is as large as 8~\kms. Clearly, in such a case, a companion could only be detected if it would yield orbital variations of
at least
the same order (compare with the data of Table~\ref{Tab:K} to see which kind of companions would be detectable). 
 
 \citet{Hinkle-1997} have shown that the radial-velocity semi-amplitude associated with Mira and semi-regular pulsators correlates well with the visual amplitude and with the pulsation period, reaching 15~\kms\ in the most extreme cases. 
 
 The binary post-AGB star 
 IRAS~08544-4431, hosting a RV~Tau variable (a class of luminous variables in the Cepheid instability strip, characterized by alternating deep and shallow minima), is a rare case  where intrinsic and orbital radial-velocity variations are superimposed (Maas et al. 2003 \citep{Maas-2003:a} present in fact quite a number of similar cases).   Here the orbital period is 499~d and the orbital semi-amplitude is 8~\kms, to be compared to 4~\kms\ for the pulsations of period 90~d (Fig.~\ref{Fig:RVTau}). Hence the two kinds of variations may be separated, but at the expense of a dense observational coverage.

\begin{figure}
 \includegraphics[height=.3\textheight]{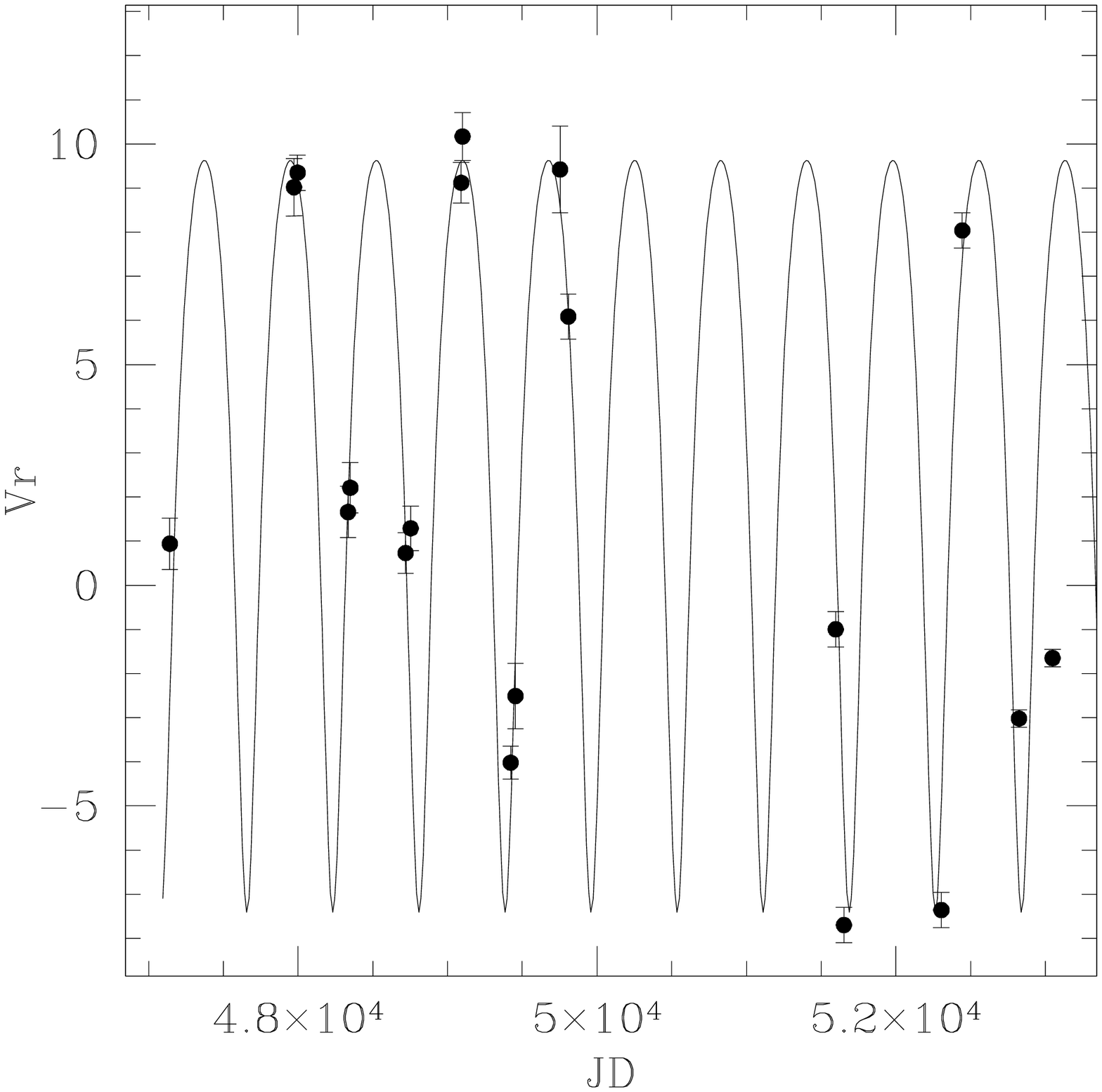}
 \includegraphics[height=.3\textheight]{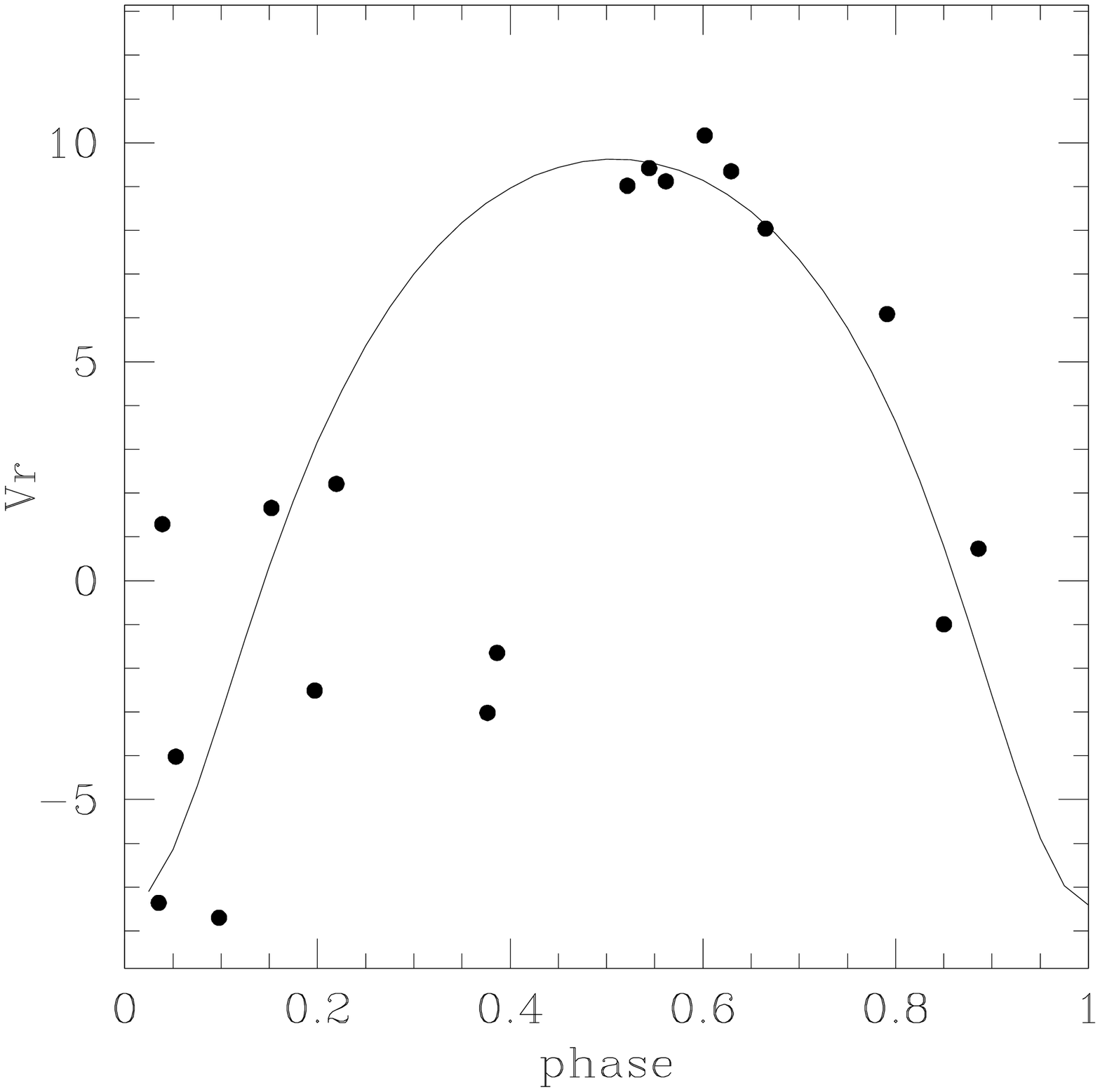}
\caption{\label{Fig:SUMa}
{\bf Left panel}: The radial-velocity curve of the 222~d-period Mira variable S~UMa. A solution with a period $P = 576$~d and an eccentricity of 0.29 provides a good match to
all the data points but the last two. Since the 1/576~d$^{-1}$ frequency is close to the 1/222.0 - 1/365.25 alias of the 1-yr frequency and of the pulsation frequency,
the intrinsic origin of the radial-velocity variations
becomes the best explanation.
From \citep{Famaey-2008:b}.
{\bf Right panel:} The radial-velocity data of S~UMa phased with the 222~d pulsation period. The solid line is an (unsatisfactory) keplerian solution of eccentricity 0.29.  
}
\end{figure}

Sometimes, however, the situation is not as clear as for the two cases discussed above. The 225.9~d Mira variable S~UMa exhibits radial velocity variations
that could at first be interpreted as orbital motion
with an apparent  period of 576~d and an eccentricity of 0.29 \citep{Udry-98a}, based on a rather scarcely sampled data set (Fig.~\ref{Fig:SUMa}). However, the last two data points deviate markedly from the solution based on earlier data points, casting doubts on that solution.
A period analysis of this data set \citep{Famaey-2008:b} reveals that the 576~d period is probably an alias of the pulsation period
(present in the radial-velocity data as 222.0 d)
and of 1 yr (1/222.0 - 1/365.25 = 1/566.0), thus
strongly reinforcing the 
suspicion that the radial-velocity variations have an intrinsic origin.

\begin{figure}[t]
 \includegraphics[height=.5\textheight]{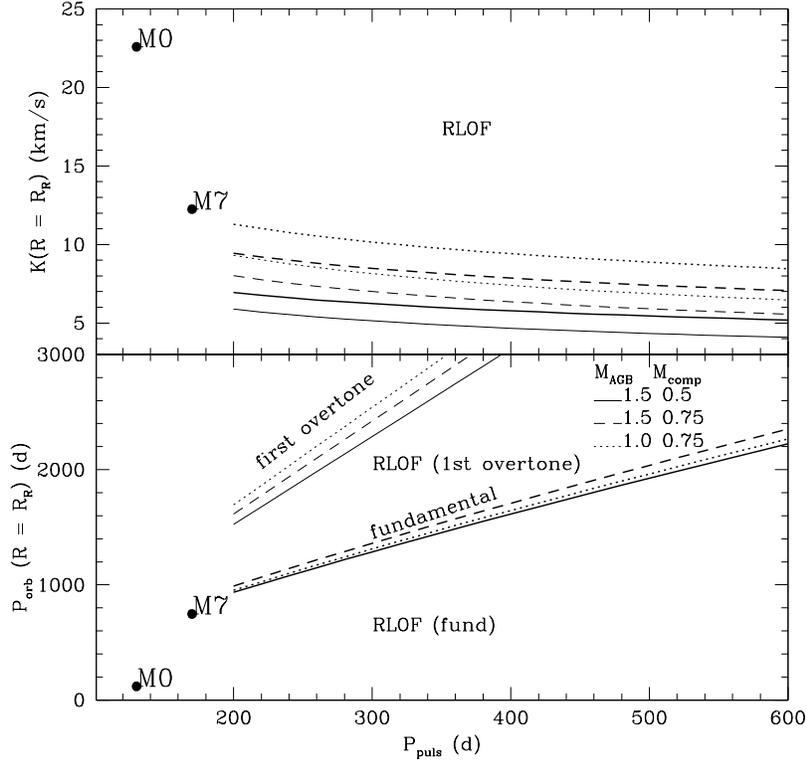}  
\caption{\label{Fig:RR=R}
{\bf Bottom panel}: Orbital
periods versus pulsation periods for Miras in semi-detached systems,
derived from the period--radius relations (Eq.~\ref{Eq:R}) for the
fundamental and first overtone modes. Detached systems are located
above the corresponding lines. The filled circles labeled M0 and M7
indicate the critical periods corresponding to the median radii of
non-Mira M0 and M7 giants \citep{DummSchild98}, respectively (adopting masses of 1.0 and
0.75~\Msun\ for the giant and its companion, respectively).
{\bf Top panel}: Orbital radial-velocity semi-amplitude versus pulsation periods
(thick lines: fundamental mode; thin lines: first overtone) for Miras in
circular semi-detached systems with $\sin i = 1$ (where $i$ is the
orbital inclination). Binary systems with larger semi-amplitudes would
involve Roche-lobe-filling AGB stars. (From \citep{Jorissen-03}) 
}
\end{figure}

In any case, periods as short as a few hundred days can in no way be associated with an orbital motion involving a star as large as a Mira, since the Mira would then fill its Roche lobe, and the system would exhibit strong signatures of mass transfer (like for instance an X-ray flux, or strong emission lines). 
To hold within a binary system, the radius $R$ of  a Mira variable derived from 
the relationships \citep{VanLeeuwen-1997} 
\begin{eqnarray}
\label{Eq:R}
\log R_{\rm  Mira} =
(\log P + 0.9 \log M + 2.07)/1.9 & \,{\rm fundamental\;
mode,}\nonumber 
\\
 & \\
\log R_{\rm  Mira} = (\log P + 0.5 \log M +
1.40)/1.5 & \,{\rm first\; overtone\; mode,}
\nonumber
\end{eqnarray}
where $P$ is expressed in days, and $R$ and $M$
in solar units, 
must be
smaller than the critical Roche lobe radius $R_R$ expressed by Eq.~\ref{Eq:RR2} \citep{Eggleton-83,Dermine-Jorissen-2008}.
The
orbital period for which a Mira of a given pulsation period (in either
the fundamental or first overtone mode) fills its Roche lobe is
displayed in Figure~\ref{Fig:RR=R}. It is derived from the above formulae
(\ref{Eq:RR2}) and (\ref{Eq:R}), and from Kepler's third law
(assuming
typical masses for the Mira star and its companion, as indicated on the
figure). 

The orbital periods
allowed by this criterion
are quite long (more than 1000~d), and are always much longer than the
pulsation periods. The corresponding radial-velocity semi-amplitudes
$K$ (Fig.~\ref{Fig:RR=R}) may be computed from Eq.~\ref{Eq:K}. This quantity must be compared with the intrinsic radial-velocity variations  of Mira variables (due to
pulsations)
which can be as high as 15 km s$^{-1}$ \citep{Hinkle-1997}. 
The detection of spectroscopic binaries among Mira variables thus appears to be almost impossible.

\subsubsection{Wood's sequence D and long secondary periods}

\begin{figure}
 \includegraphics[height=.5\textheight]{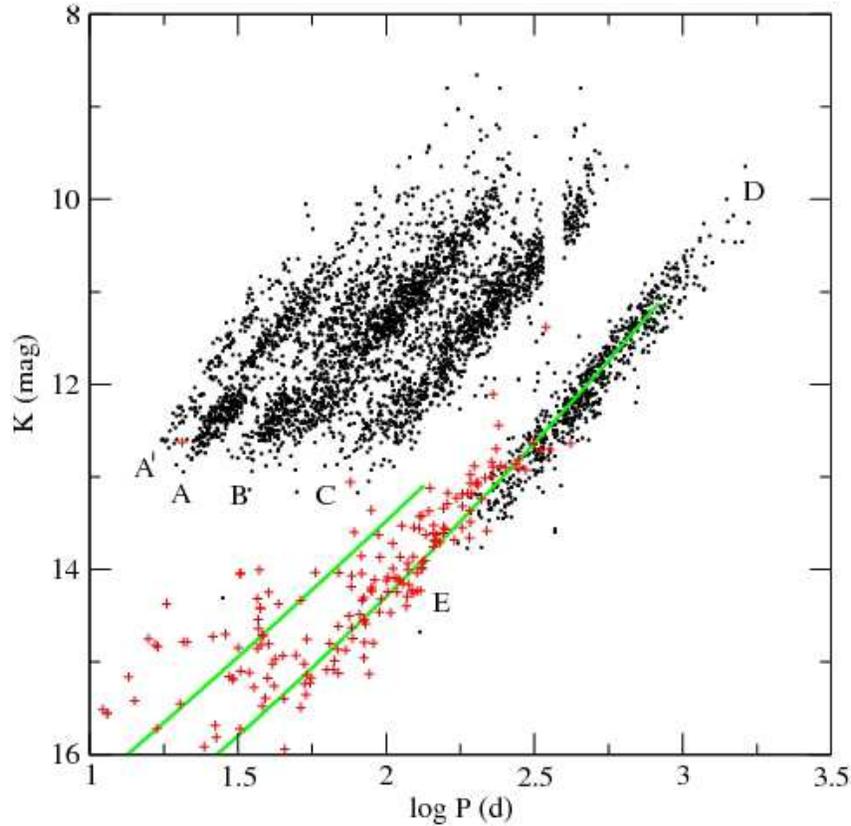}  
\caption{\label{Fig:LSP}
Period-luminosity relations for red giant pulsators discovered by the MACHO project in the Large Magellanic Cloud.  
Ellipsoidal and contact eclipsing binaries  are shown as plus signs. The solid lines  show the critical orbital periods for systems of unit mass ratio filling their Roche lobe, with primary masses of 0.85 and 2.5~\Msun. (From \citep{Derekas-2006})
}
\end{figure}

Variable stars with 'long secondary periods' (LSP, characterized by periods between 200 and 4000 days, i.e., a factor 5 to 15 longer than the primary period) and with 
$V$-band amplitudes up to 1 mag have been 
known for decades \citep{Payne-Gaposchkin-1954,Houk-1963}, 
but the interest for it has been renewed since Wood \citep{Wood-1999,Wood-2000} showed that these
secondary periods 
follow a period-luminosity (PL) relation (the so-called 'sequence D') in the PL diagram of long-period variables (LPVs) in the LMC.
Besides the expected sequence corresponding to the fundamental pulsation mode of Miras (the so-called 'sequence C') and its higher harmonics (sequences A, A' and B, mainly populated by semi-regular variables), a sequence very clearly appeared at {\em longer} periods (the so-called 'sequence D'; Fig.~\ref{Fig:LSP}), involving about 25\% of the LPVs in the LMC. Various hypotheses have been proposed to explain 
the origin of the LSP variability: rotation of a spotted star, episodic dust ejection or obscuration, radial or non-radial pulsations, stellar or substellar companions. \citet{Wood-1999} suggested that stars on
sequence D are components of semi-detached binary systems.  
But is it possible that 25\% of all the LPVs in the LMC are member of binary systems, let alone {\em semi-detached} systems?

\begin{figure}
 \includegraphics[height=.5\textheight]{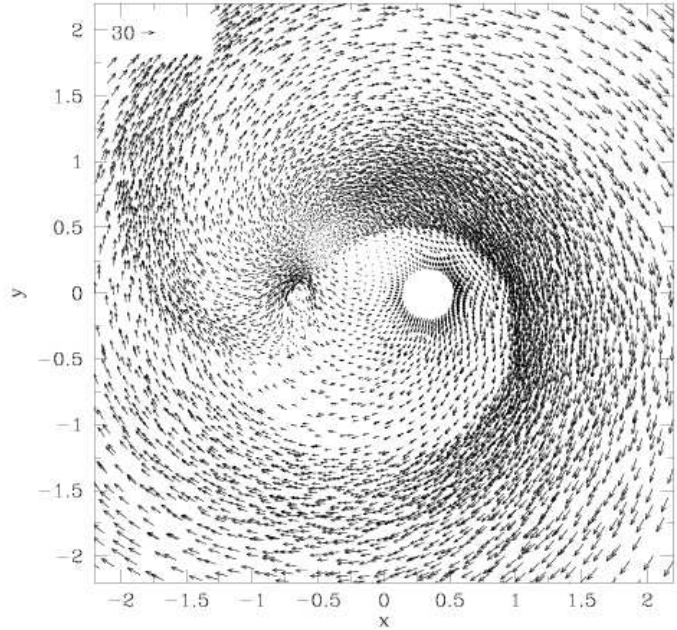}  
\caption{\label{Fig:Theuns}
The structure of the flow (represented by its velocity field) from a mass-losing AGB star (located at $x = 0.33, y = 0$) around the accreting star 
(located at $x = -0.66, y = 0$) in a frame rotating with the system. The double-spiral structure is clearly visible in this simulation of an adiabatic flow.
 (From \citep{Theuns-96})
}
\end{figure}

\citet{Derekas-2006} and \citet{Soszynski-2004b}  noticed
that the so-called 'sequence E', containing eclipsing and ellipsoidal binaries, is the extension of sequence D for fainter (i.e. non LPV) stars. \citet{Soszynski-2007} obtained another clear result calling for the role of binarity in the LSP phenomenon: by considering the quantity $1.3 R_M - B_M$ (where $R_M$ and $B_M$ are the MACHO magnitudes in the red and blue bands, respectively),  the LSP phenomenon should disappear since \citet{Derekas-2006} found that LSP stars have a $B_M/R_M$ amplitude ratio of 1.3 on average, as compared to 1.1 for ellipsoidal variability, which should be grey. \citet{Soszynski-2007} found that in 5\% of the stars belonging to sequence~D, a double-humped structure with a period equal to the LSP one is observable in the $1.3 R_M - B_M$ residual lightcurve, but with a slight phase lag with respect to the LSP lightcurve. This finding gives credit to the binary hypothesis, and the phase shift even hints at dust obscuration as being responsible for the double-humped lightcurve, since hydrodynamical simulations of the mass transfer involving a mass-losing AGB star reveals that the accretion column forms a spiral arm behind the companion \citep[Fig.~\ref{Fig:Theuns};][]{Theuns-Jorissen-93,Theuns-96,Mastrodemos-98,Nagae-2004,Jahanara-2005}.
Therefore, the eclipse of the star by this matter trailing behind the companion
occurs slightly before  the geometric conjunction.

This will be discussed further below in Sect.~\ref{Sect:obscuration}, where we will show that such dust obscuration events are rather common among binaries involving mass-losing giants. The dust cloud present in the LSP stars is not very conspicuous, though, since 
mid-infrared colors of LSPs and non-LSPs are similar \citep{Olivier-Wood-2003,Wood-2004} and there are no LSPs showing the large [60]-[25] $\mu$m IRAS excesses exhibited by some R Coronae Borealis stars, a class of stars  known for forming dust in large quantities.

\begin{figure}
\includegraphics[height=.4\textheight]{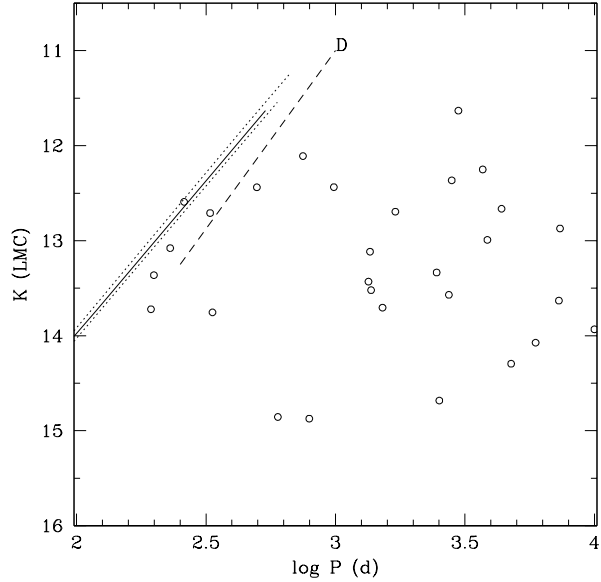}
\caption{\label{Fig:MbolP}
The orbital-period as a function of (absolute) $K$ magnitude for
galactic M giants in binary
systems. The ordinate
axis corresponds to the $K$ magnitude that the
M giants would have if they were put at the distance of the LMC (see
text). 
The dashed line labelled 'D' corresponds to Wood's
sequence of long secondary periods. 
The solid and dotted lines correspond to the limit on the
orbital period imposed by the Roche lobe radius (for $M_1 = 1.3$~\Msun\ and $M_2 =
0.6$~\Msun), for stellar radii comprised between 40 and 120~\Rsun.  
(From \citep{Jorissen-2008}). 
}
\end{figure}

Another piece of evidence in favour of binarity was presented by \citet{Derekas-2006}, \citet{Soszynski-2004b},  \citet{Soszynski-2007} and \citet{Jorissen-2008} who showed that Wood's sequence D closely matches the sequence expected for binary stars involving giant stars filling their Roche lobe. This is clearly illustrated by the solid lines in Fig.~\ref{Fig:LSP} and by Fig.~\ref{Fig:MbolP}, which is a diagram orbital-period -- luminosity for the exhaustive sample of
spectroscopic binary M giants from \citet{Jorissen-2008}.  The ordinate
axis of Fig.~\ref{Fig:MbolP} corresponds to the $K$ magnitude that these
M giants would have if they were at the distance of the LMC, i.e., $K
({\rm LMC}) = M_K + 18.50$, where the \citet{McNamara-2007} LMC distance modulus   
has been adopted, and the absolute $K$ magnitude is derived from the 2MASS value combined with the distance from the maximum-likelihood estimator of \citet{Famaey-2005}. 
It is very clear that Wood's
sequence D does indeed come close to the upper envelope of
the region occupied by the galactic binary M giants, which is defined by the condition that they hold within their Roche lobe.  

If Wood's sequence D is indeed related to LPVs filling their Roche lobe, the remaining question is: how come that there are so many? 
\citet{Soszynski-2007} suggests that variability along sequence D originates in
giants with  substellar companions, the latter being supposedly much more numerous than the 15\% of spectroscopic binaries (not even restricting to semi-detached systems) observed among M giants \citep{Frankowski-2008}. 
Although the mass functions of the orbital solutions obtained by 
\citet{Wood-2004} and \citet{Hinkle-2002}  are indeed compatible with substellar companions, 
the eccentricities and longitudes of periastron derived for 5 among the 6 orbits presented by \citep{Hinkle-2002} are surprisingly similar and may cast doubts on the orbital origin of the observed variations.

\section{Photometry}
\label{Sect:photometry}

As stated in the previous section on spectroscopic binaries,  not many spectroscopic binaries are known among Mira variables, because the orbital radial-velocity semi-amplitude $K_{\rm orb}$ corresponding to a long-period binary is generally smaller than $K_{\rm puls}$. Other methods must thus be used to detect binaries involving Mira variables! Fortunately, there are specific methods using photometry to do so.

%Eclipsing binaries: http://www.midnightkite.com/binstar.html

\subsection{Miras with flat-bottom light curves}

Miras exhibiting a light curve with a flat bottom {\it and} a small-amplitude, despite their Mira-like period, are good candidates for being 
binaries with a faint companion, which dominates the system light around minimum light and is responsible for the flat bottom. A good example thereof is
Z~Tau ($P \sim  450$ d), whose AAVSO (see Sect.~\ref{Sect:resources}) light curve exhibits a flat bottom  at $V = 14$.

\subsection{Ellipsoidal or eclipsing variables}

Ellipsoidal variables are  characterized by a light cycle which is exactly half the orbital period, caused by the tidal deformation of the star nearly filling its Roche lobe. 
The detection  of ellipsoidal variations or of eclipses has often been the first evidence of binarity for many systems. It offers an easy way to infer the  binary period \citep[see ][for the case of symbiotic stars]{Belczynski00}. Ellipsoidal variations should have the same amplitude in all photometric bands, as it is a purely geometrical effect \citep{Hall-1990}.

\subsection{Dust obscuration and circumbinary disks}
\label{Sect:obscuration}

Hydrodynamical simulations of a wind-driven accretion flow in binary systems \citep{Theuns-Jorissen-93,Mastrodemos-98,Nagae-2004,Jahanara-2005,Edgar-2008} predict the formation of a spiral pattern corresponding to the accretion wake bended by the Coriolis force (Fig.~\ref{Fig:Theuns}). 
This prediction has been nicely confirmed by the direct imaging in visible light of the proto-planetary nebula AFGL 3068, using the ACS camera on board the HST, which reveals a spiral pattern winding several times around the central star \citep{Mauron-2006}.
Periodic obscuration of the central star by dust clumps trapped in
the accretion structure -- be it spiral wave, disk around the companion, or
a circumbinary disk --
seems common in binary systems involving a mass-losing giant star. This phenomenon very likely plays a role in the photometric variability observed in the following classes of stars:
\medskip\\
\begin{tabular}{lp{6cm}l}
\multicolumn{3}{c}{Binary systems with evidence for dust obscuration} \\
\hline
Stellar class & Examples & References\\
\hline
LSP Wood's sequence D & many &\citep{Soszynski-2007}
\smallskip\\
short $P_{\rm orb}$ Ba or S stars & HD 35155, HD 46407,  HD 121447 & \citep{Jorissen-92c,Jorissen-1991,Jorissen-92b}
\smallskip\\
AGB C star & V Hya & \citep{Knapp-1999:b}\smallskip\\
Post-AGB & HR 4049 	& \citep{Waelkens91}\smallskip\\
Supergiants &  $\epsilon$ Aur &  \citep{Lissauer-1996}\\
\hline\\
\end{tabular}

\begin{figure}
\includegraphics[height=.5\textheight]{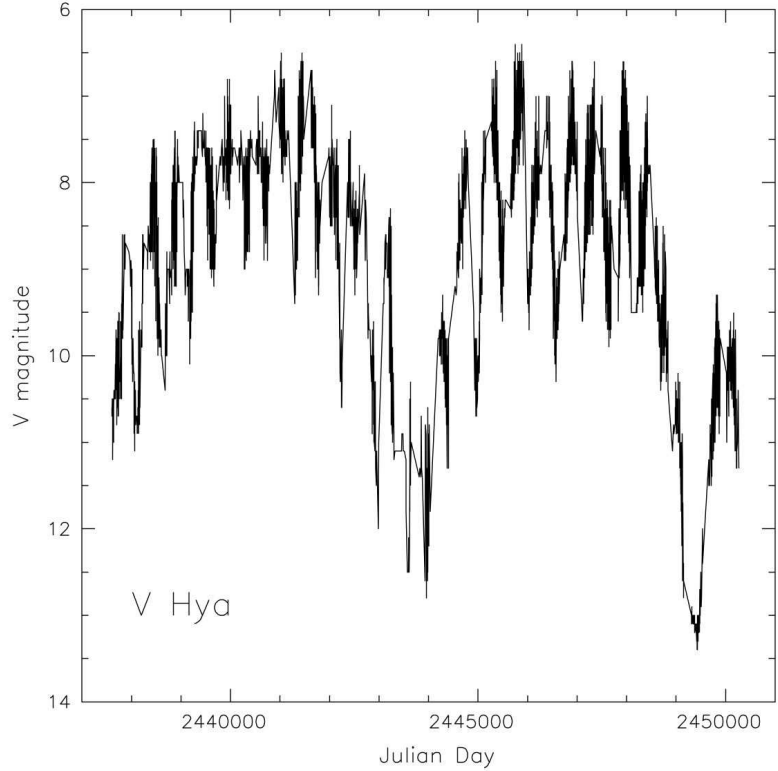}
\caption{\label{Fig:VHya}
The $V$ band light curve of V Hya between October 1961 and July 
1996 from the AAVSO archives.  These data show a more or less regular variability with 
a period of 533~d and a peak-to-peak variation of 
 2 magnitudes at $V$, plus a longer-term variation with a period of about  6500 days (17--18 years) 
with deep minima (5 to 6 magnitudes). (From \citep{Knapp-1999:b})
}
\end{figure}

\begin{figure}
\begin{minipage}{\textwidth}
\includegraphics[height=.3\textheight]{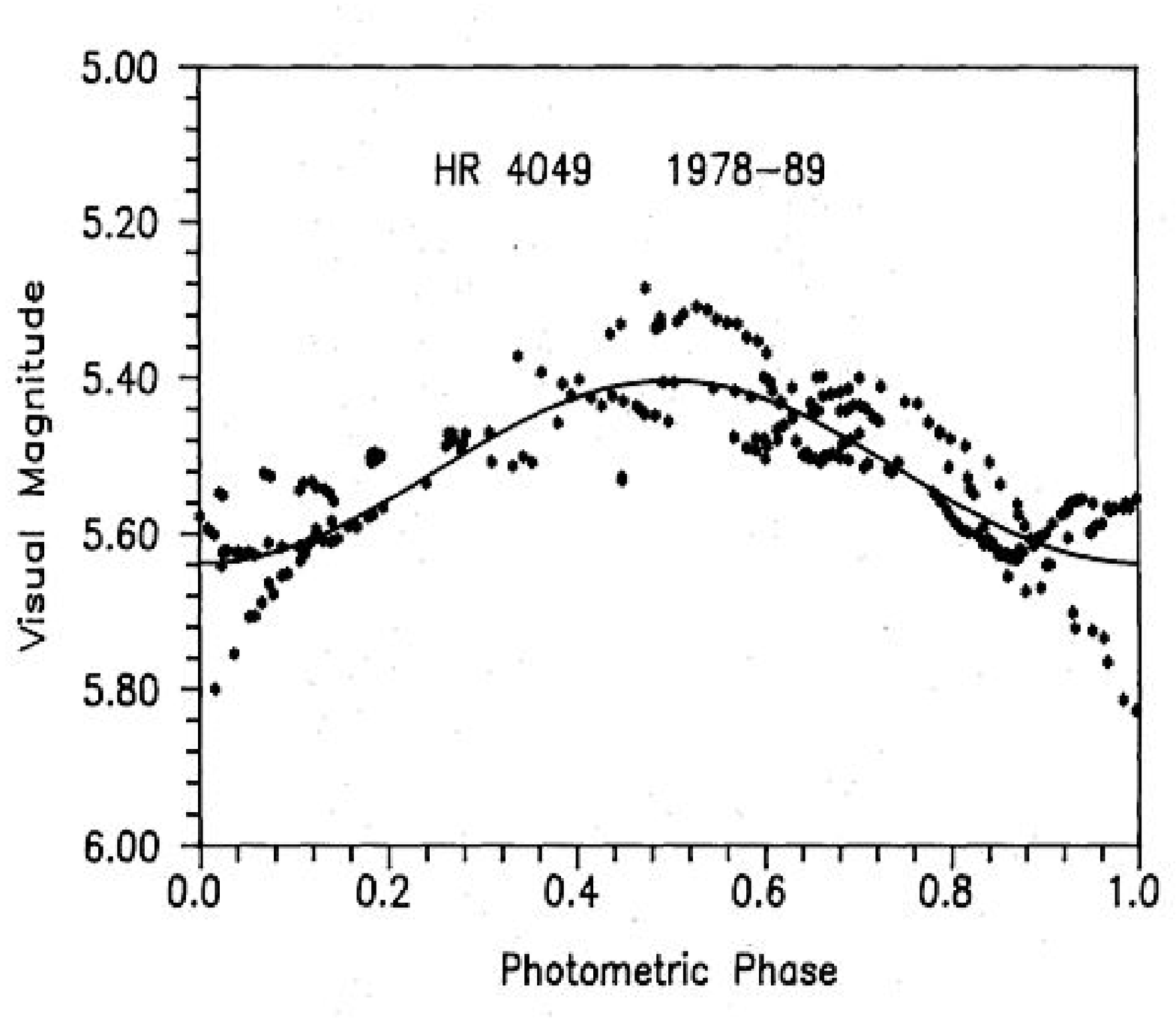}
\includegraphics[height=.3\textheight]{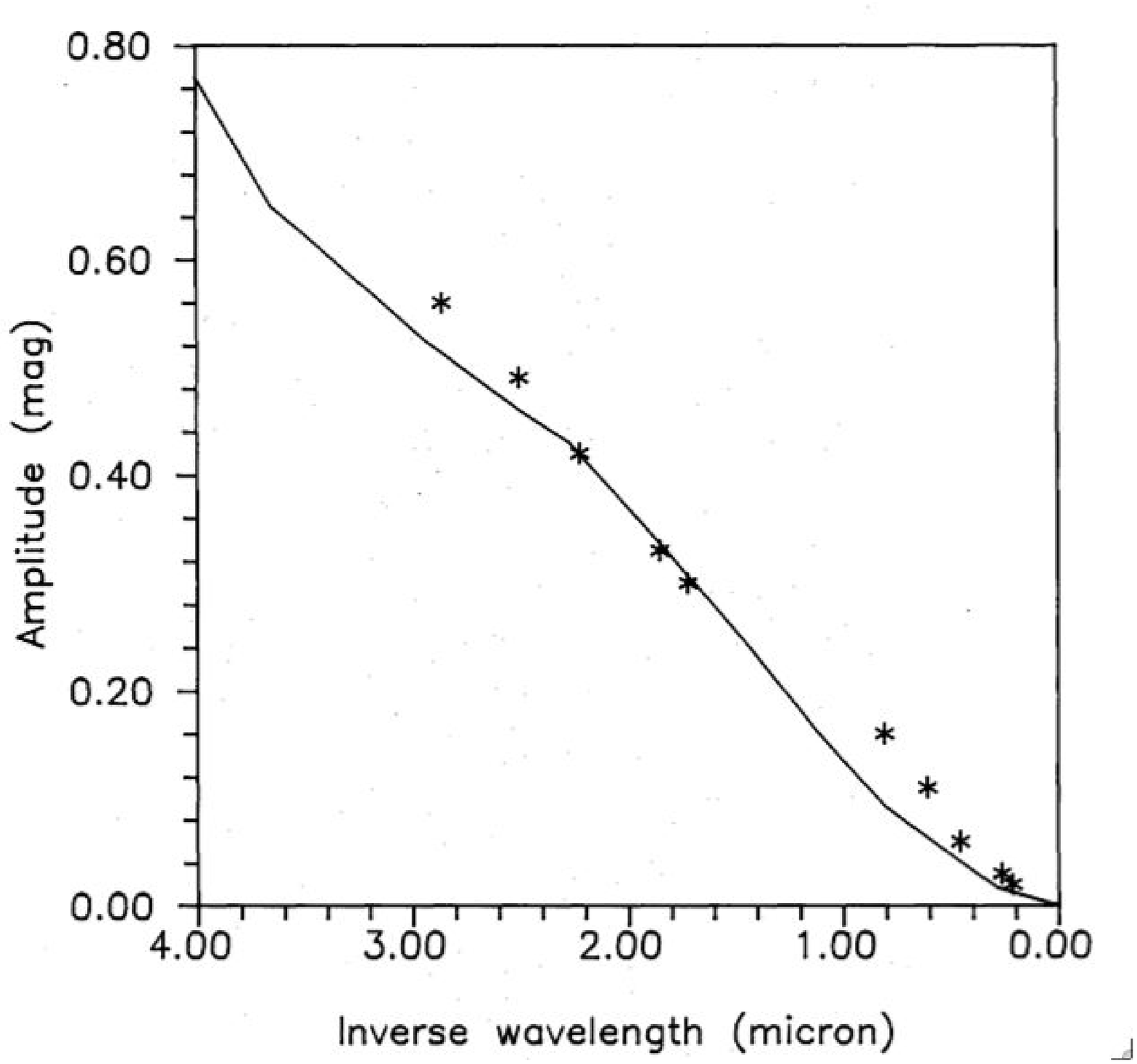}
\vspace{1pt}\\
\includegraphics[height=.3\textheight]{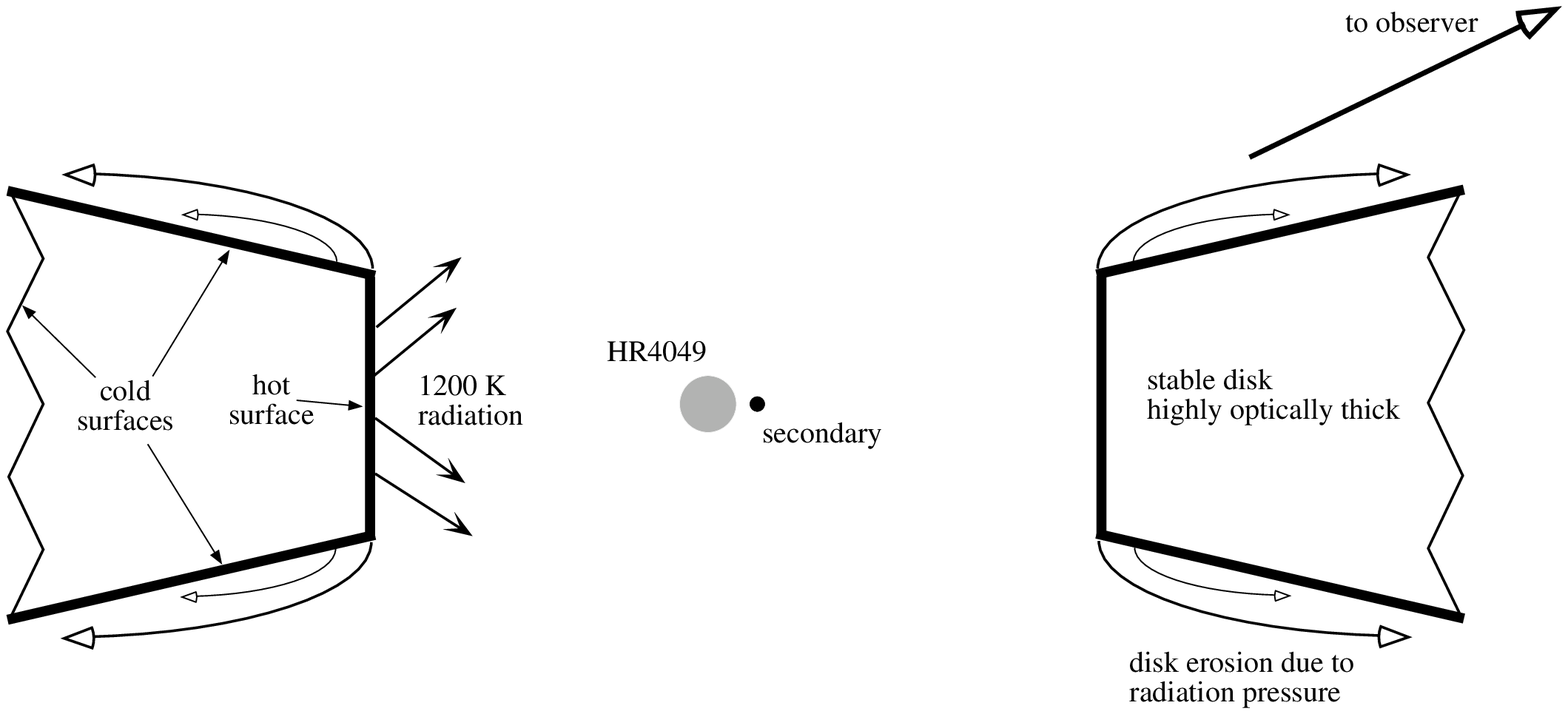}
\end{minipage}
\caption{\label{Fig:HR4049}
The $V$ light curve of the post-AGB star HR~4049 (upper left panel). The modulation is caused by the line of sight going through the dusty circumbinary disk once per orbit, according to the model depicted in the bottom panel. The upper right panel compares the amplitude of the variations in various photometric bands with the extinction law in the interstellar medium.   
 (From \citep{Waelkens91} and \citep{Dominik-2003})}
\end{figure}

The case is especially clear for the  C star V Hya (Fig.~\ref{Fig:VHya}). Quoting \citet{Knapp-1999:b}: {\it The morphology of the 17~y variation resembles that of an 
eclipsing binary, but with an eclipse duration which is far 
longer than can be produced by a stellar companion and of 
an amplitude which shows that essentially the entire stellar 
photosphere is occulted. We suggest that the regular long-period dimming of V Hya is due to a thick dust cloud orbiting 
the star and attached to a binary companion.} 
The orbital period is inferred from the light curve displayed on Fig.~\ref{Fig:VHya}, and there is no spectroscopic confirmation so far (difficult because of the long period, hence small semi-amplitude!).

The 27-yr period observed in the F0 supergiant $\epsilon$ Aur 
\citep{Lissauer-1996}  may be a similar case, with the long duration of the eclipse showing that 
the secondary cannot be a star but must be a large (510~AU) 
dark body, which observations strongly suggest is a dust disk orbiting a companion \citep{Lissauer-1996,Huang-1965}. Eclipses best explained by dust clouds 
attached to a binary companion are also seen in a small number of planetary nebula nuclei, e.g. in NGC 2346 \citep{Costero-1993}. Finally, the prototypical post-AGB star HR 4049 exhibits eclipses by a dust disk, which has been imaged directly by VLTI
\citep[Fig.~\ref{Fig:HR4049};][]{Waelkens91,Dominik-2003,Antoniucci-2005}.
The geometry of the binary system (bottom panel of Fig.~\ref{Fig:HR4049}) is such that the line of sight goes through the circumbinary disk once per orbital cycle, thus  the resulting light curve has
%*AF*
% why ellipsoidal variables are invoked here? Waelkens91 do not even
% seriously consider this possibility, why should we mention it?
only one minimum per cycle (upper left  panel of Fig.~\ref{Fig:HR4049})
rather than two as in ellipsoidal variables.
%%%
The color dependence of the amplitude of the variations is the same as the interstellar extinction law (upper right  panel of Fig.~\ref{Fig:HR4049}), thus confirming that the optical variability is caused by scattering on dust grains of similar size as in the interstellar medium. 

Such circumbinary disks are almost a defining property of binary post-AGB stars, since the presence of a 21~$\mu$m excess, caused by the presence of cold dust in a circumbinary disk extending far away from the heat source (the post-AGB star),  may be used for identifying {\it binary} post-AGB stars \citep{VanWinckel-03,DeRuyter-2006,VanWinckel-2006}.

The M4III + A system SS Lep (= HD 41511) shares with 
post-AGB stars the presence of a circumbinary disk \citep{Jura-2001,Verhoelst-2007}, fed by non conservative RLOF. Using interferometry, \citet{Verhoelst-2007} have indeed shown that the M giant fills its Roche lobe, and that  the A type companion has a radius about 10 times larger than that expected for a main sequence star. It has very likely swollen as a result of accretion.  The system HR~1129 (G2Ib + B7) shows as well an infrared signature of dust most likely associated with a circumbinary disk \citep{Griffin-2006}.

\begin{figure}
\includegraphics[height=.4\textheight]{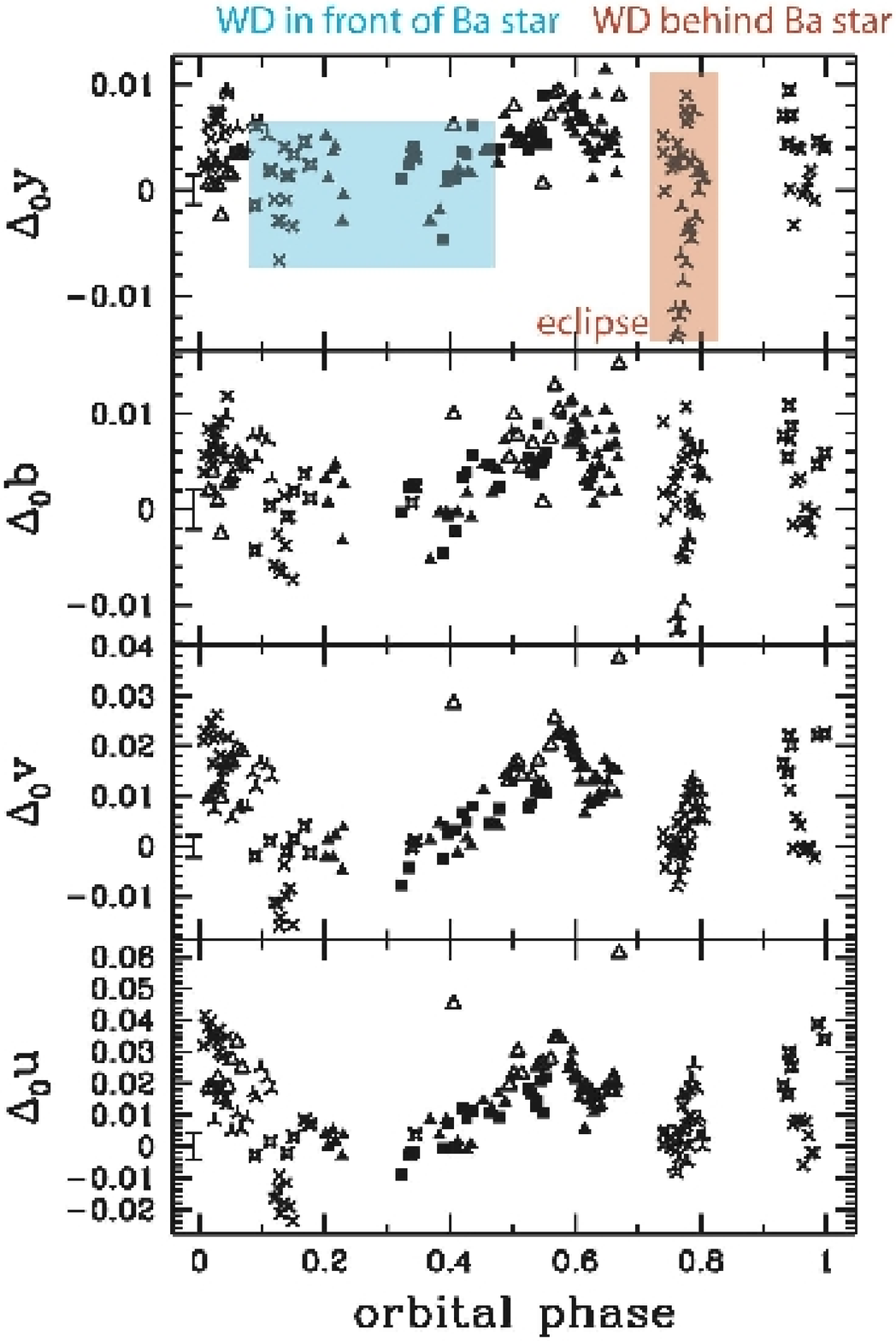}
\includegraphics[height=.4\textheight]{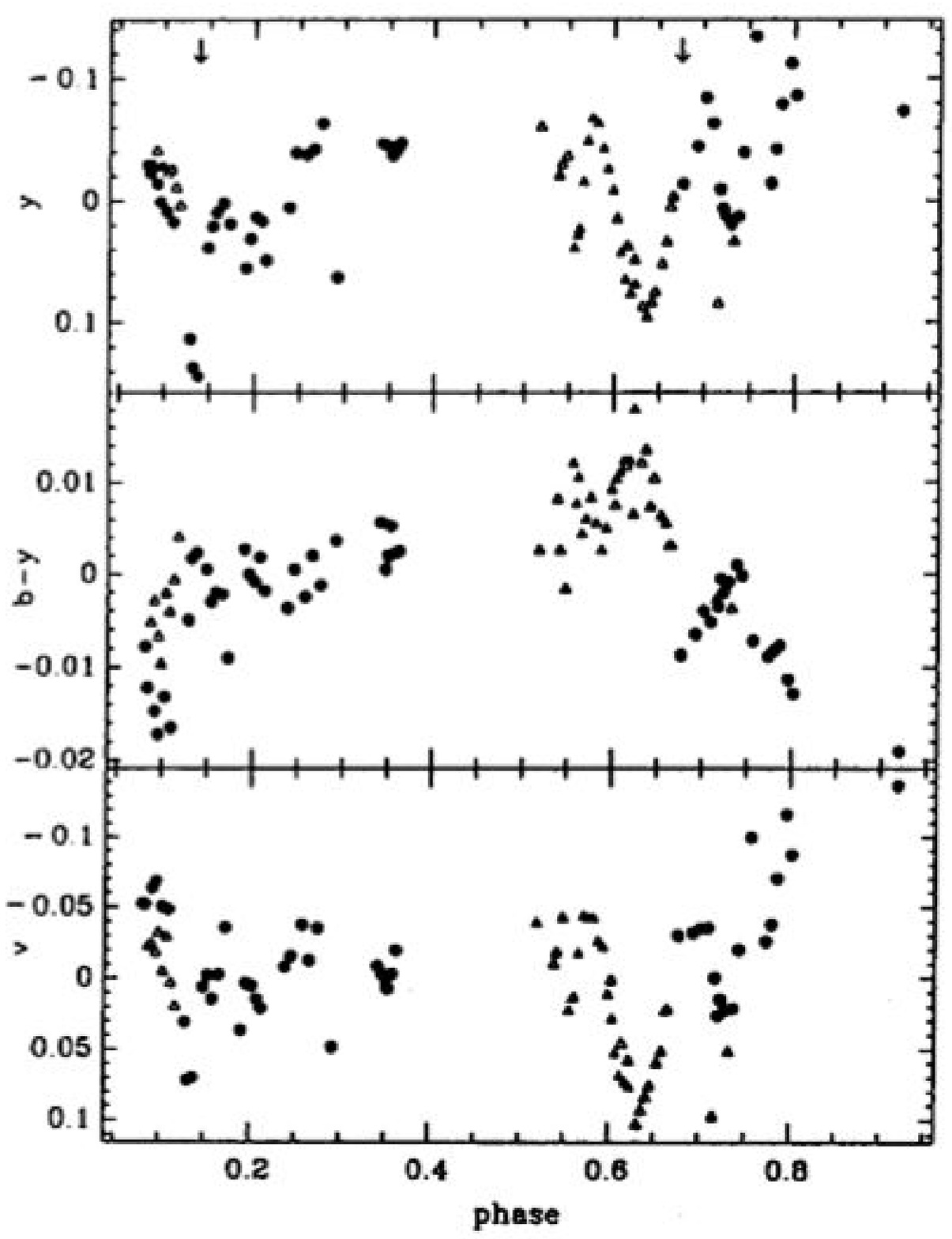}
\caption{\label{Fig:46407}
The light curves phased with the
orbital period for the barium star HD~46407 (left panel; $P = 457$~d)
and the S star HD~35155 (right panel; $P = 640$~d) which are among the
systems with the shortest periods in these classes. Data for HD~46407 \citep{Jorissen-1991,Jorissen-92b}
and HD~35155 \citep{Jorissen-92c} in the Str\"omgren system from the
Long-Term Photometry of Variables program \citep{Manfroid-1991}. The
small arrows in the right panel identify the phases corresponding to
the eclipse (0.17) and transit (0.65) of the companion. HD~35155
exhibits as well intrinsic photometric variations.  
}
\end{figure}

\begin{figure}
\includegraphics[height=.3\textheight]{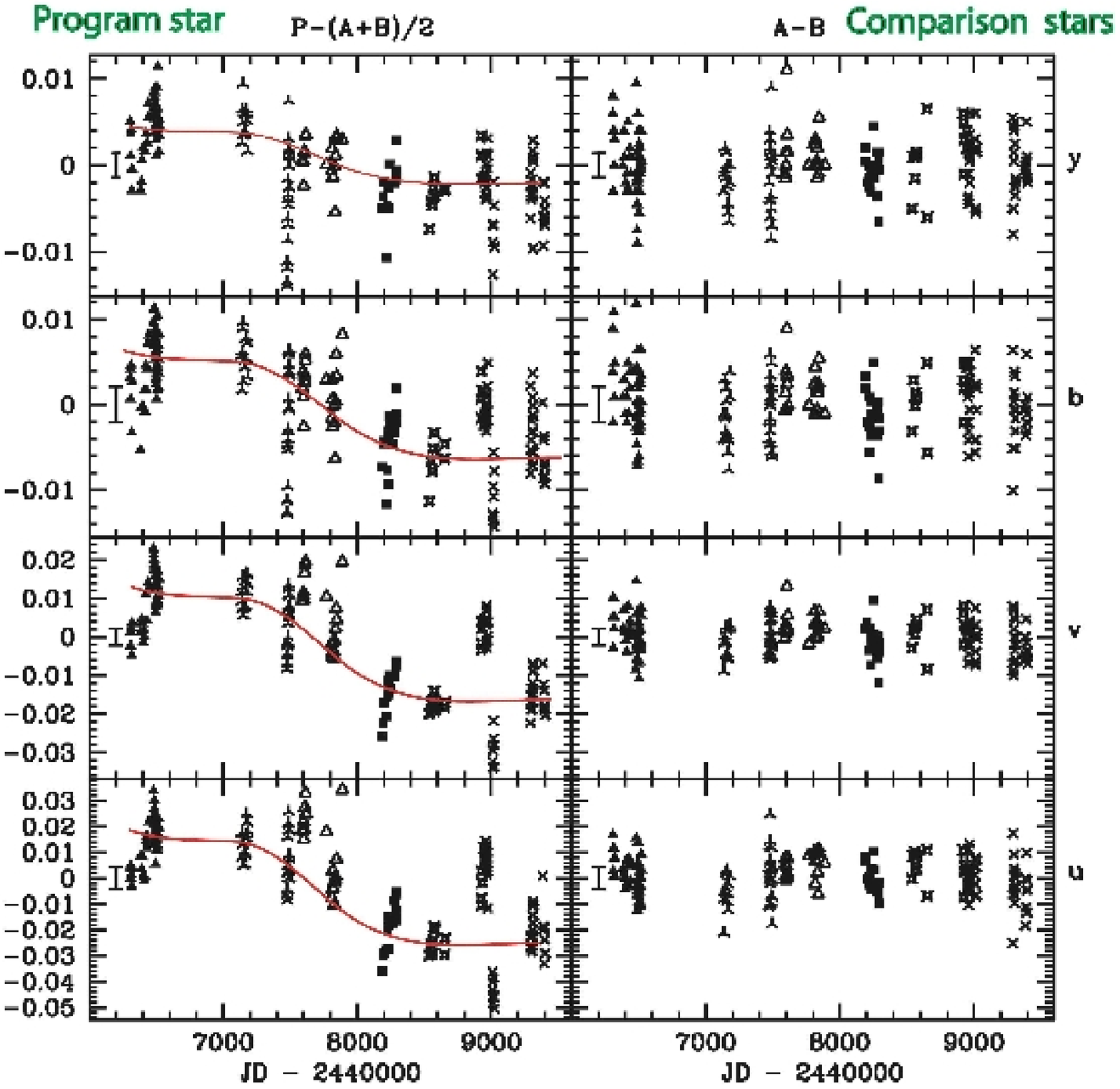}
\includegraphics[height=.3\textheight]{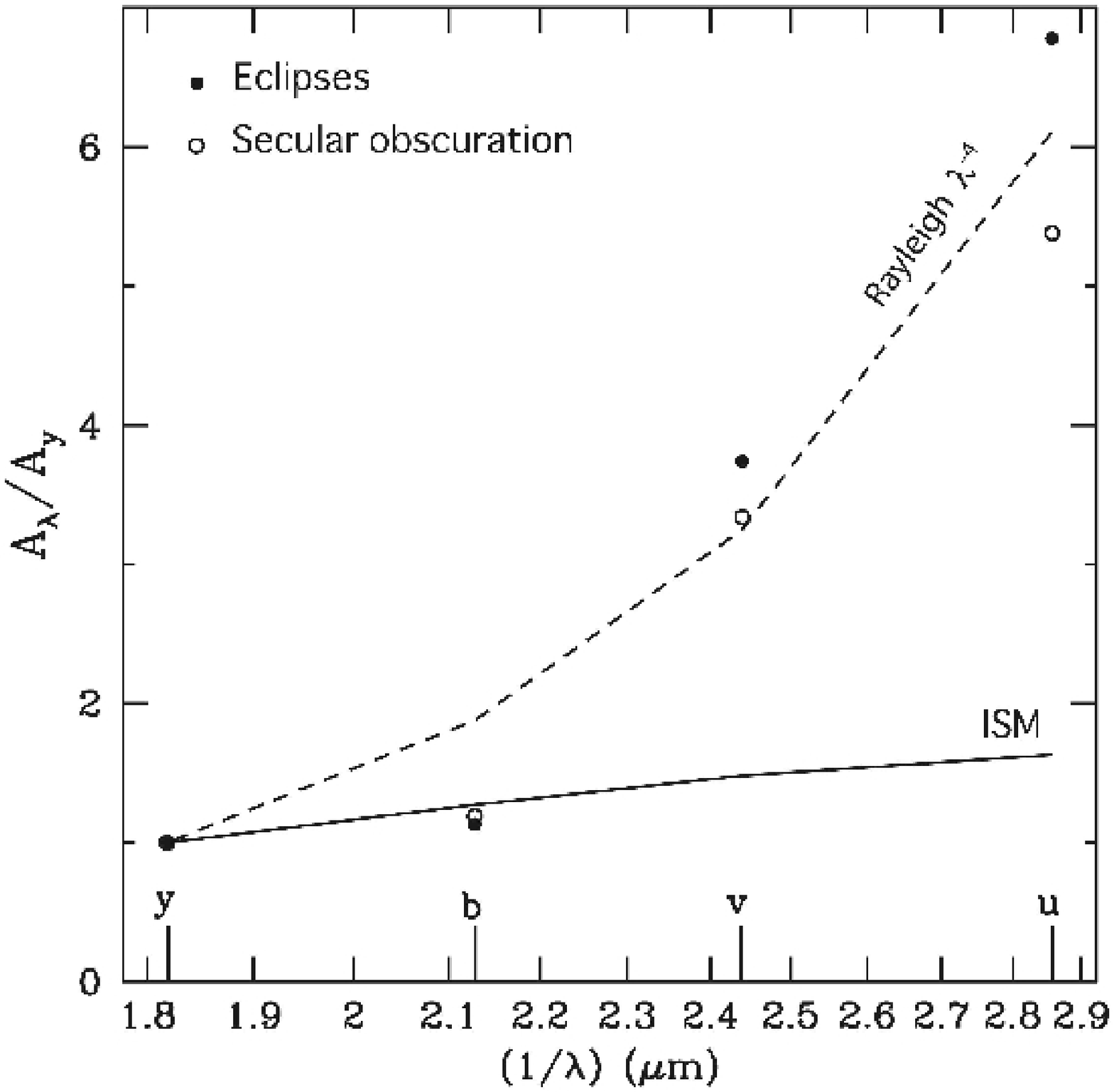}
\caption{\label{Fig:secular}
{\bf Left panel:} The long-term light curve (in the Str\"omgrem bands $uvby$, from bottom to top) of  HD~46407 with respect to the average of the comparison stars (left), and the difference between the two comparison stars (right). The solid line roughly illustrates the secular change in the average  brightness. See \citet{Jorissen-92b} for details. {\bf Right panel:} The amplitude of the secular and eclipse variations in the $u$, $v$ and $b$ bands normalized by the amplitude in the $y$ band, and comparison with the interstellar-medium extinction law and the Rayleigh $\lambda^{-4}$ law. 
}
\end{figure}

The lightcurves of the Ba star HD~46407 and the S star HD~35155  
(see Fig.~\ref{Fig:46407}) are a bit more puzzling.
Barium stars are post-mass-transfer objects with WD companions \citep{McClure-84b,Jorissen-Mayor-88,BohmVitense00} which are too faint to yield  detectable eclipses in the visual.  And yet, the reality of the eclipses in the short-period Ba star HD~46407 ($P = 457$~d)
cannot be doubted, since the eclipses have been observed during different cycles, though with variable depths and phase lags \citep{Jorissen-1991,Jorissen-92b}. A very important clue as to the origin of these eclipses comes from the fact that the deepest eclipse (observed around JD $2\,447\,400$) has occurred just before an episode of {\it secular} obscuration (left panel of Fig.~\ref{Fig:secular}).  The fact that the color dependency of the amplitude for both the eclipse and secular variations follows a $\lambda^{-4}$ law (right panel of Fig.~\ref{Fig:secular}), reminiscent of Rayleigh scattering, at least in the Str\"omgren $u$ and $v$ bands, is a further indication that the phenomenon is caused by (small) dust grains. In $b$ and $y$, it is closer to the ISM color absorption law. 
Despite the fact that barium stars host K giants which do not lose large amounts of mass, the secular obscuration event calls for the presence of {\it recent} dust in the system, rather than for the remnant of the 
spiral arm dating back to the epoch where the current WD companion was a mass-losing AGB star.

A similar eclipsing behaviour has been reported for the S star HD~35155 with 640~d period by \citep{Jorissen-92c}. Despite the fact that \citet{Adelman-2007} did not confirm the occurrence of the eclipse, the two IUE spectra of HD~35155 obtained by chance at the time of eclipse have the smallest flux among the 10 available IUE     spectra, though neither the continuum nor the emission lines disappear completely \citep{Ake-91,Jorissen-96}. 

HD~121447 is the coolest barium star (K7III) with the second shortest orbital period (186~d). 
Ellipsoidal variability was first suspected for this star \citep{Jorissen-95}, but the absence of synchronous rotation and a strong color dependence of the light curve 
\citep{Adelman-2007} 
do not support this conclusion, so that light variations caused by scattering on dust formed by the compression of the red-giant wind in the accretion process, becomes an attractive possibility.

%% \begin{figure}
%% \begin{minipage}{0.5\textwidth}
%% \includegraphics[height=.12\textheight]{ITVir_1.eps}
%% %\end{minipage}
%% \mbox{}\\
%% %\vspace{1pt}\\
%% %\begin{minipage}[b]{0.5\textwidth}
%% \includegraphics[height=.1\textheight]{ITVir_2.eps}
%% %\end{minipage}
%% \mbox{}\\
%% %\vspace{1pt}\\
%% %\begin{minipage}{0.5\textwidth}
%% \includegraphics[height=.1\textheight]{ITVir_3.eps}
%% \end{minipage}
%% \caption{\label{Fig:121447}
%% The light curve of the barium star HD~121447 phased with the orbital period ($P = 186$~d). (From \citep{Adelman-2007}).}
%% %*AF*
%% % I would consider removing this figure:
%% % Due to the vertical extension this figure wastes a lot of space.
%% % At the same time, it does not bring that much new information (the
%% % previous figure already gives Ba star lightcurve).
%% % The "strong color dependence" advertised in the text isn't that obvious
%% % to me from this graph.
%% %%%
%% \end{figure}

To summarize, disks (often circumbinary)
 are observed in many different classes of binary stars involving a mass-losing star, and seem to be very common in such circumstances.
They play an important role in the evolution of such systems:\\
- they control the evolution of the eccentricity through tidal effects and angular momentum exchange with the orbit \citep{Artymowicz91}; \\
- they trigger a dust/gas segregation, so that the abundance pattern observed at the surface of the post-AGB star is shaped by 
the condensation temperature of the various elements \citep{Maas-2005}.
It is believed that this physical segregation operates through re-accretion by the post-AGB star of gas depleted in 
refractory elements which stayed in the dust phase. It leads to very Fe-depleted post-AGB atmospheres (down to [Fe/H] $=-4.8$) \citep{Waelkens-1991:a}.
%%%AJ
Intriguingly, a similar scenario has recently been proposed \citep{Venn-2008} as the possible cause of the ultra low metallicities observed in several stars from the Hamburg/ESO survey \citep{Reimers-1997}, like the record holder HE~1327-2326 with [Fe/H] $=-5.6$ \citep{Venn-2008,Frebel-2005}. This scenario would require these stars to be binaries. There is no definite evidence thereof so far, but the available radial-velocity data is not necessarily conclusive \citep{Venn-2008}.
%%%- they provide a stable environnement to anneal amorphous dust grains into %crystalline ones %\citep{Molster-1999,DeRuyter-2006,VanWinckel-2006,Gielen-2007,Deroo-2007,Edgar-%2008}. 

\section{Astrometry}
\label{Sect:astrometry}

\subsection{The basics}

Very long-period systems are hard to detect with spectroscopy or photometry.
This is where astrometry comes to rescue.
Astrometric binaries differ from the more traditional visual binaries
in the sense that for visual binaries, both components are observed so that
it is the {\it relative} orbital motion (of one component with respect to
the other) which is readily detected.
For astrometric binaries instead, it is the {\it absolute} (non-linear) motion of one component on the sky which is detected. Space astrometry, which started with the Hipparcos satellite \citep{ESA-1997} launched in 1989, has reached the milliarcsecond (mas) accuracy level for stars down to about $V = 10$, and the coming Gaia satellite (to be launched in 2011) should improve this accuracy by a factor of 100. Many achievements have already been made in the field of binary stars by Hipparcos, as reviewed by \citet{Perryman-2008}, and many more must be expected from Gaia. We will discuss just a few here, illustrating the potential of astrometry in the discovery of specific kinds of binaries \citep[A very extensive review of this potential
is presented in ][]{Perryman-2008}. For instance, since astrometric methods  are independent of spectral types (as opposed to spectroscopic methods, since the detection of spectroscopic binaries is dependent upon the possibillity to follow accurately the variations of the position of spectral lines), astrometry offers a way to derive  unbiased frequencies of binaries of different spectral types \citep[Table~\ref{Tab:freq};][]{Frankowski-2007}. Substellar companions may also be detected thanks to astrometry, if accurate enough \citep{Kaplan-Makarov-2003,Makarov-2004,Makarov-Kaplan-2005}. Astrometry offers as well good prospects to detect binaries involving Mira variables (which are especially difficult to find with spectroscopy), provided however that spots or an inhomogeneous surface brightness or asymmetries in the shape of the stellar disk of these very extended stars do not cause variations of the position of the photocentre that  confuse the parallactic and orbital motion \citep{Bastian-2005,Eriksson-2007}.

\begin{figure}
\includegraphics[height=.3\textheight]{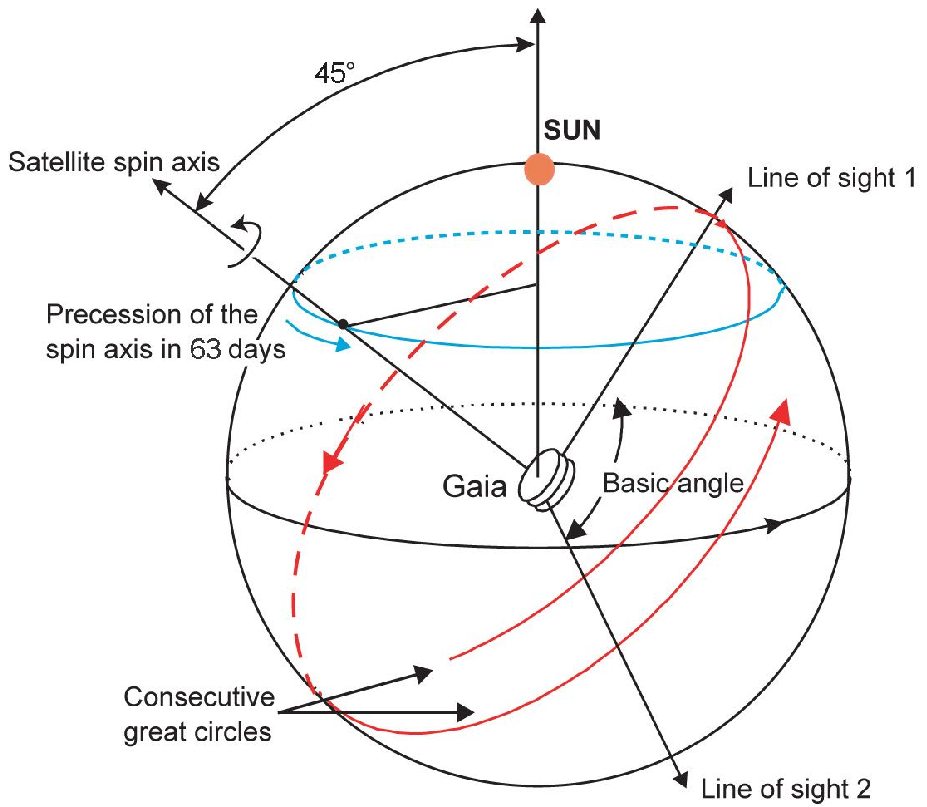}
\includegraphics[height=.3\textheight]{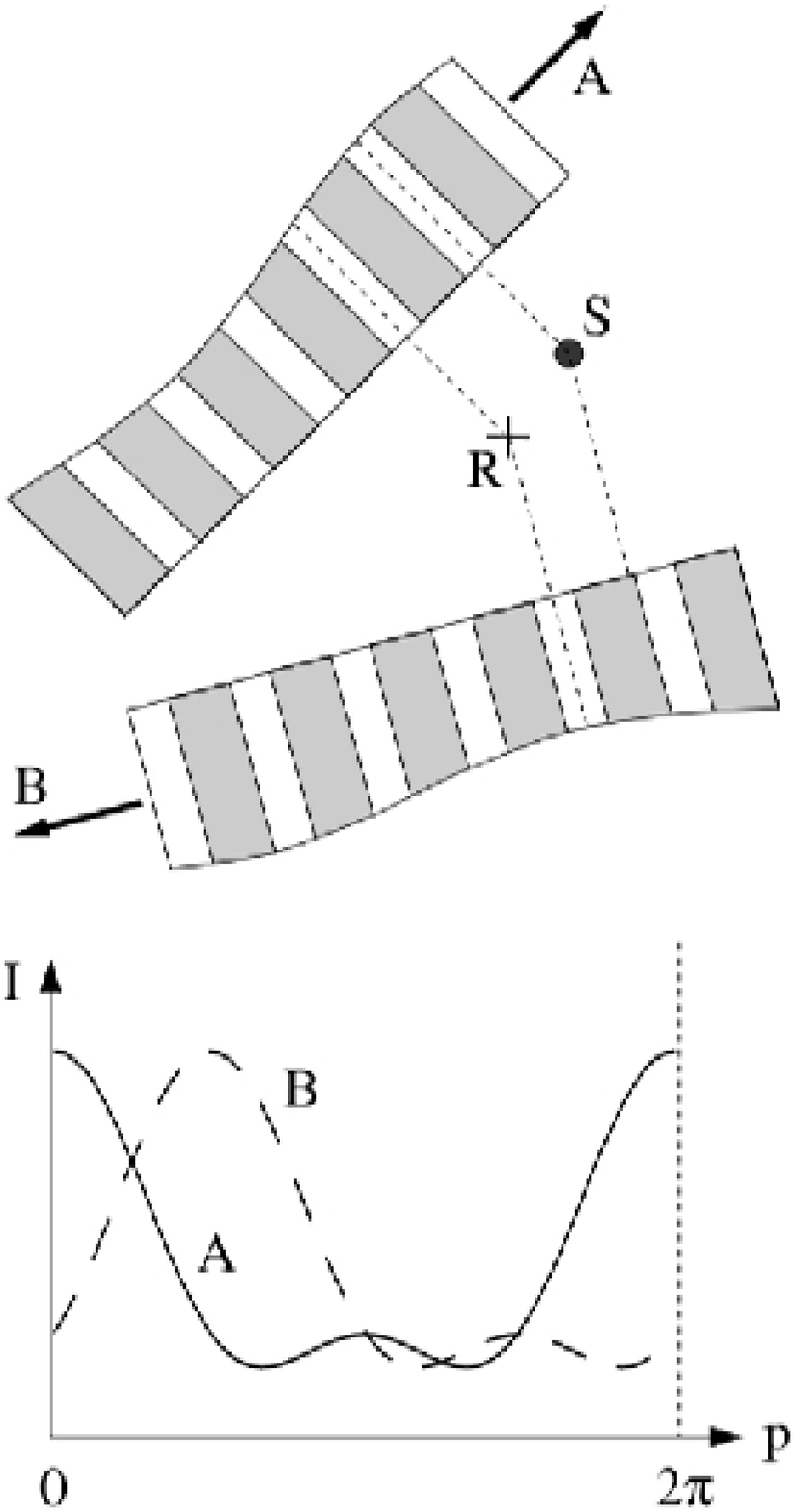}
\caption{\label{Fig:Hipparcos}
{\bf Left panel:} The scanning law of Gaia. (From the ESA-Gaia web site)
{\bf Right panel:} The modulating grid in the Hipparcos focal plane (top), for two different scanning directions A and B, producing the modulated signals displayed on the bottom panel.  The phase $p$ for the source $S$ is measured differentially with respect to that of the reference point $R$, modulo the grid step. In reality, the grid extends over the images of $S$ and $R$ in the focal plane, although this was not displayed for clarity. (From \citep{Quist-1999})
}
\end{figure}

\begin{figure}
\includegraphics[height=.3\textheight]{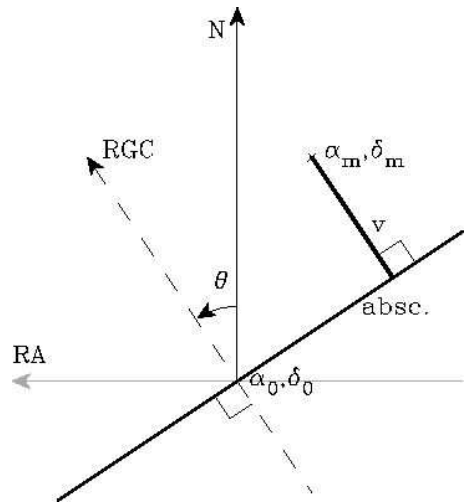} 
\caption{\label{Fig:Hipparcos2}
Definition of the abscissa offset $v$ for a predicted position $(\alpha_m, \delta_m)$, with respect to the observed position somewhere along the thick line perpendicular to the reference great circle labeled 'RGC'. Remember that the measurement is one-dimensional. 
 (From \citep{vanLeeuwen-1998:a})
}
\end{figure}

The principle of the astrometric measurement by Hipparcos and Gaia needs to be explained first.
In order to achieve a good accuracy overall on the celestial sphere,
Hipparcos had two fields of view, 0.8 degree square each, widely separated
on the sky (by about 58 degrees of arc) and superimposed on the detector.
In the case of Gaia, the numbers will be 0.37 degree square and 106.5
degrees, respectively (left panel of Fig.~\ref{Fig:Hipparcos}). 
%*AF*
% the figure referenced here is a bit cryptic to me, as noted above.
Since the satellite is spinning, this signal from stars crossing the
instrument fields of view is modulated by a one-dimensional grid in
the focal surface (right panel of Fig.~\ref{Fig:Hipparcos}).
%%%
As the satellite scanned 
the sky in a complex series of precessing great circles, maintaining a constant inclination 
to the Sun's direction, a continuous pattern of one-dimensional measurements was 
built up. In order words, it is the 'abscissa' along the great circle which is being measured, the position of the star along the perpendicular to the reference great circle being unknown. These one-dimensional abscissa measurements are then confronted to the expected positions of the star on the sky as a function of time, for a given reference position, parallax and proper motion for that star. The difference between these one-dimensional positions is called 'abscissa residual', and is noted $v$ on Fig.~\ref{Fig:Hipparcos2} \citep{vanLeeuwen-1998:a}. These abscissa residuals are then combined with the variance-covariance matrix of the observational errors to yield a $\chi^2$ value characterizing the quality of the solution \citep{vanLeeuwen-1998:a,Pourbaix-2000:b}.

\begin{figure}
\includegraphics[height=.3\textheight]{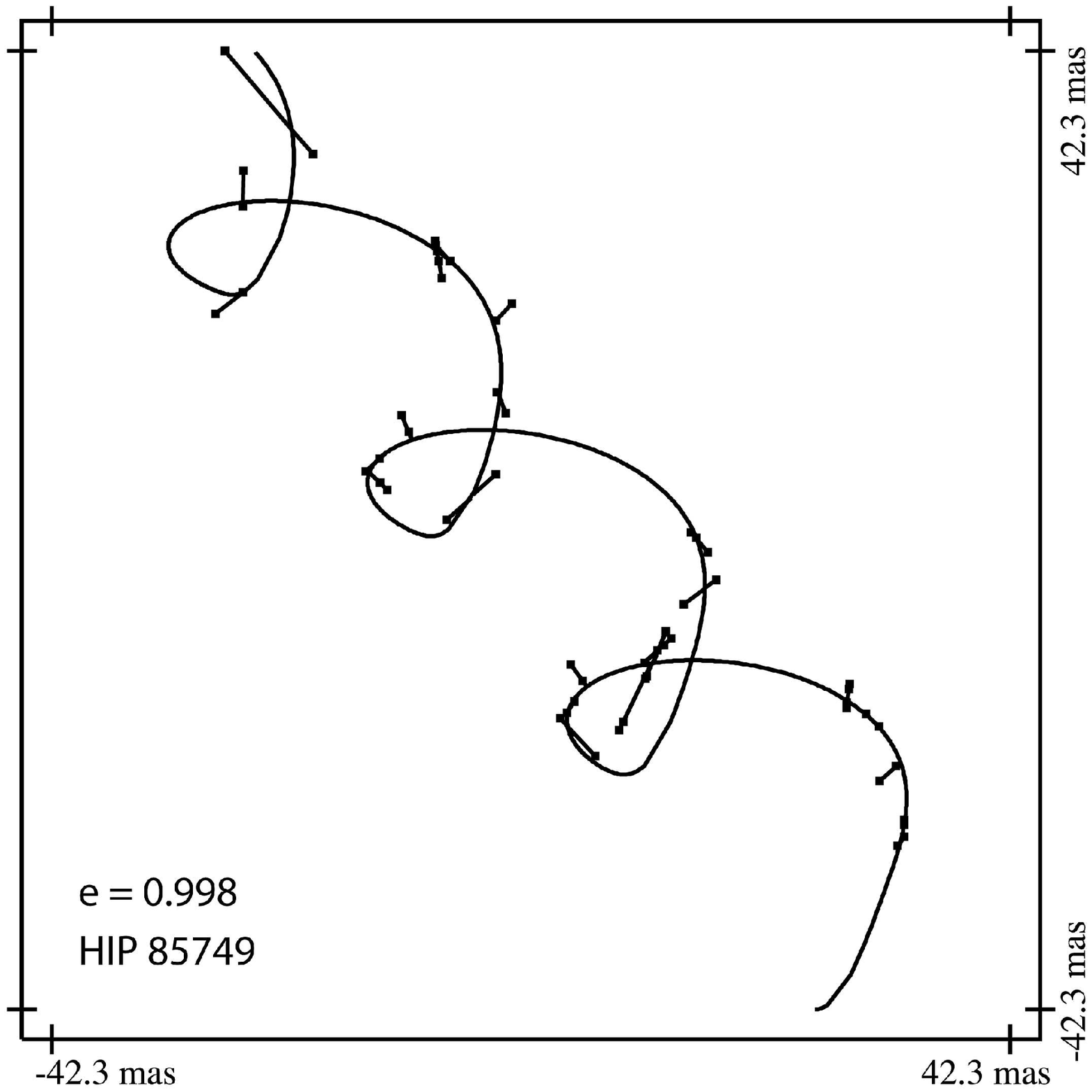}
\includegraphics[height=.3\textheight]{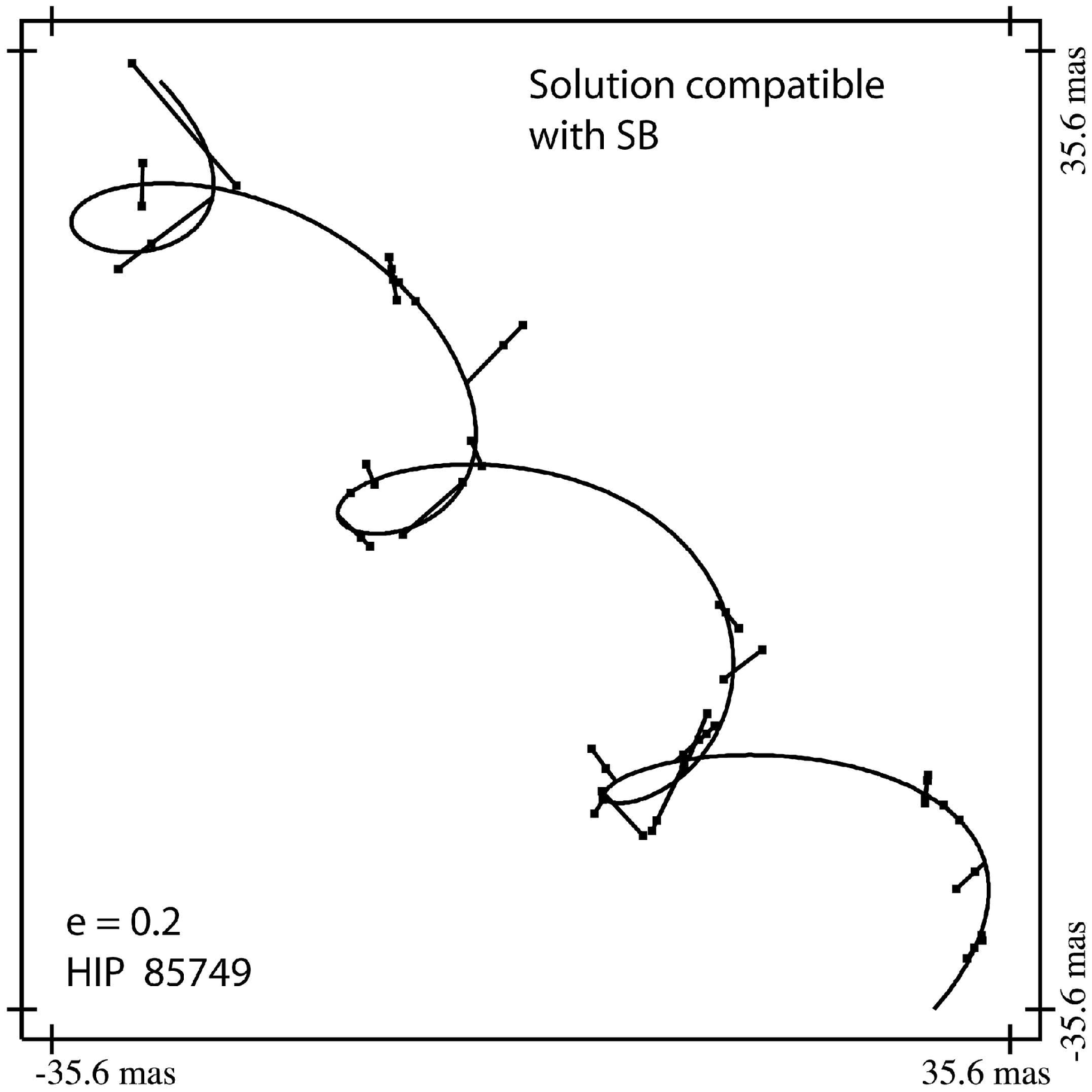}
\caption{\label{Fig:inc}
{\bf Left panel:} A solution fitting the astrometric motion of HIP~85749 with a quasi-parabolic orbit ($e = 0.998$) aligned with the line of sight. {\bf Right panel:} The orbital solution with $e = 0.2$, consistent with the spectroscopic orbital elements. Both solutions have similar goodness-of-fit values. The small line segments connect the predicted position with the observed great-circle abscissa (see Fig.~\ref{Fig:Hipparcos}). 
(From \citep{Pourbaix-2002})
}
\end{figure}

Right away, two difficulties become apparent in the context of binary stars:\\ (i) It is the position of the {\it photocentre} of the system which is measured, reducing to the position of the brightest component if the difference in brightness is large. It may be shown \citep{Binnendijk-1960} that the relation between the semi-major axis $a_0$ of the orbit of the photocentre around the centre-of-mass and semi-major axis $a$ of the relative orbit writes
\begin{equation}
\label{Eq:ap}
a_0 = a (\kappa - \beta),
\end{equation}
where $\kappa = M_2/(M_1+M_2)$ and $\beta = 1/(1+10^{0.4\Delta m})$, where  
$\Delta m = m_2 - m_1$, so that when component 2 is much fainter than component 1 (i.e., $\Delta m \rightarrow \infty$),
$\beta = 0$ and $a_0 = a\kappa$ is then the semi-major axis of the orbit of component 1 around the centre-of-mass. 
Therefore the access to the masses is not straightforward from
astrometric orbits. A summary of what may be known about the masses
for the various kinds of binaries (and combinations thereof) is listed
in Table~\ref{Tab:masses} in Sect.~\ref{Sect:summary}. 
\medskip\\
(ii) The one-dimensional nature of the astrometric (Hipparcos or Gaia) measurement imposes further limitations on the derivability of the orbital elements. As discussed by \citet{Pourbaix-2002}, two-dimensional observations 
make it possible to draw an orbit projected on the plane of the sky, and to derive the areal constant $\Gamma'$
of Kepler's second law
corresponding to the motion of the star on that projected orbit. Since the law of areas holds in both the true and projected orbits, one may write $\Gamma' = \Gamma \cos i$, where $\Gamma$ is the areal constant in the true orbit, which is related to the orbital elements through the relation $\Gamma = a^2 (1-e^2)^{1/2} 2\pi / P$ and may thus be derived from these. The ratio $\Gamma'/\Gamma$ then yields the inclination. With one-dimensional measurements, $\Gamma'$, and therefore $i$,  cannot be derived accurately. In those circumstances, the derived $\Gamma'$ value is often close to 0, so that $i$ close to 90$^\circ$ follows (edge-on orbits). To accomodate the limited arc span on the sky with this spurious edge-on orbit, there is no other possibility than an eccentricity very close to unity (quasi-parabolic orbit), a  very large semi-major axis and a longitude of periastron close to $90^\circ$ (apsidal line aligned with the line of sight). This is illustrated 
in Fig.~\ref{Fig:inc}, which compares two possible solutions (having similar goodness-of-fit values) fitting the astrometric motion of HIP~85749. The left panel is a solution with a quasi-parabolic orbit ($e = 0.998$), an inclination close to $90^\circ$ and a large semi-major axis, whereas the right panel displays the solution with $e = 0.2$ consistent with the spectroscopic orbital elements.

\begin{figure}
\includegraphics[height=.3\textheight]{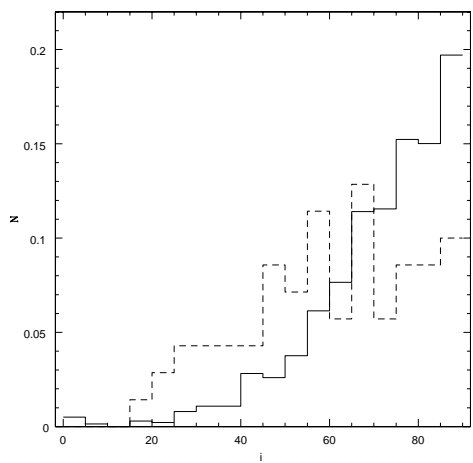}
\caption{\label{Fig:i}
Solid line: The distribution of orbital inclinations (from the Thiele-Innes set of orbital elements) for the 1385 orbits from the Ninth Catalogue of Spectroscopic Binary Orbits \citep{Pourbaix-04a} with an Hipparcos entry. Many of those astrometric orbital solutions are spurious, with inclinations close to $90^\circ$. Dashed line: The same for the solutions passing all consistency tests defined by \citet{Jancart-2005}.
}
\end{figure}

Even when spectroscopic orbital elements are known, spurious astrometric orbits with an inclination close to $90^\circ$ may still arise, as found by \citet[][see Fig.~\ref{Fig:i}]{Jancart-2005}.  These authors have  designed statistical tests to identify those spurious solutions. A more detailed description of these astrometric methods to detect binaries is given in the next section.

\subsection{Detecting binaries from astrometric data}
\label{Sect:IAD}

A tailored reprocessing of the Hipparcos  (and in the future, Gaia) {\it Intermediate Astrometric Data}
\citep[hereafter IAD; ][]{vanLeeuwen-1998:a} makes it
possible to look for orbital signatures in the astrometric motion,
following the method outlined in
\citep{Pourbaix-2000:b}, \citep{Pourbaix-Boffin:2003},
\citep{Pourbaix-2004} and applied to barium stars in \citep{Jorissen-Zacs-2005} and \citep{Jorissen-2004}.

The basic idea is to quantify the likelihood of the fit of the Hipparcos
IAD with an orbital model. For that purpose,
\citet{Pourbaix-2001:b} \citep[see also][]{Jancart-2005}
introduced several  statistical indicators  
to decide whether to keep or to discard an orbital solution.
The relevant criteria are as follows
(we keep the notation from \citep{Jancart-2005}):
\begin{itemize}
\item The addition of 4 supplementary parameters (the four Thiele-Innes orbital 
constants; see Eq.~\ref{Eq:xy} and Sect.~\ref{Sect:elements}) describing the orbital motion should result in a statistically
significant decrease of the $\chi^2$ for the fit of the $N$ IAD with an orbital
model with 9 free parameters ($\chi^2_T$), 
as compared to a fit with a single-star solution with 5 free parameters ($\chi^2_S$, the 5 parameters being the the positions $\alpha$, $\delta$, the proper motions $\mu_\alpha$, $\mu_\delta$ and the parallax $\varpi$).
This criterion is expressed by an $F$-test:
\begin{equation}
Pr_2 = Pr[F(4,N-9) > \hat{F}],
\end{equation}
where
\begin{equation} 
\hat{F} =  
\frac{N-9}{4}\;\; \frac{\chi^2_S - \chi^2_T}{\chi^2_T}.
\end{equation}
$Pr_2$ is thus the first kind risk associated with the rejection of the null 
hypothesis: ``{\em there is no orbital wobble present in the data}''.

\item Getting a substantial reduction of the $\chi^2$ with the Thiele-Innes model 
does not necessarily   imply that the four Thiele-Innes constants $A,B,F,G$ are
significantly different from 0. The first kind risk associated with the rejection
of the null hypothesis ``{\em the orbital semi-major axis is equal to zero}'' may
be expressed as
\begin{equation}
Pr_3 = Pr[\chi^2(4) > \chi^2_{ABFG}],
\end{equation}
where
\begin{equation}
\chi^2_{ABFG} =  \chi^2_{S} -  \chi^2_{T}
\end{equation}
and $\chi^2(4)$ is a random variable following a $\chi^2$ distribution with
four degrees of freedom.

\item For the orbital solution to be   a significant one, its parameters should
not be strongly correlated with the  other astrometric parameters
(e.g., the proper motion). In other words, the covariance matrix ${\bf V}$ of
the astrometric solution should be dominated by its diagonal terms, as
measured by the {\it efficiency}
$\epsilon$ of the matrix being close to 1 \citep{Eichhorn-1989}. The efficiency is expressed by 
\begin{equation}
\label{Eq:epsilon}
\epsilon = \sqrt[m]{\frac{\Pi_{k=1}^m \lambda_k}{\Pi_{k=1}^m {\bf V}_{kk}}},
\end{equation}
where $\lambda_k$ and ${\bf V}_{kk}$ are respectively the eigenvalues and the 
diagonal terms of the covariance matrix ${\bf V}$.
\end{itemize}

In other words, an orbital solution will be most significant if
$Pr_2$ and $Pr_3$ are small and $\epsilon$ large.
With the above  notations, the requirement for a star to
qualify as a binary
was defined as
\begin{equation}
\label{Eq:alpha}
\alpha \equiv (Pr_2 + Pr_3)/\epsilon \le 0.02,
\end{equation}
where the threshold value of 0.02 has been chosen to minimize false
detections, as derived from the application of the method to barium stars
\citep[see below and ][]{Jorissen-2004}. 

Hipparcos data are, however, seldom precise enough to derive
the orbital elements from scratch. Therefore, when a spectroscopic orbit is
available beforehand, it is advantageous to import $e, P, T_0$ from the
spectroscopic orbit and to derive the remaining astrometric elements
\citep[as done in ][]{Pourbaix-2000:b,Jancart-2005,Pourbaix-Boffin:2003}. 
The above scheme has been used by \citet{Jancart-2005} to search for an orbital signature in the astrometric motion of binary systems belonging to the {\it Ninth Catalogue of Orbits of Spectroscopic Binaries} \citep[\SB9; ][]{Pourbaix-04a}, and gives best results for systems with orbital periods in the range 100 -- 3000~d and  parallaxes in excess of 5~mas.

\begin{figure}
\includegraphics[height=.3\textheight]{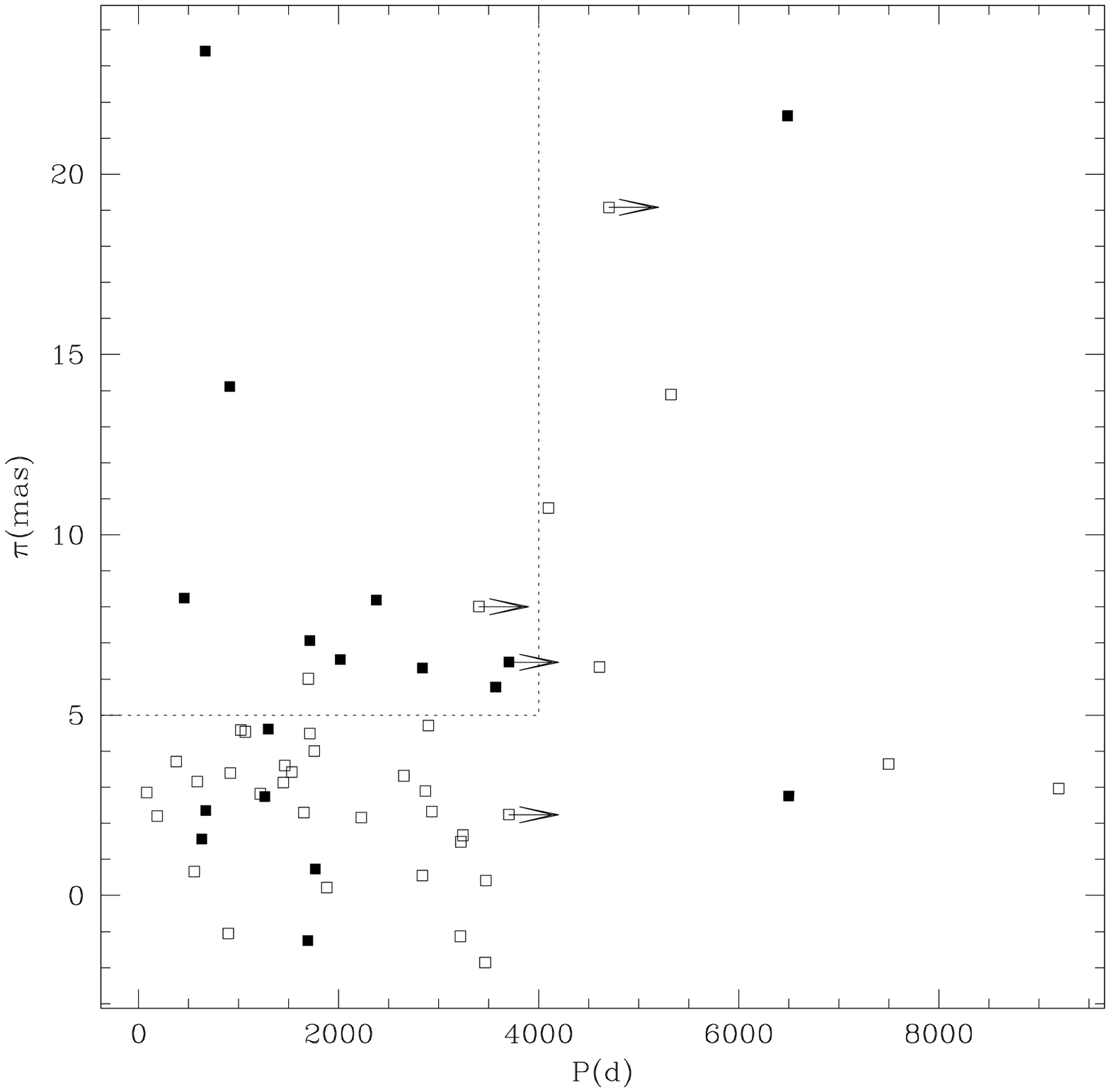}
\includegraphics[height=.3\textheight]{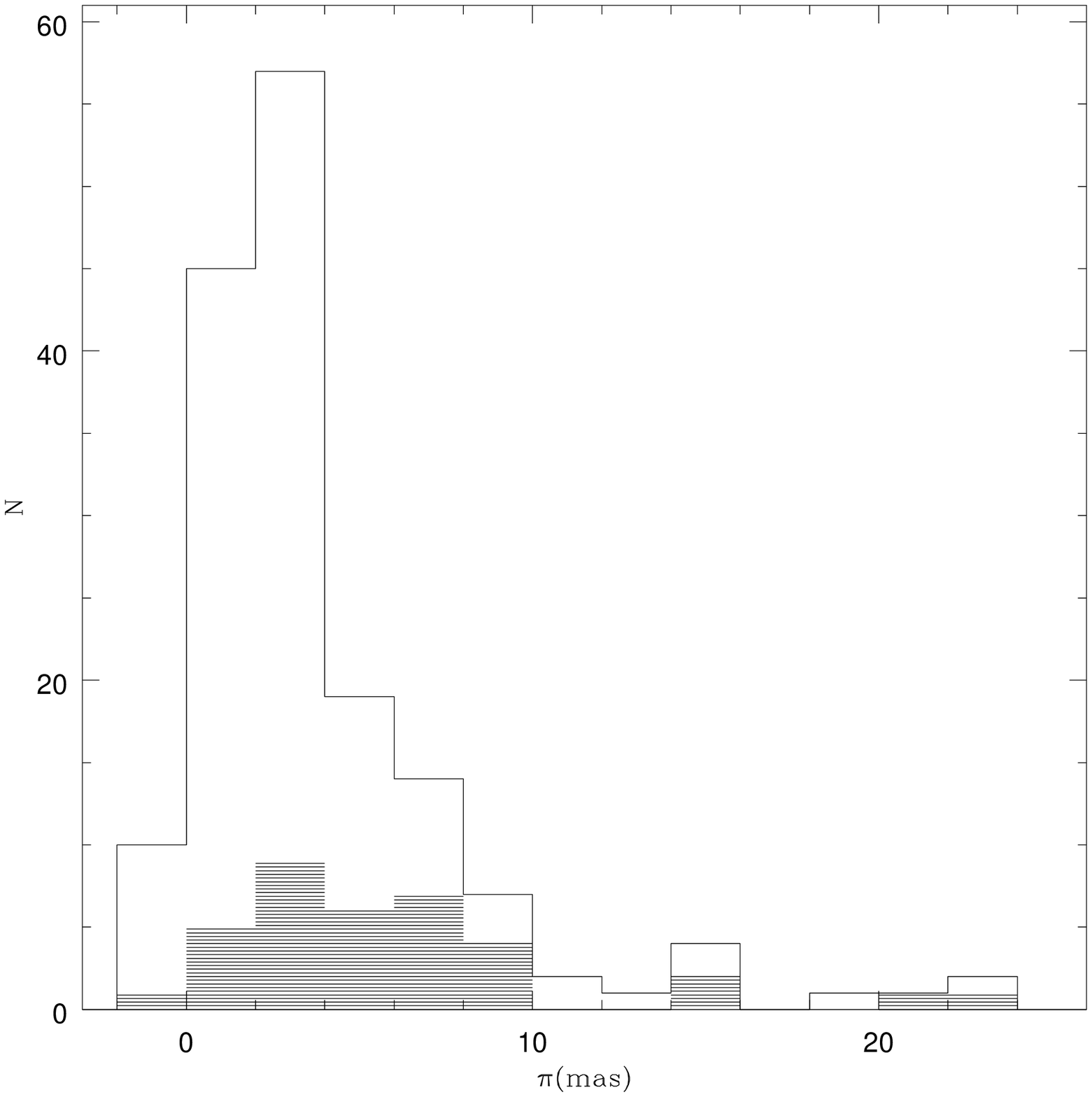}
\caption{\label{Fig:Dubrovnik}
{\bf Left panel:} 
Barium stars (previously known to be SBs) 
flagged as astrometric binaries by the algorithm are represented by
  black symbols.  
Arrows denote stars with only a lower limit available on the period. 
{\bf Right panel:} Fraction of stars flagged as binaries (shaded
histogram) compared to 
total number of stars, as a function
of parallax.  
As expected, the detection rate becomes high for $\varpi > 5$~mas.
(From \citep{Jorissen-2004})
}
\end{figure}

If a spectroscopic
orbit is not available, trial $(e, P, T_0)$ triplets scanning a regular grid (with
$10 \le P (\rm d) \le 5000$ imposed by the Hipparcos scanning law and the mission
duration) may be used. The quality factor
$\alpha$ (Eq.~\ref{Eq:alpha}) is then computed for each trial $(e, P, T_0)$ triplet, and if
there exist triplets yielding $\alpha < 0.02$, 
the star is flagged as a binary (In fact, scanning only $P$ with $e$ set to zero allows to save considerable computing time and would only miss very eccentric binaries).
To test its success rate, this method has been applied by \citet{Jorissen-2004} on a sample of barium stars. 
Barium stars constitute an ideal sample to test this algorithm, because
they are all members of binary systems \citep{Jorissen-VE-98,McClure-83}, with periods ranging 
from about 
100~d to more than 6000~d. The catalogue of \citet{Lu-83} contains  163 {\it bona fide}
barium stars with an Hipparcos entry
(excluding the supergiants HD~65699 and HD~206778 = $\epsilon$~Peg).
When  
$\varpi > 5$~mas and  $100 < P (\rm d)< 4000$, 
the (astrometric) binary detection rate is close to 100\%, {\it i.e.}, the astrometric
method recovers all known spectroscopic binaries (Fig.~\ref{Fig:Dubrovnik}).
When considering the 
whole sample, the detection rate falls to 22\% (= 36/163) 
 because many barium stars have small parallaxes or very long periods. 
Astrometric orbits with $P > 4000$~d ($>11$~yrs)
can generally not be extracted from the Hipparcos IAD, which span only 3~yrs 
(see left panel of Fig.~\ref{Fig:Dubrovnik}).
Similarly, when $\varpi < 5$~mas, as have most barium stars  (right panel of Fig.~\ref{Fig:Dubrovnik}), 
the Hipparcos IAD  are not precise enough to extract
the  orbital motion.

\begin{figure}
\includegraphics[height=.3\textheight]{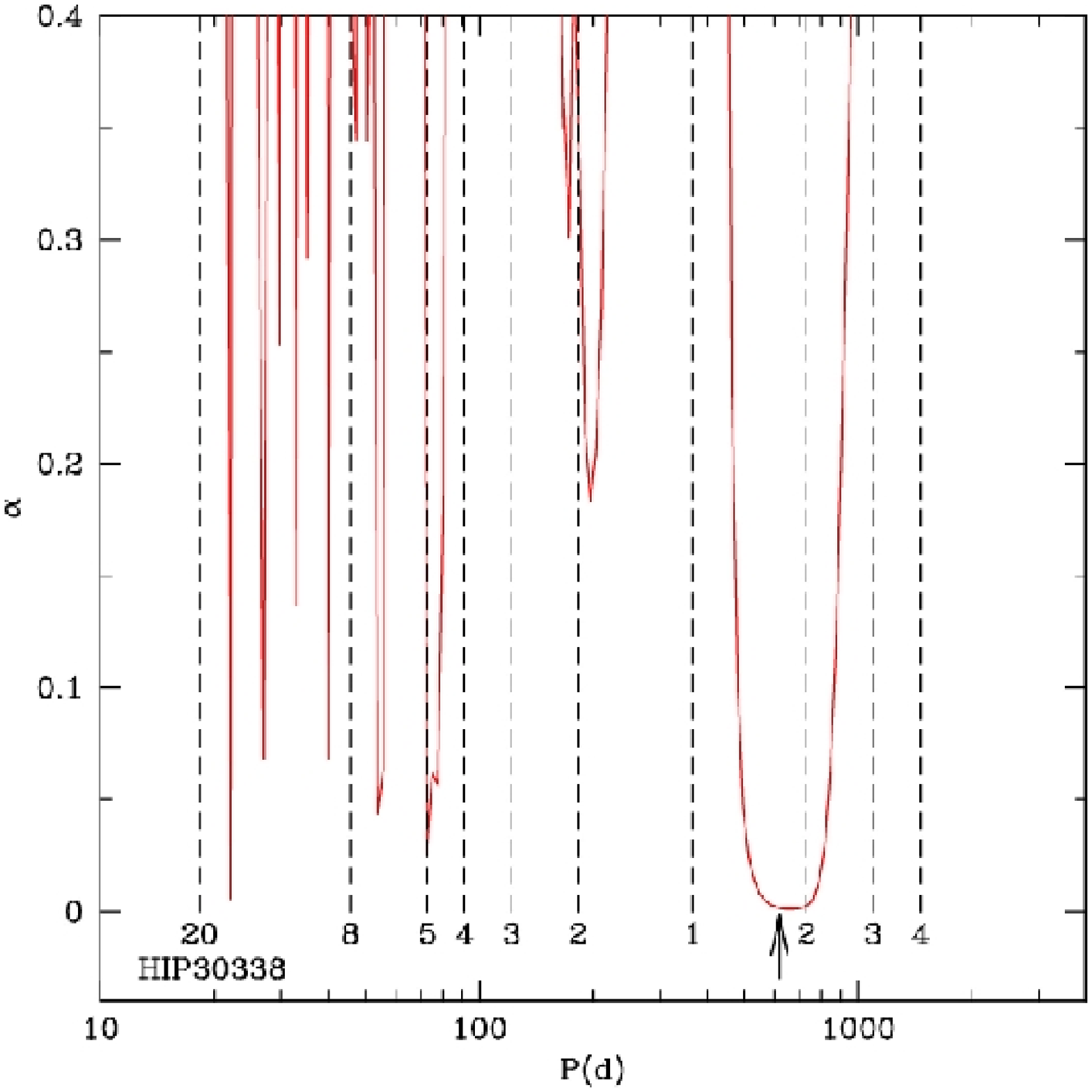}
\includegraphics[height=.315\textheight]{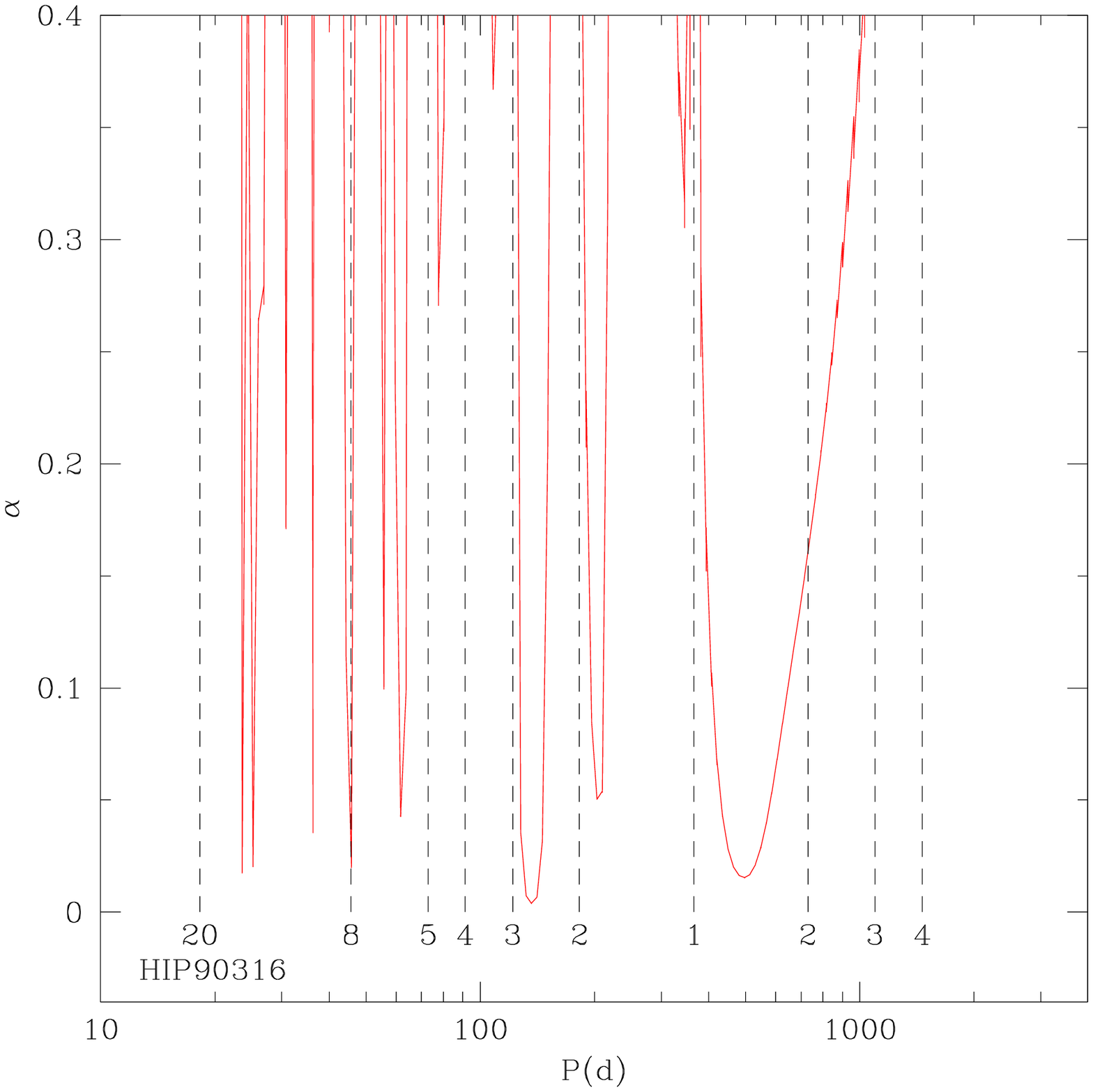}
\caption{\label{Fig:periods}
The $\alpha$ statistics (see Eq.~\ref{Eq:alpha}) 
as a function of the trial orbital period (assuming $e
= 0$) for the $P = 629$~d spectroscopic binary HIP 30338 (left panel). The vertical arrow indicates the spectroscopic period, which falls in the region of small $\alpha$. By comparison, a period of $\sim 500$~d may be inferred for HIP~90316 (right panel).
The vertical dashed lines represent multiple, or integer fractions, of 1~yr.
At those periods, there 
is a strong correlation between the parallactic and orbital signals, which degrades the $\alpha$ statistics 
and makes binaries difficult to find.
} 
\end{figure}

When the orbit is not known beforehand, the method makes it even
possible to find a good estimate for the orbital period, provided that
the true period is not an integer fraction, or a multiple, of one year. 
Because parallactic motion has a 1-yr period, any orbital motion with
a period corresponding to a (sub)multiple of 1~yr will not be easily
disentangled from the parallactic motion. Therefore, for those
periods,
the parallax will be strongly correlated with the orbital parameters
so that the efficiency $\epsilon$ (Eq.~\ref{Eq:epsilon}) will be
small, causing $\alpha$ to be large 
(see Fig.~\ref{Fig:periods}). 

\begin{figure}
\includegraphics[height=.2\textheight]{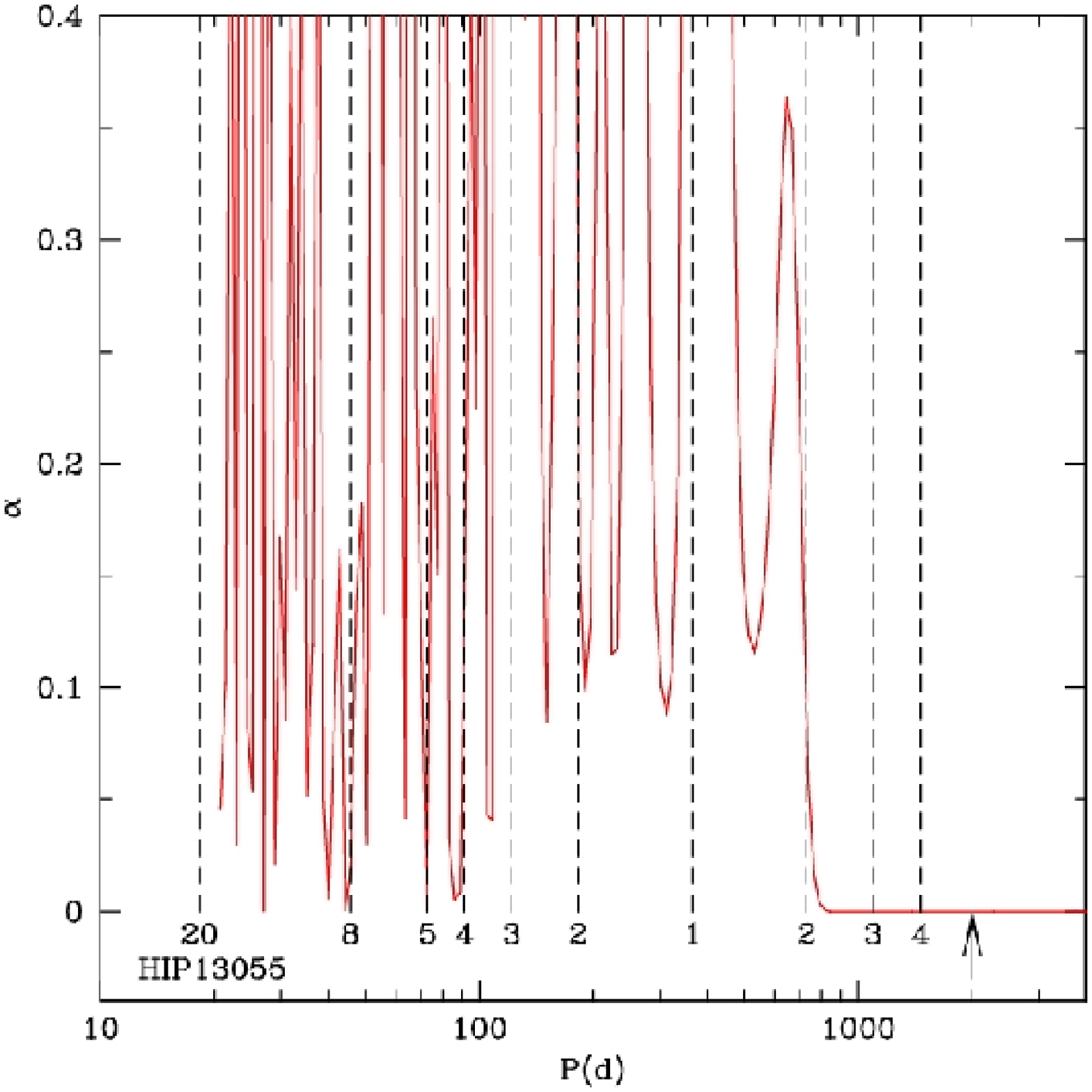}
\includegraphics[height=.201\textheight]{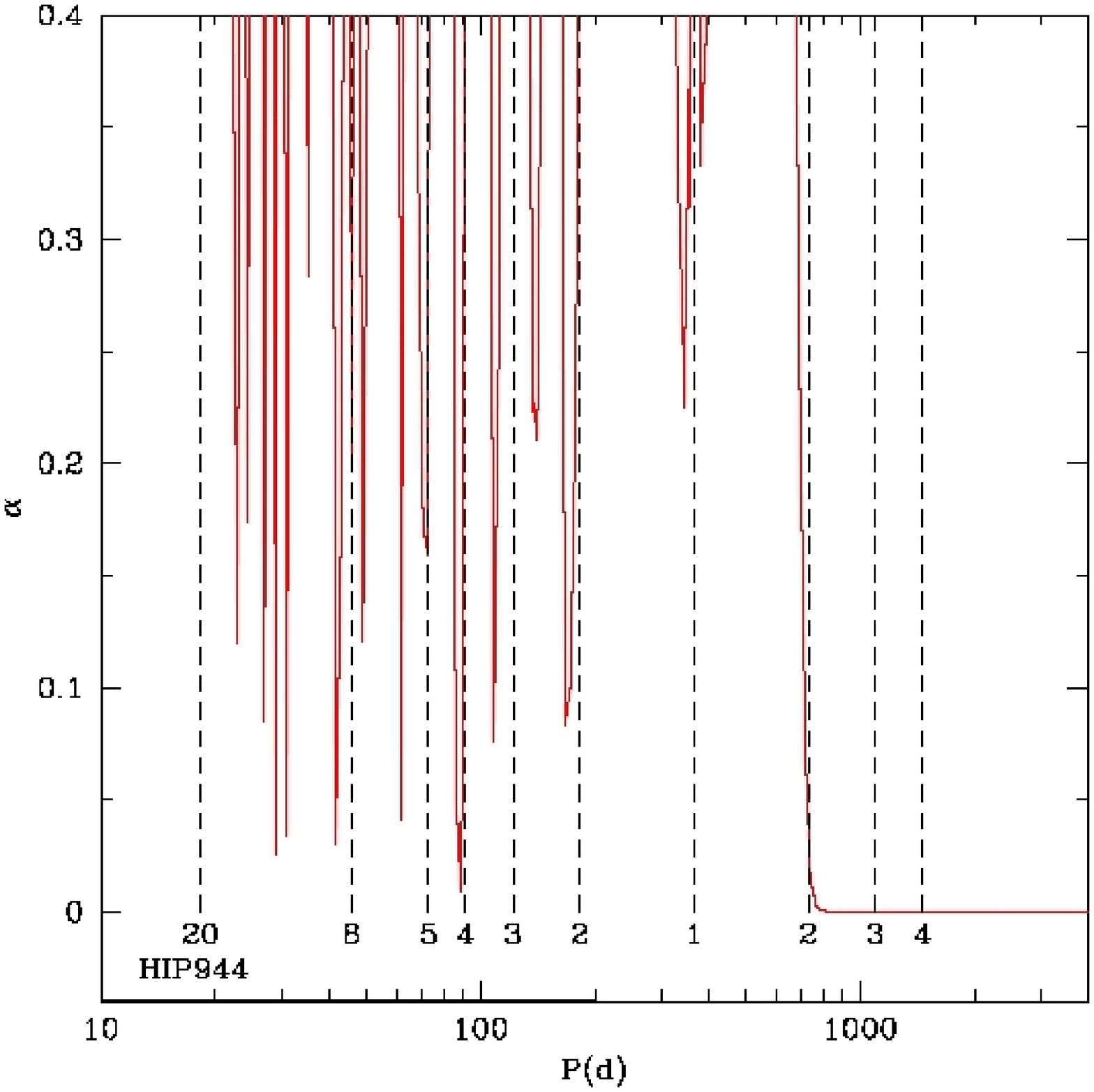}
\includegraphics[height=.21\textheight]{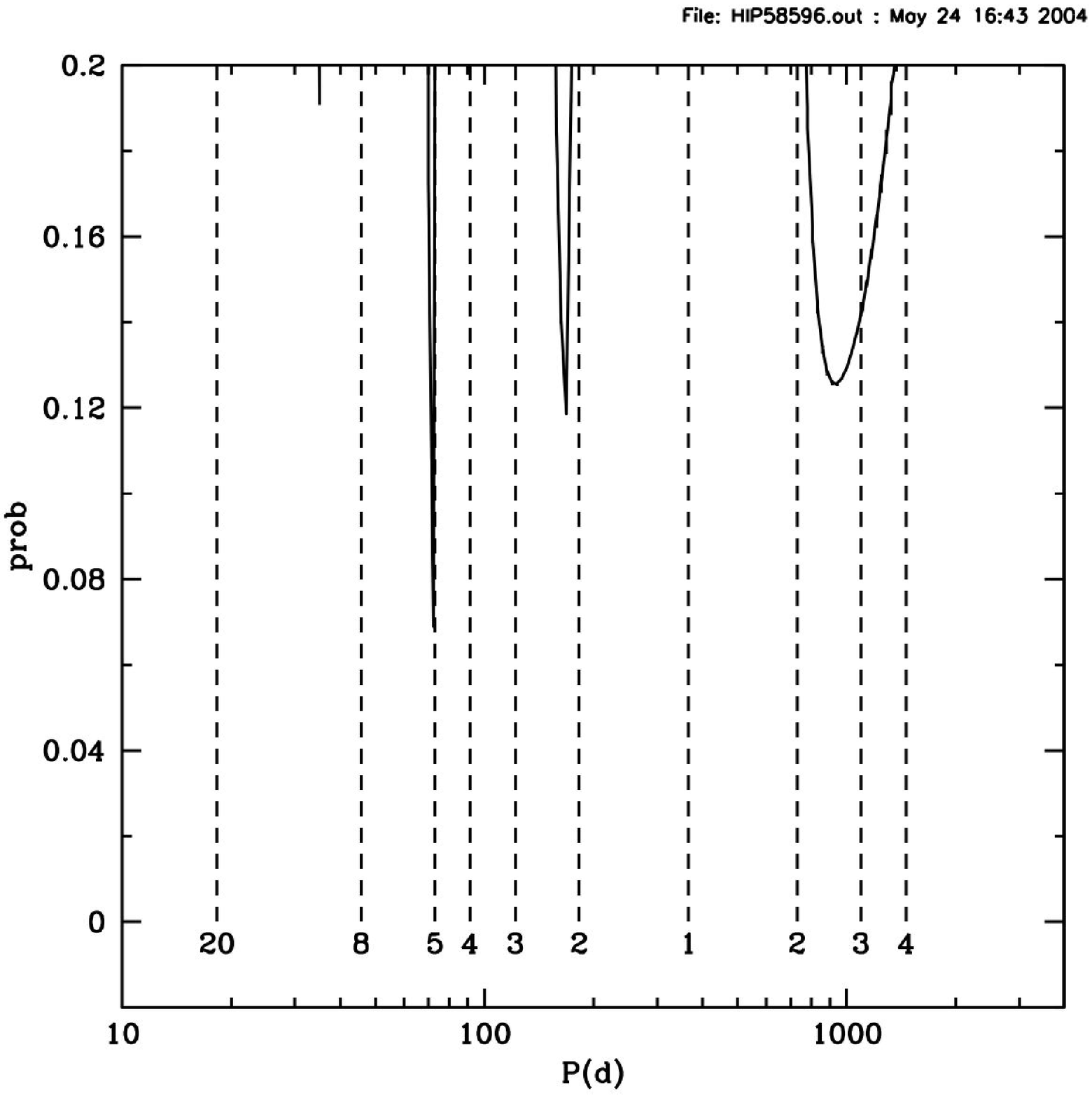}
\caption{\label{Fig:periods2}
Same as Fig.~\ref{Fig:periods} for the $P = 2018$~d spectroscopic binary HIP~13055 (left panel) and for HIP~944 (middle panel) for which $P > 800$~d may be inferred. For comparison, the right panel displays the $\alpha$ statistics for HIP~58596, a star with no evidence for binarity, neither from astrometry nor from spectroscopy \citep{Jorissen-Zacs-2005,Drake-2008}.
}  
\end{figure}

Inspection of the run of $\alpha$ versus
the trial periods $P$ clearly shows that $\alpha$ is minimum in the vicinity of the true (spectroscopic) orbital period (see left panel of Fig.~\ref{Fig:periods}). The period may thus be guessed by looking at the region where $\alpha$ is small. Conversely, for stars with no evidence of binarity from a radial-velocity monitoring,  neither does  astrometry find a period range where $\alpha$ becomes small (right panel of Fig.~\ref{Fig:periods2}). 

Thus, an interesting astrophysical outcome of the algorithm
is a list of classical and metal-deficient barium stars shown to be astrometric 
binaries by the analysis of the Hipparcos
IAD \citep{Jorissen-Zacs-2005,Jorissen-2004}, and which were not
subject to radial-velocity monitoring so far.

\subsection{$\Delta \mu$ binaries: Comparing Hipparcos and Tycho-2  proper
motions} 
\label{Sect:Tycho}

\begin{figure}
\includegraphics[height=.3\textheight]{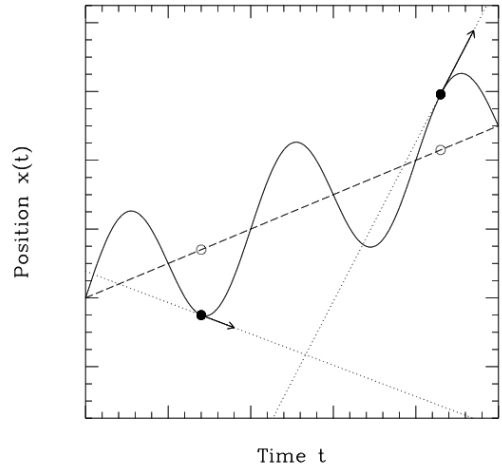}
\caption{\label{Fig:Wielen}
The orbital motion (solid line) of a binary, when sampled with a time span short with respect to the orbital period, may remain unnoticed, being an almost linear motion (represented by the arrows), adding to the proper motion of the centre of mass. On the contrary, when based on positional measurements covering a time span  much longer than the orbital period, the proper motion will not be affected by the orbital motion, which averages out. 
(From \citep{Wielen-1997})
}
\end{figure}

\citet{Wielen-1997,Wielen-1999} and \citet{Kaplan-Makarov-2003}   suggested that the comparison of 
Hipparcos and Tycho-2 \citep{Hog-2000:a,Hog-2000:b} proper motions offers a way to detect binaries with long periods
(typically from 1500 to 30000~d).
The Hipparcos proper motion,
being based on observations  spanning only 3~yrs, may be altered by the orbital
motion, especially for systems with periods in the above range 
whose orbital motion was not recognized by Hipparcos. On the other hand,
this effect should average out in the Tycho-2 proper motion, which is
derived from observations covering a much longer time span (Fig.~\ref{Fig:Wielen}). 
This method has been used by \citet{Wielen-1999}, \citet{Makarov-2004}, \citet{Pourbaix-2004} and \citet{Frankowski-2007}.  

The method evaluates the quantity
\begin{equation}
\chi^2_{\rm obs} = (\vec{\mu}_{\rm HIP} - \vec{\mu}_{\rm Tyc})^t\;  {\bf W}^{-1}\;  (\vec{\mu}_{\rm HIP} - \vec{\mu}_{\rm Tyc}),
\end{equation}
where $\vec{\mu}_{\rm HIP}$ and $\vec{\mu}_{\rm Tyc}$ are the vectors of $\alpha$ and $\delta$ components of the Hipparcos and
Tycho-2 proper motions, respectively, and  {\bf W} is the associated $2\times2$ variance-covariance matrix. 
          
Since the above quantity follows a $\chi^2$ probability distribution function with 2 degrees of freedom, 
it is then possible
to compute the probability Prob$(\chi^2 > \chi^2_{\rm obs})$, giving the first kind risk of rejecting the null hypothesis 
$\vec{\mu}_{\rm Tycho} = \vec{\mu}_{\rm HIP}$ while it is actually true. 

\begin{figure}
\includegraphics[height=.5\textheight]{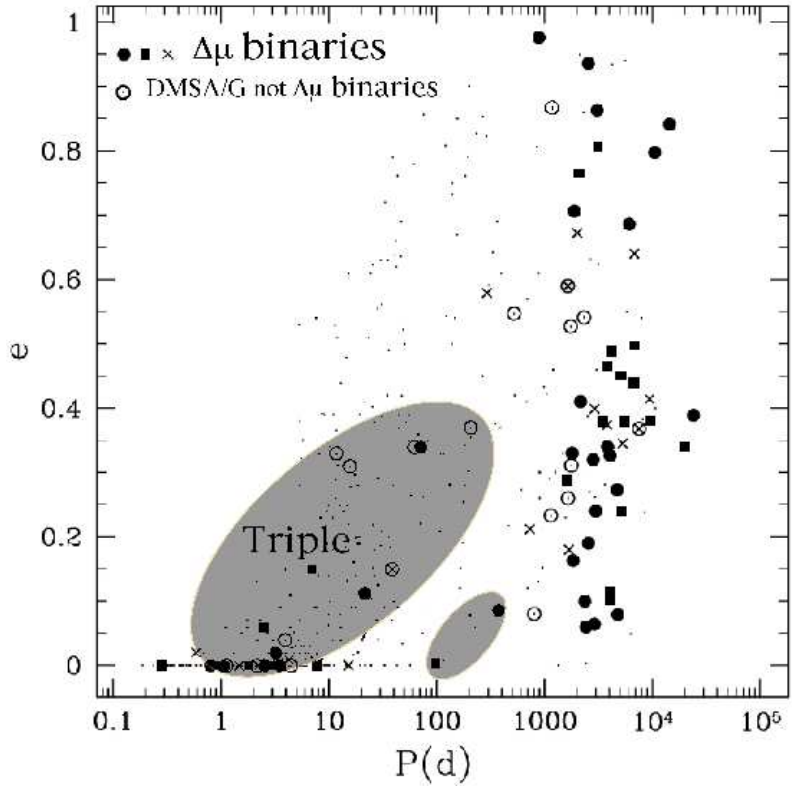}
\caption{\label{Fig:P_Delmu}
Eccentricity-period diagram for stars from the \protect\SB9\ catalogue
with parallaxes larger than 10~mas.
Symbols are as follows: large filled symbols:
proper-motion binaries detected at the 0.9999 confidence level, defined as 
$1 - $Prob$(\chi^2 > \chi^2_{\rm obs})$ (circles and squares 
denote those stars which are or are not, respectively, 
flagged as DMSA/G in the Hipparcos Catalogue, i.e., as stars with a non-linear proper motion);
crosses: systems not flagged as DMSA/G, lying in the confidence-level range
0.99 -- 0.9999; open circles: DMSA/G systems not flagged as proper-motion binaries at the 0.9999 level.
(From \citep{Frankowski-2007})
}
\end{figure}

\begin{figure}
\includegraphics[height=.3\textheight]{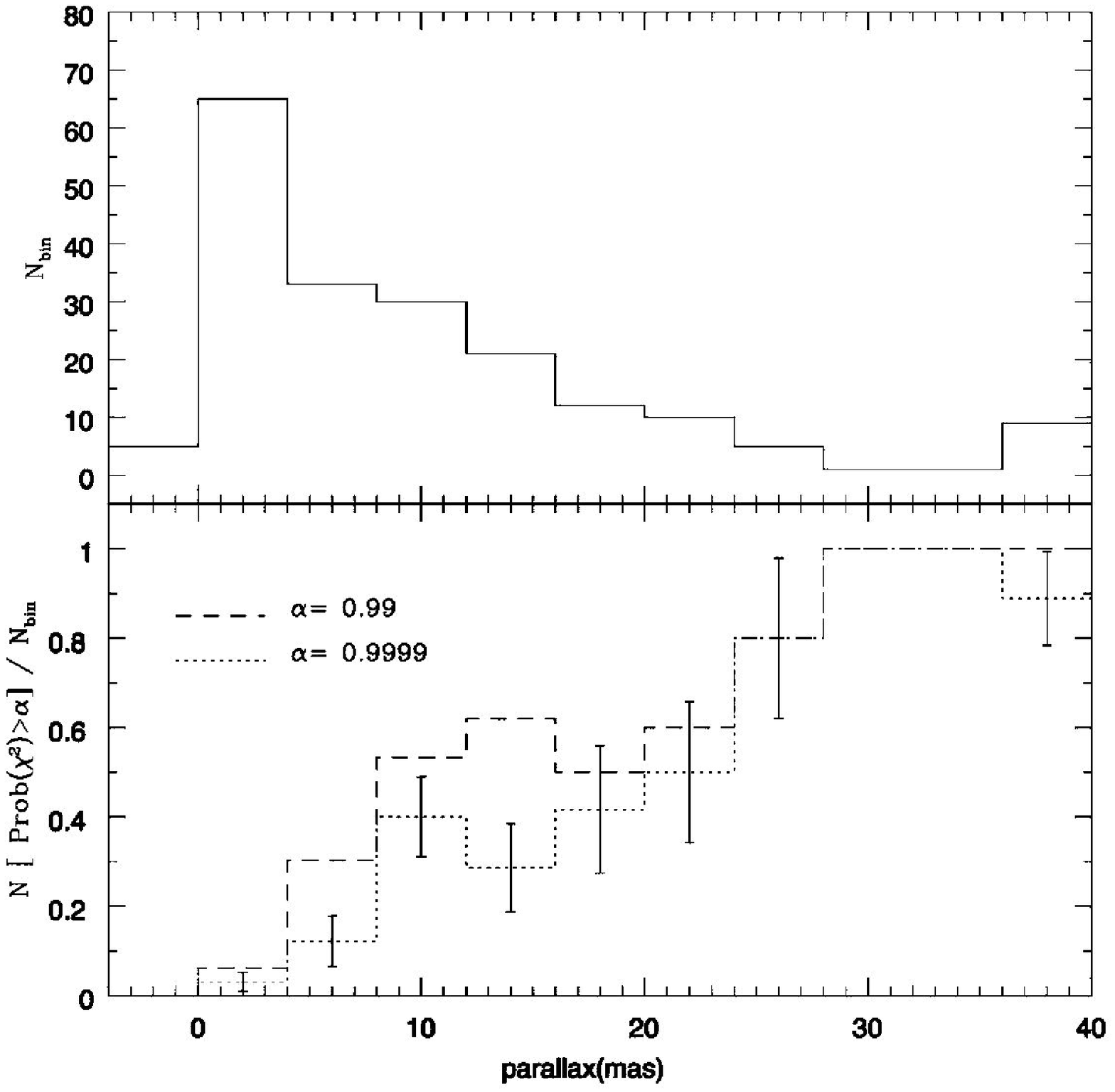}
\includegraphics[height=.3\textheight]{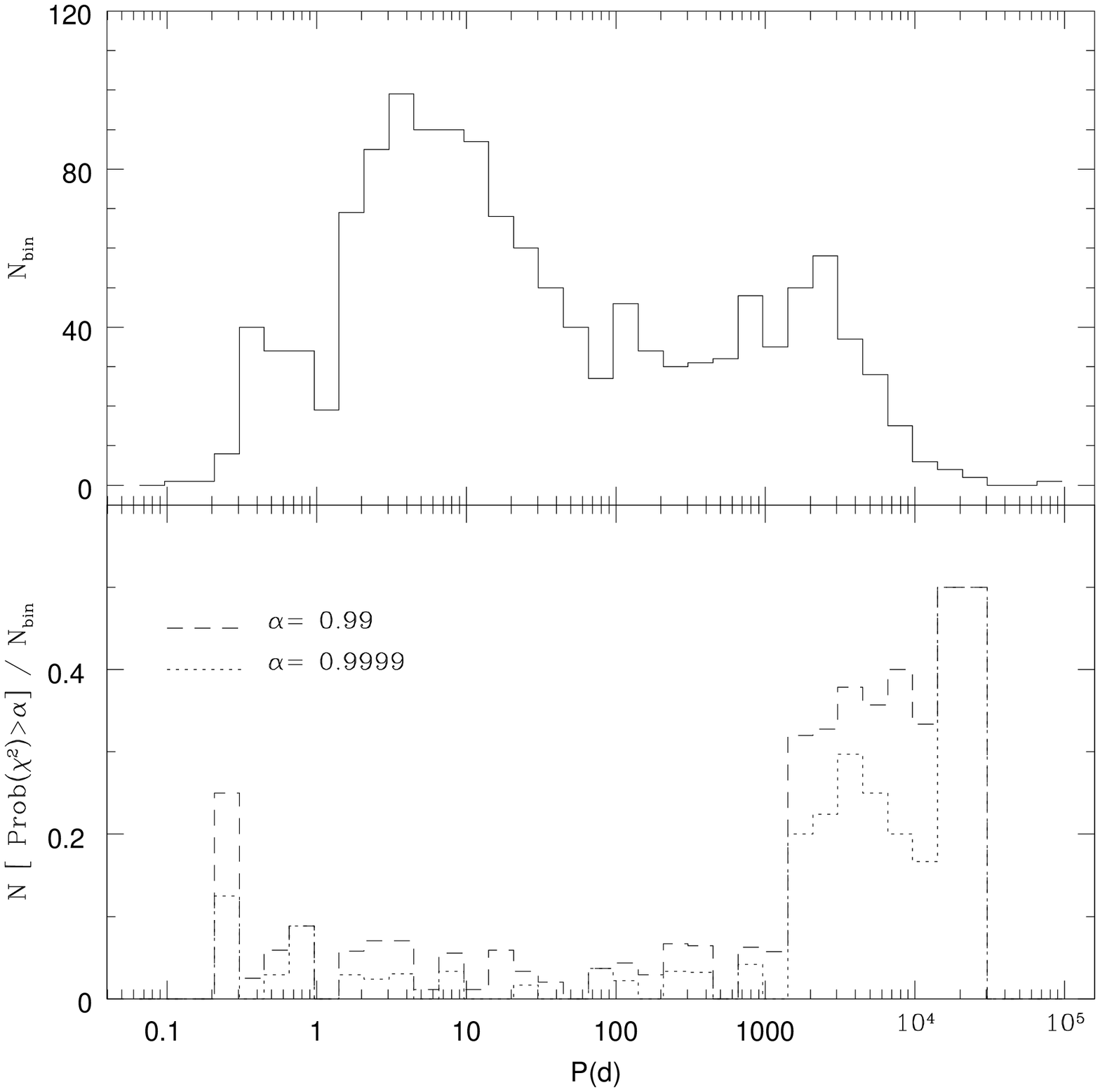}
\caption{\label{Fig:P_Delmu2}
{\bf Upper left panel:}
Parallax distribution of \protect\SB9\ stars with orbital periods
larger than 1500~d. {\bf Lower left panel:} Fraction of those \protect\SB9\ stars
detected as proper-motion binaries at confidence levels $\alpha$ of 0.99 and
0.9999. For parallaxes between 10 and 25~mas, about 50\% of the
\protect\SB9\ 
stars with periods larger than 1500~d are detected, and that fraction
becomes larger than 80\% above 25~mas. 
{\bf Upper right panel:} Orbital-period distribution for stars from the 
\protect\SB9\ catalogue. 
{\bf Lower right panel:} Fraction of stars in
a given bin for which the Hipparcos and Tycho-2 proper motions differ at
confidence levels of 0.9999 (dotted line) and 0.99 (dashed line).
(From \citep{Frankowski-2007})
}
\end{figure}

To evaluate the efficiency of the method, it has been applied by
\citet{Frankowski-2007} to all spectroscopic binary stars from the \SB9\
catalogue \citep{Pourbaix-04a} with both an Hipparcos and a Tycho-2 entry. Figs.~\ref{Fig:P_Delmu} and \ref{Fig:P_Delmu2} show that the detection efficiency is very good in the period range 1500 - 30000~d for systems with parallaxes in excess of 10 -- 20~mas.
On the
other hand, when the proper-motion method detects systems with orbital periods shorter
than about 400~d, there is a good chance that the system is triple, the proper-motion method being sensitive to the long-period pair. This suspicion has been confirmed by \citet{Frankowski-2007}, and by \citet{Fekel-2005,Fekel-2006} for the two 'textbook' cases HIP~72939 and HIP~88848. \citet{Makarov-Kaplan-2005} and \citet{Frankowski-2007} flagged these two stars as proper-motion binaries, despite the fact that the orbital periods known at the time are quite short,  3.55 and 1.81~d
respectively. The discovery by \citet{Fekel-2005,Fekel-2006}  that these two systems are in fact triple, with long periods 
of 1641 and 2092~d, respectively, came as a nice {\it a posteriori} validation of the proper-motion method for identifying long-period binaries.  The reprocessing of the Hipparcos data, accounting for the long-period pair, then yielded values for the Hipparcos proper motions perfectly consistent with the Tycho-2 ones, as it should. The situation is summarized in Table~\ref{Tab:Deltamu}.

\begin{table}[t]
\begin{tabular}{lrrr}
\multicolumn{4}{c}{HIP 72939} \\
\multicolumn{4}{c}{$P = 3.55$ and 1641~d}  \\
\multicolumn{4}{c}{\citet{Fekel-2006}} \\
\hline
& Hipparcos & Tycho-2 & Hip. reproc.  \\
\hline
$\mu_\alpha \cos \delta$ (mas/yr) & $-67.5 \pm 0.9$ & $-66.2\pm0.8$  &  $ -67.0 \pm 1.9$ \\
$\mu_\delta$ (mas/yr) &  $55.9\pm0.9$ & $62.1\pm0.9$ & $63.5\pm1.5$ 
\medskip\\
\hline
 \multicolumn{4}{c}{HIP 88848}\\
 \multicolumn{4}{c}{$P = 1.81$ and 2092~d}\\
 \multicolumn{4}{c}{\citet{Fekel-2005}}\\
\hline
& Hipparcos & Tycho-2 & Hip. reproc.  \\
\hline
$\mu_\alpha \cos \delta$ (mas/yr) & $138.1 \pm 1.9$ & $108.5\pm1.3$ & $107.$\\
$\mu_\delta$ (mas/yr) & $-18.6\pm1.7$ & $-25.4\pm1.2$ & $-31$ \\
\hline
\medskip\\
\end{tabular}
\caption{Two systems whose triple nature was first suspected by the proper-motion method and later confirmed by radial-velocity studies.}
\label{Tab:Deltamu}
\end{table}

A very interesting side product of the proper-motion binary detection method is that, being totally independent of the star spectrum, it provides an unbiased estimate of the binary frequency among the different spectral classes. 
Table~\ref{Tab:freq} reveals that the frequency of proper-motion binaries is constant within the error bars (at least in the range F to M). 
According to \citet{Frankowski-2007}, proper-motion binaries represent about 35\% of the total number of spectroscopic binaries, because of their more restricted period range. Thus, the frequencies of proper-motion binaries listed in Table~\ref{Tab:freq}
can be multiplied by 1/0.35 = 2.86 to yield the total fraction of spectroscopic binaries, namely 28.8\%. This value is close to the spectroscopic-binary frequency of $30.8\pm1.7$\% for cluster K giants found by \citet{Mermilliod-2007b}.

\begin{table}[]
\caption{\label{Tab:freq}
The frequency of proper-motion binaries among different spectral classes, from \cite{Frankowski-2007}. $N$ is the total number of stars of the corresponding spectral type
in the sample.
}
\begin{tabular}{lrr}
\hline
Spectral type &\multicolumn{1}{c}{\%} & \multicolumn{1}{c}{$N$} \\
\hline
B & $14.3\pm13.2$ & 8664 \\
A & $15.2\pm3.0$ & 15662 \\
F & $11.4\pm1.0$ & 21340 \\
G & $10.5\pm0.8$ & 19628 \\
K & $9.3\pm0.7$ & 29349 \\
M & $9.8\pm1.3$ & 4217 \\
All & $10.1\pm0.4$ & 103304 \\
\hline
\end{tabular}
\end{table}
  		
\subsection{Variability-Induced Movers and Color-induced displacement binaries}

The two methods presented in this section are unusual in the sense that they combine photometry and astrometry to diagnose a star as a binary.

The first category, Variability-Induced Movers, is defined in the Double and Multiple System Annex (DMSA) of the Hipparcos Catalogue, where it is known as  DMSA/V. 
Stars flagged as DMSA/V are generally large-amplitude variables whose  residuals with respect to a single-star astrometric solution are correlated with the star apparent brightness. Such a situation may arise in a binary system, when the position of the photocentre of the system varies as a result of the light variability of one of its
components: when the variable component is at maximum light, it dominates the system light, and the photocentre position is closer to the position of that star
\citep[][see also Eq.~\ref{Eq:ap}]{Wielen-1996}. \citet{Pourbaix-2003} have shown, however, that many of the long-period variables DMSA/V listed in the Hipparcos Catalogue are not binaries, the correlation of the astrometric residuals with the stellar brightness resulting in fact from the colour variation accompanying the light variation of a long-period variable. This colour variation was not accounted for in the standard processing applied by the Hipparcos reduction consortia, which adopted a constant colour for every star. Thus, the so-called chromaticity effect (for a given position on the sky, the position of a star on the Hipparcos focal plane depends on the star colour, because of various chromatic aberrations in the Hipparcos optical system) was not appropriately corrected 
in the case of long-period variables. When this effect is incorporated in the astrometric processing \citep[by using {\it epoch} colours;][]{Platais-2003}, a single-star solution appears to appropriately fit the observations for 161 among the 188 stars originally flagged as DMAS/V in the Hipparcos Catalogue, as shown by  \citet{Pourbaix-2003}. The 27 remaining DMSA/V may be true binaries, and are listed in Table~\ref{Tab:VIM}.

\begin{table}
\caption{\label{Tab:VIM}
The 27 stars keeping their DMSA/V status after the appropriate re-processing by \citet{Pourbaix-2003}. They may be true binaries. The column labelled 'Var' lists the variability type (M = Mira; SR = semi-regular; L = irregular). 
}
\begin{tabular}{rlll}
\hline
HIP & GCVS & Var & Rem\\
\hline
781 & SS Cas & M \\
2215 & AG Cet & SR \\
11093 & S Per & SRc \\
23520 & EL Aur & Lb \\
31108 & HX Gem & Lb:\\
36288 & Y Lyn & L \\
36669 & Z Pup & M \\
41058 & T Lyn & M \\
43575 & BO Cnc & Lb:\\
45915 & CG UMa & Lb \\
46806 & R Car & M & $V = 11.3$ companion located 1.8'' away\\ 
54951 & FN Leo & L \\
67410 & R CVn & M \\
68815 & $\theta$ Aps & SRb \\
75727 & GO Lup & SRb\\
76377 & R Nor & M \\
79233 & RU Her & M & composite spectrum ?\\
80259 & RY CrB & SRb \\
90709 & SS Sgr & SRb \\
93605 & SU Sgr & SR \\
94706 & T Sgr & M & composite spectrum\\
95676 & SW Tel & M\\
99653 & RS Cyg & SRa\\
100404 & BC Cyg & L\\
109089 & RZ Peg & M \\
111043 & $\delta^2$ Gru & Lb: \\
112961 & $\lambda$ Aqr & L \\
\hline
\end{tabular}
\end{table}

\begin{figure}
\includegraphics[height=.3\textheight]{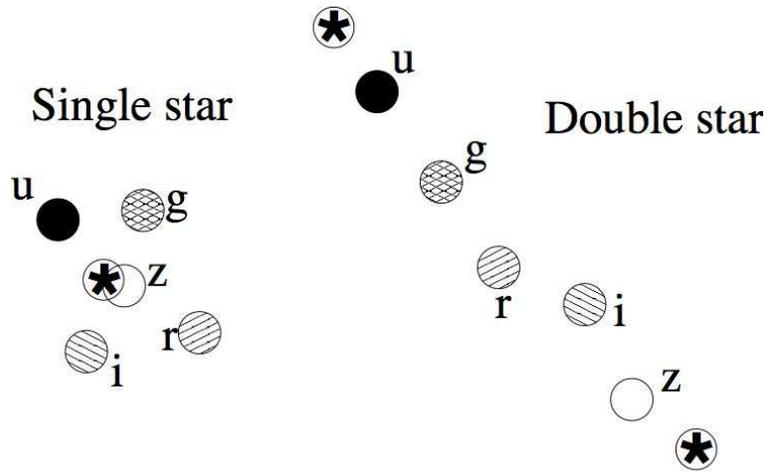}		
  			\caption{  			
  			\label{Fig:SDSS}		
Schematic position of the photocentre in the different 
SDSS bands. {\bf Right panel:}  For double stars, the positions are aligned along the line joining the 
two stars and are ordered according to the central wavelength of the filter. 
The true positions of the stars 
are represented as a five-branch star symbol.
{\bf Left panel:} For single stars, measurement errors prevent the positions from being perfectly superimposed.
(From \citep{Pourbaix-2004:a})}
 \end{figure}

\begin{figure}
\includegraphics[height=.5\textheight]{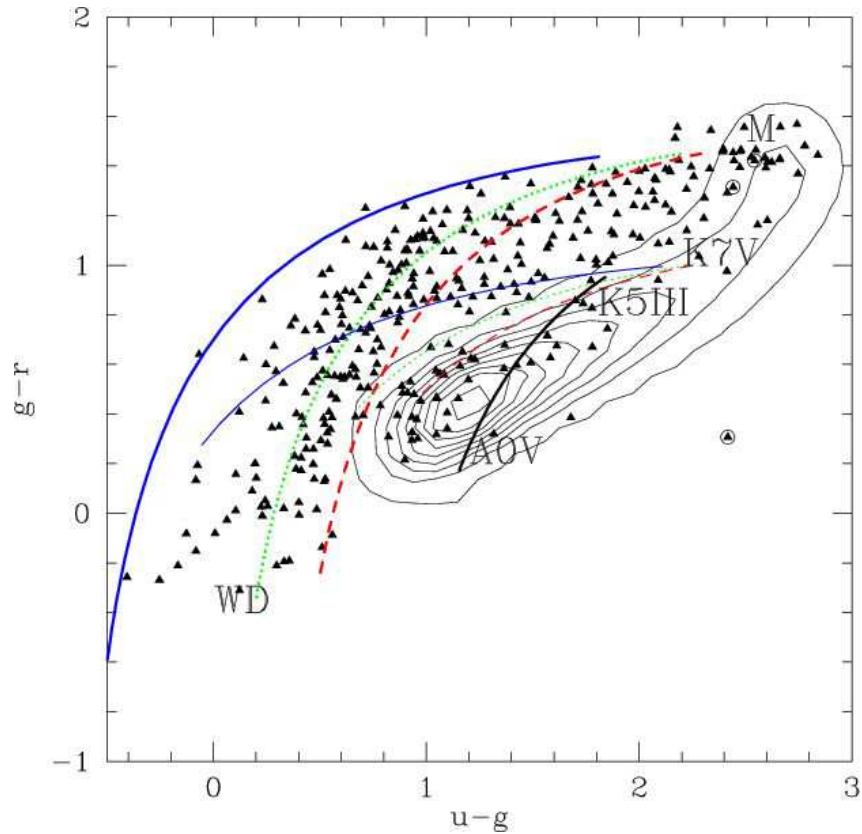}
\caption{    			\label{Fig:SDDS_results}
  			 Color-color diagram of the putative binaries (triangles) superimposed on the original parent population from the second SDSS data release (contours).
The thick lines on the left correspond to systems with an M dwarf
and three different kinds of WDs.  Thin lines are the same, but for a K7V star instead of
an M dwarf.
The short thick solid line in the centre, overlapping the density contours, corresponds to A0V/K5III pairs.  
(From \citep{Pourbaix-2004:a})}
\end{figure}

Colour-induced displacement binaries are detected when the position of the photocentre depends on the considered photometric band, after correcting for any instrumental chromatic effect \citep{Christy-1983}. This situation is analogous to the previous one, except that the light variability of one component is here replaced by the colour difference between the two stars: if the two components of a binary system have very different colours, the respective contribution of each component to the integrated system light will be different in the different photometric bands, and so will
be the position of the photocentre. This method thus requires accurate astrometry in various filters and is able to detect only binary systems with components of very different spectral types. The former condition is met by the Sloan Digital Sky Survey \citep[SDSS;][]{Adelman-SDSS-2007}, which provides  positions in the Gunn system $u,g,r,i,z$ (Fig.~\ref{Fig:SDSS}). The latter condition is met by white-dwarf/red-dwarf pairs. 
\citet{Pourbaix-2004:a} found 346 pairs among the $4.1\; 10^6$ stars with $u,g < 21$ in the second SDSS data release \citep{Adelman-SDSS-2007} with a distance between the $u$ and $z$ positions  larger than $0.5''$. Most (about 90\%) of these must indeed correspond to red-dwarf/white-dwarf pairs, as indicated on Fig.~\ref{Fig:SDDS_results}, the remaining 10\% being probably early main-sequence/red-giant pairs.

\section{Other detection methods}
\label{Sect:diverse}

\subsection{Detecting binaries from rapid rotation}

Another method for finding binaries (among old, late-type stars) involves the identification of fast rotators, since evolved late-type stars are not expected to rotate fast \citep[see e.g.][]{DeMedeiros-1996,DeMedeiros-1999}. Fast rotation can  be ascribed to {\it spin-up processes operating either in tidally interacting systems} \citep[like RS~CVn systems among K giants; ][]{DeMedeiros-2002}, or in {\it mass-transfer systems}, through transfer of spin angular momentum. 

In recent years, the link between rapid rotation of old, late-type stars and spin-accretion during (wind) mass transfer in binaries has been strengthened, most notably by the discovery of the family of WIRRing stars \citep[standing for 'Wind-Induced Rapidly Rotating';][]{Jeffries-Stevens-96}.
Originally, this class 
was defined from  a small
group of rapidly-rotating, magnetically-active K dwarfs with hot WD
companions discovered by the ROSAT Wide Field Camera and {\it Extreme
Ultraviolet Explorer} (EUVE) surveys \citep{Jeffries-Stevens-96,Jeffries-Smalley-96,Vennes-97}. 
Since these stars exhibit no short-term
radial-velocity variations, it may be concluded that any orbital period 
must be a few months long at least. Moreover, several arguments, based on
proper motion, WD cooling time scale, and lack of photospheric Li,
indicate that the rapid rotation of the K dwarf cannot be ascribed to
youth. \citet{Jeffries-Stevens-96} therefore suggested that the K
dwarfs in these wide systems were spun up by the accretion of the wind
from their companion, when the latter was a mass-losing AGB star.  The
possibility of accreting a substantial amount of spin from the
companion's wind
has been predicted \cite{Theuns-Jorissen-93,Theuns-96,Mastrodemos-98} by {\it smooth-particle
hydrodynamics} simulations of wind accretion in detached binary systems.
A clear signature that mass transfer has been
operative in the WIRRing system 2RE~J0357+283 is provided by the
detection of an overabundance of barium \cite{Jeffries-Smalley-96}.
Interestingly enough, the class of WIRRing stars is no more restricted to
binaries with a late dwarf primary, since
in recent years, WIRRing systems were found  among many different classes of stars: 
\begin{itemize}
\item A number of {\bf barium stars} (post-mass-transfer systems involving a K giant and a WD) have been found to rotate fast, with in some cases \citep[HD~165141 and 56~Peg;][]{Jorissen-96,Frankowski-2006} clear indications that the mass transfer responsible for the chemical pollution of the atmosphere of the giant has occurred recently enough for the magnetic braking not to have slowed down the giant appreciably. These systems thus exhibit signatures of X-ray activity comparable to RS~CVn systems (HD~165141, 56~Peg) or host a warm, relatively young WD (HD~165141). HD~77247 is another barium star with broad spectral lines  \citep{Jorissen-VE-98}. It could thus be a fast rotator, but it has no  other peculiarity, except for its outlying location in the eccentricity -- period diagram ($P =  80.5$~d; $e = 0.09$). In this case, the fast rotation could thus rather be the result of tidal synchronization.
\item Among {\bf binary M (or C) giants}, V~Hya, HD~190658 and HD~219654 are fast rotators (respectively with $V\sin i = 16$~\kms, $15.1\pm0.1$~\kms\ and $13.6\pm0.8$~\kms)
\citep{Famaey-2008:b,Knapp-1999:b}. HD~190658 is the M-giant binary with
the second shortest period known so far
(199~d; \citep[][]{Famaey-2008:b}),
and  exhibits as well ellipsoidal variations \citep{Samus-1997}.
Fast rotation is therefore  expected in such a close binary with strong  tidal interactions. The radius derived under the assumption of a rotation synchronized with the orbital motion  is then $R \sin i = 59$~\Rsun, corresponding to
$R/a_1 = 1.16$, where $a_1$ is the semi-major axis of the giant's
orbit around the centre of mass of the system
\citep{Lucke-Mayor-1982}. The radius deduced from Stefan-Boltzmann law is 62.2~\Rsun\ \citep{Frankowski-2008}, which implies an orbit seen very close to edge-on ($\sin i = 0.948$ or $i = 71.4^\circ$).  Adopting typical masses of 1.7~\Msun\ for
the giant and 1.0~\Msun\ for the companion, one obtains $R/a = 0.43$
and $R/R_R = 1$ ! Thus, the star apparently fills its Roche lobe. 
\item {\bf Binary nuclei of planetary nebulae of the Abell-35 kind}  (containing only four members so far:  Abell 35 = 
BD$-22^\circ$3467 = LW Hya, LoTr~1, LoTr~5 = HD 112313 = IN Com =
2RE~J1255+255, and WeBo~1 = PN G135.6+01.0) have  
their optical spectra dominated by late-type (G-K)
stars \cite[with demonstrated barium anomaly in the case of WeBo\,1 only;][]{Gatti97,Thevenin-Jasniewicz-97,Bond02,Pereira03}, but the UV spectra reveal the presence of extremely hot 
($>$\,10$^5$\,K), hence young, WD companions. A related  
system is HD\,128220, consisting of an sdO and a G star in a
binary system of period 872\,d \cite{HowarthHeber90}.
In all these cases, the
late-type star is chromospherically active and rapidly rotating, and this rapid rotation is likely to result from a recent episode of mass transfer. 
\item Rapid rotation seems to be a common property of the giant component in the few known {\bf symbiotics of type  d'} \citep{Jorissen-Zacs-2005,Zamanov-2006}.
The evolutionary status of this rare set of yellow d' symbiotic systems (SyS),
which were all shown to be of solar metallicity, has recently 
been clarified \citep{Jorissen-Zacs-2005}
with the realisation that in these systems, the
companion is {\em intrinsically} hot (because it recently 
evolved off the AGB), rather than being heated by accretion or
nuclear burning as in the other brands of SyS. Several arguments support this claim: 
(i) d'~SyS host G-type giants whose mass loss is not strong enough to
heat the companion through accretion and/or nuclear burning;  
(ii) the cool dust observed in d'~SyS
\citep{Schmid-Nussbaumer-93} is a relic from the mass lost by the AGB
star; 
(iii) the optical nebulae observed in d'~SyS are most likely 
genuine PN rather than the 
nebulae associated with the ionized wind of the cool component
\citep{Corradi-99}. d'~SyS often appear in PN catalogues.    
AS~201 for instance actually hosts {\em two} nebulae \citep{Schwarz-91}: 
a large fossil planetary nebula detected by direct imaging, and a small
nebula formed in the wind of the current cool component;
(iv) rapid rotation  has likely been caused by spin accretion from the
former AGB wind like in WIRRing systems. The fact that the cool star has
not yet been slowed down by magnetic braking is another indication
that the mass transfer occurred fairly recently. 
\citet{Corradi-1997}
obtained 4000~y for the age of the nebula around AS~201, 
and 40000~y for that around V417~Cen.

\item The {\bf blue stragglers} were first identified by \citet{Sandage-1953} in the globular cluster M3. Since their discovery, blue 
stragglers have been found in many other globular clusters and 
in open clusters as well. The frequency of blue stragglers is 
also correlated with position in clusters \citep[see][and references therein]{Carney-2005}. Depending on the cluster, this frequency may either increase or decrease with the distance from the centre, thus calling for two different mechanisms for the creation of blue stragglers in clusters. 
In the central regions, dynamical effects involving collisions 
may enhance the probabilities for creating blue stragglers \citep{Sills-1999}. In the outer regions, where binary systems are most likely to survive, mass transfer is a likely explanation \citep{McCrea-1964,Mateo-1990}.
Despite being initially found in the globular clusters, blue stragglers are now found as well in the (halo) field where they are known as  either 'blue metal-poor stars'  \citep{Preston-94,Preston-Sneden-00,Sneden-03a} or 'ultra-Li-deficient stars' \citep{Ryan-2001,Ryan-2002}.
The very large binary frequency among these classes of stars \citep{Carney-2005,Preston-Sneden-00,Sneden-03a,Carney-01} strongly suggests
that field blue stragglers result from mass transfer in a binary,  rather than from the various merger 
processes that are currently believed to produce blue stragglers in the inner regions of globular clusters. 

Finally coming to the issue at stake in this section,  \citet{Carney-2005}, \citet{Preston-Sneden-00} and \citet{Ryan-2002} find that the   field blue straggler  stars that are binaries have higher rotational velocities  
than stars of comparable temperature but showing no evidence of binarity. Moreover, the orbital periods are too long 
for tidal effects to be important, implying that spin-up during mass transfer when the orbital separations and 
periods were smaller is the cause of the enhanced rotation.
  This conclusion 
is supported by the abnormally high proportion of blue-metal-poor binaries with long periods and small orbital 
eccentricities, properties that these binaries share with barium stars and related binaries \citep{Preston-Sneden-00}. \citet{Fuhrmann-Bernkopf-99}  likewise draw a connection 
between rapid rotation, Li depletion, and mass transfer 
from a companion in rather long-period systems (several hundred days).
\end{itemize}

\subsection{X-rays}
  				    
\begin{figure}
\includegraphics[height=.25\textheight]{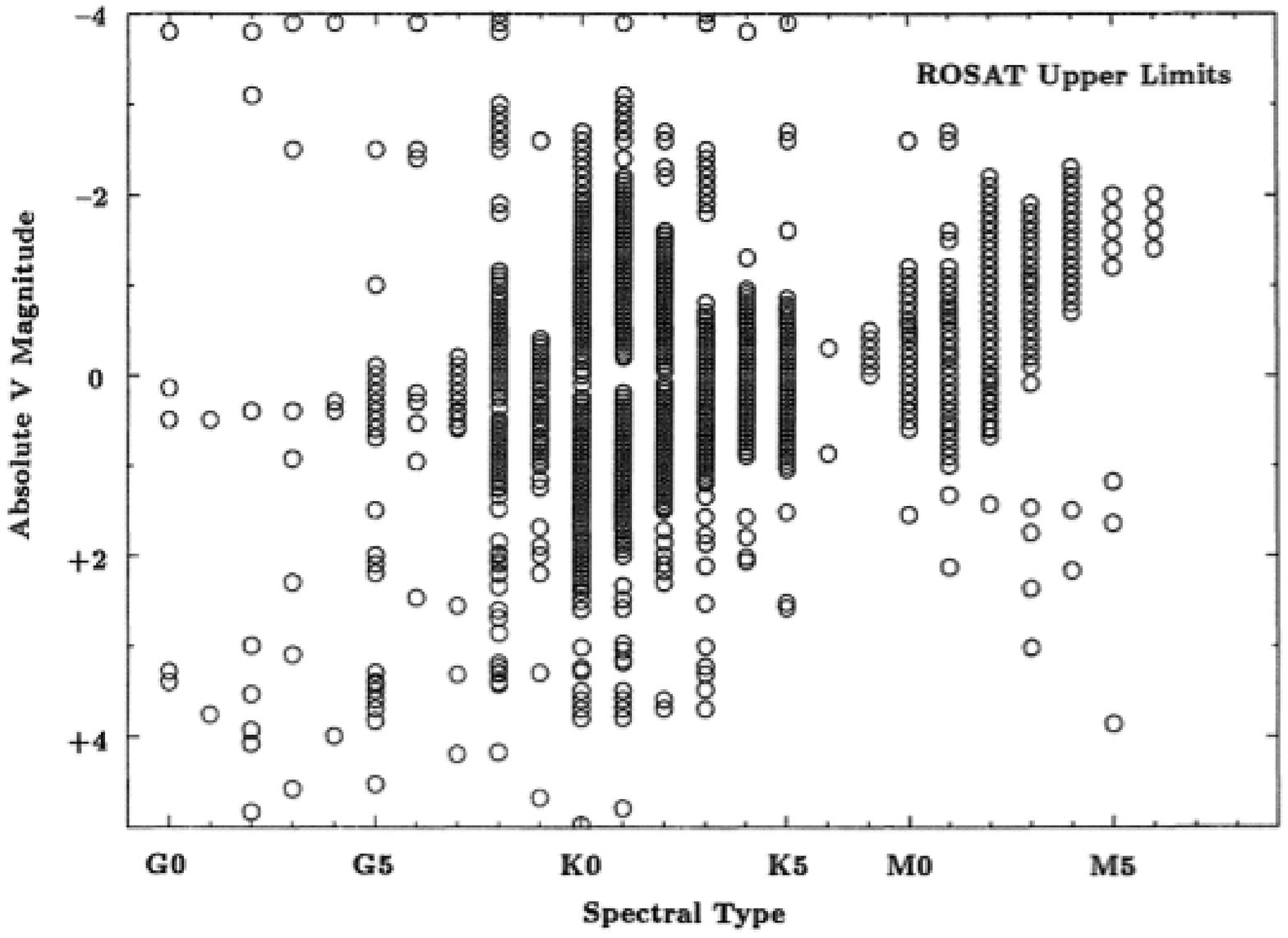}	
\includegraphics[height=.25\textheight]{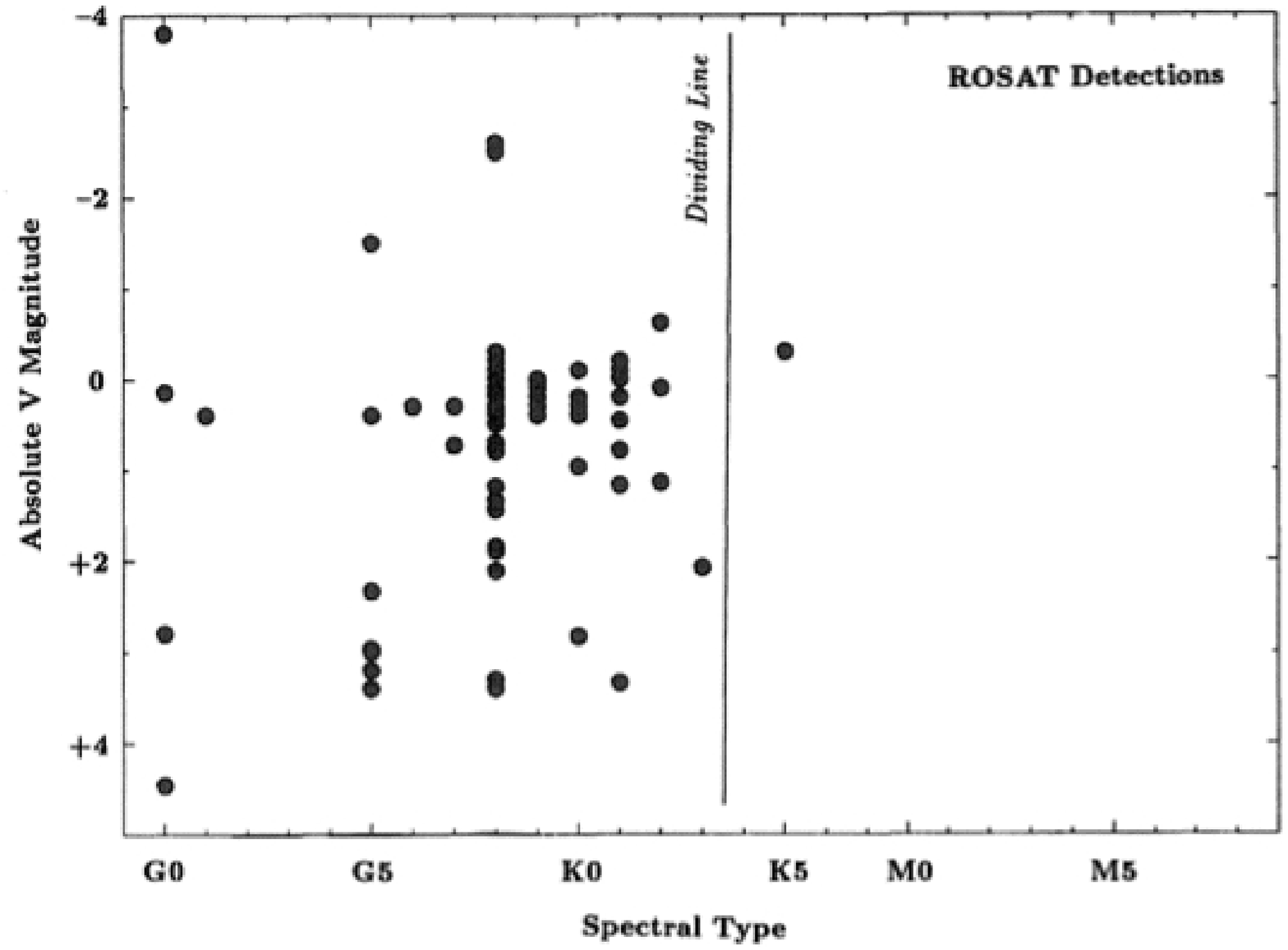}	
  			\caption{  			
  			\label{Fig:dividing}		
Comparison of the location in the Hertzsprung-Russell diagram of stars with and without X-rays detected by the ROSAT satellite (right and left panels, respectively). The 'dividing line' is clearly apparent on the right panel.
(From \citep{Haisch-1991})
}
\end{figure}

Several physical processes, which will be discusssed below, are responsible for the emission of X-rays  in  binaries, so that their detection provides a useful diagnostic of binarity. 
In single stars, X-rays are only expected in stars located to the left of the Linsky-Haisch line in the Hertzsprung-Russell diagram \citep{Linsky-1979,Haisch-1991}, separating warm stars with hot coronae from cool stars with wind mass loss (Fig.~\ref{Fig:dividing}). Exceptions to this rule are, however, 
M dwarfs rotating rapidly because they are young, thus producing a dynamo effect which may heat the corona.

X-rays (corresponding to photons with $h\nu > 0.1$~keV requiring $T > 10^6$~K) may come from:
\begin{itemize}
\item {\bf hot stellar coronae} heated by a dynamo triggered by fast rotation \citep{Pallavicini-1981}, either due to youth, or to synchronous rotation in a close binary system \citep[as in RS~CVn systems;][]{Eaton-1979}. 
\item {\bf nuclear fusion} at the surface of a WD (as in novae or symbiotic stars). In classical novae, explosive H-burning (releasing $10^{33} - 10^{34}$
erg~s$^{-1}$) occurs, whereas in 'super soft sources' (as are symbiotic stars), it is probably quiescent H-burning. 
\item {\bf accretion} in binary systems like 
\begin{itemize}
\item cataclysmic variables (CVs, consisting of a dwarf star and a WD in a  semi-detached system with periods of a few hours), which themselves subdivide into
\begin{itemize}
\item Classical novae, some can be detected in X-rays outside of eruptions;
\item Dwarf novae, with outburst powered by accretion-disk instability;
\item Polars (AM Her systems), involving highly magnetized WDs around which no accretion disk can form because of the very strong magnetic field;
\item Intermediate Polars (DQ Her systems), involving highly magnetized WDs in a rather wide system which allows some room for an accretion disk restricted to the outer region of the Roche lobe; in the inner region, no accretion disk can form because of the very strong magnetic field; 
\end{itemize}
\item 	Algols (which are semi-detached systems consisting of a subgiant component  and a more massive main sequence component);
\item symbiotic systems, consisting of a giant star and a WD in a system with a period exceeding 100~d. However, as we discuss below, the 
	accretion-driven nature of the X-rays emitted by SyS has been debated.
\end{itemize}
\item {\bf wind collision}, as in SyS.
\end{itemize}

\subsubsection{X-rays from binaries: The case of SyS}

One of the defining properties of SyS is to host a hot compact star, generally a WD,
accreting matter from a giant companion \citep{ASP303}. The 
WD is heated either directly by accretion, or indirectly 
by nuclear burning fueled by accretion \citep[][and references therein]{Jorissen-03a}. 
Symbiotic stars are X-ray sources \citep{Muerset-97}, at the level $10^{30}$ to 
$10^{33}$~erg~s$^{-1}$.  The physical process emitting these X-rays in
symbiotic stars is still  debated, though. Direct evidence for the presence of an accretion disk, 
in the form of continuum flickering and far UV continuum  is not 
usually found in symbiotic stars \citep{Sokoloski-2001,Sokoloski-2003,Sion-2003},
unlike the situation prevailing in cataclysmic variables and low-mass 
X-ray binaries \citep{Sokoloski-2001,Guerrero-2001}.  Z~And is the only SyS where the accretion-driven nature of X-rays makes no doubt, because flickering on time scales ranging from minutes to days has been observed 
\citep{Sokoloski-2003}.
Other mechanisms were therefore advocated to account for the X-ray 
emission from SyS, like 
thermal radiation from the hot component in 
the case of supersoft X-ray sources, or the shock forming in 
the collision region between the winds from the hot and cool 
components \citep{Muerset-97}. Interestingly, \citet{Soker-2002} even suggests that 
the X-ray flux from SyS exclusively arises from the fast 
rotation of the cool component spun up by wind accretion from the 
former AGB component (now a WD)! If that hypothesis is correct, the 
properties of X-rays from SyS would be undistinguishable from 
those of active (RS CVn-like) binaries. It is an intriguing hypothesis,
 because it requires a
high incidence of fast rotation among SyS, which has so far only been reported for d' symbiotics \citep[see above and][]{Zamanov-2006}.

\begin{figure}
\includegraphics[height=.5\textheight]{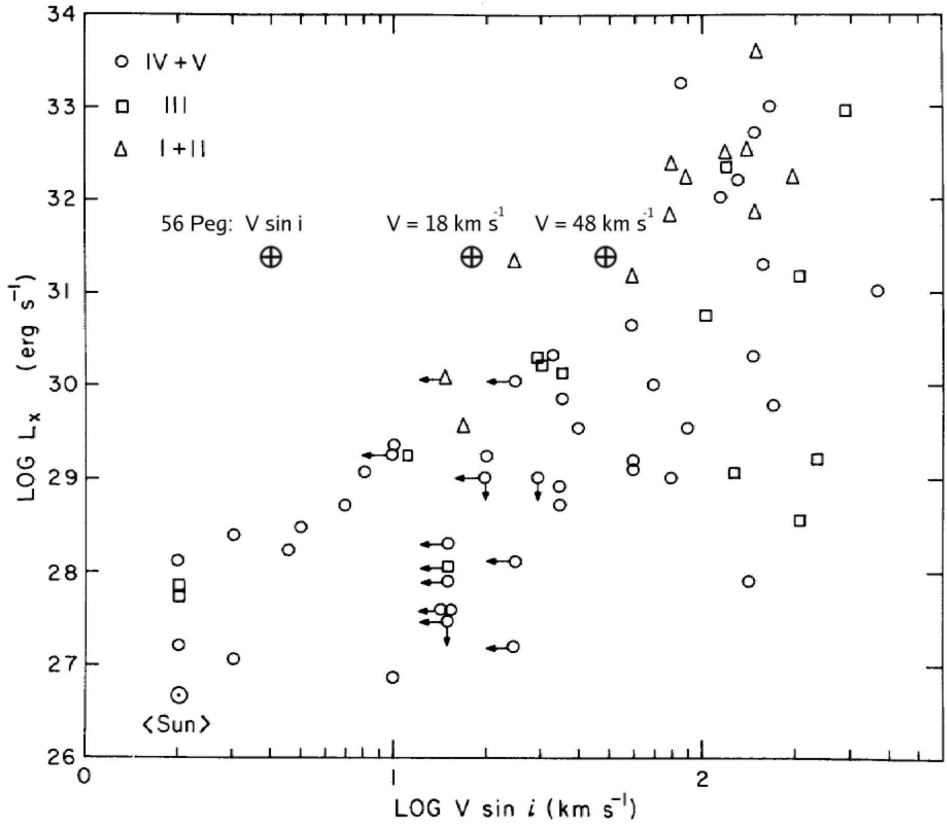}	
  			\caption{  			
  			\label{Fig:56Peg}		
X-ray luminosities vs.\ projected rotational velocities for
56 Peg (encircled crosses, for $V \sin i$ and for two values of $V$:
assuming orbital corotation or fast rotation) and stars of various spectral
types and luminosity classes
detected by the {\it Einstein} Observatory. Different symbols indicate
different luminosity classes, as indicated in the figure. 
(Adapted from  \citep{Frankowski-2006} and \citep{Pallavicini-1981}).
}
\end{figure}

The barium system 56~Peg provides an interesting illustration of the difficulty of finding the exact origin of the emitted X-rays.  
X-rays were detected with the \textit{Einstein} satellite by \citet{Schindler-82}. 
In a subsequent study, \citet{Dominy-Lambert-1983}  attributed the X-rays to an
 accretion disk around the WD companion.
\citet{Frankowski-2006}   re-interpreted the X-rays as due to  RS CVn-like activity, because the recent finding by \citet{Griffin-2006:a} that the orbital period of the system is as short as 111~d can only be reconciled with other properties of 56~ Peg if it is  a fast rotator seen almost pole-on!
Its X-ray luminosity is consistent with the rotation-activity relationship of \citet{Pallavicini-1981}, provided that it rotates at a velocity of about 50~\kms\ (Fig.~\ref{Fig:56Peg}) \citep[despite an observed value $V_{\rm rot} \sin i$ of only 4.4~\kms;][]{DeMedeiros-1999}, thus implying an almost pole-on orbit. A low inclination is also required by evolutionary considerations, in order to allow the companion of this barium star to be a WD, given the very low mass function obtained by \citet{Griffin-2006:a}: $f(M) = 3.73\; 10^{-5}$~\Msun.

Given the small orbital separation, one may still wonder whether there 
should not be as well some contribution to the X-ray flux coming 
from accretion, as advocated by \citet{Schindler-82}
and \citet{Dominy-Lambert-1983}. 
The major evidence suggesting that there may be a link between mass 
transfer and X-ray emission in active binaries is the correlation 
between the X-ray luminosity and the Roche-lobe filling fraction 
$\gamma_2$ for 
RS~CVn systems noted by \citet{Welty-Ramsey-1995}. 
\citet{Singh-1996} re-examined this issue using samples of RS~CVn 
and Algol binaries, and confirm the weak 
correlation found earlier by \citet{Welty-Ramsey-1995}. 
However, Singh et al. ``{\it regard this as a rather 
weak argument, because both the X-ray luminosities and the Roche lobe 
filling fractions are themselves correlated with the stellar radii in 
the sample of RS~CVn binaries, and thus regard this correlation as 
merely a by-product of the inherent size dependence of these 
quantities.''} 
Actually, despite the short orbital period, the filling factor of
56~Peg~A is not that close to
unity.\footnote{Adopting $i = 5^\circ$, $M_1 = 3$~\Msun, $M_2 = 0.96$~\Msun,
Griffin's orbital
elements yield an orbital separation $A = 152$~R$_{\rm \odot}$ and a Roche
lobe radius around the giant component of 73~\Rsun, well in excess of the
40~\Rsun\ representing the stellar radius itself, and translating into
$\gamma_2 = 0.55$.} 

\subsection{Composite spectra and magnitudes}
\label{Sect:composite}

\begin{figure}
\includegraphics[height=.3\textheight]{composite_hurley.ps}
\includegraphics[height=.4\textheight]{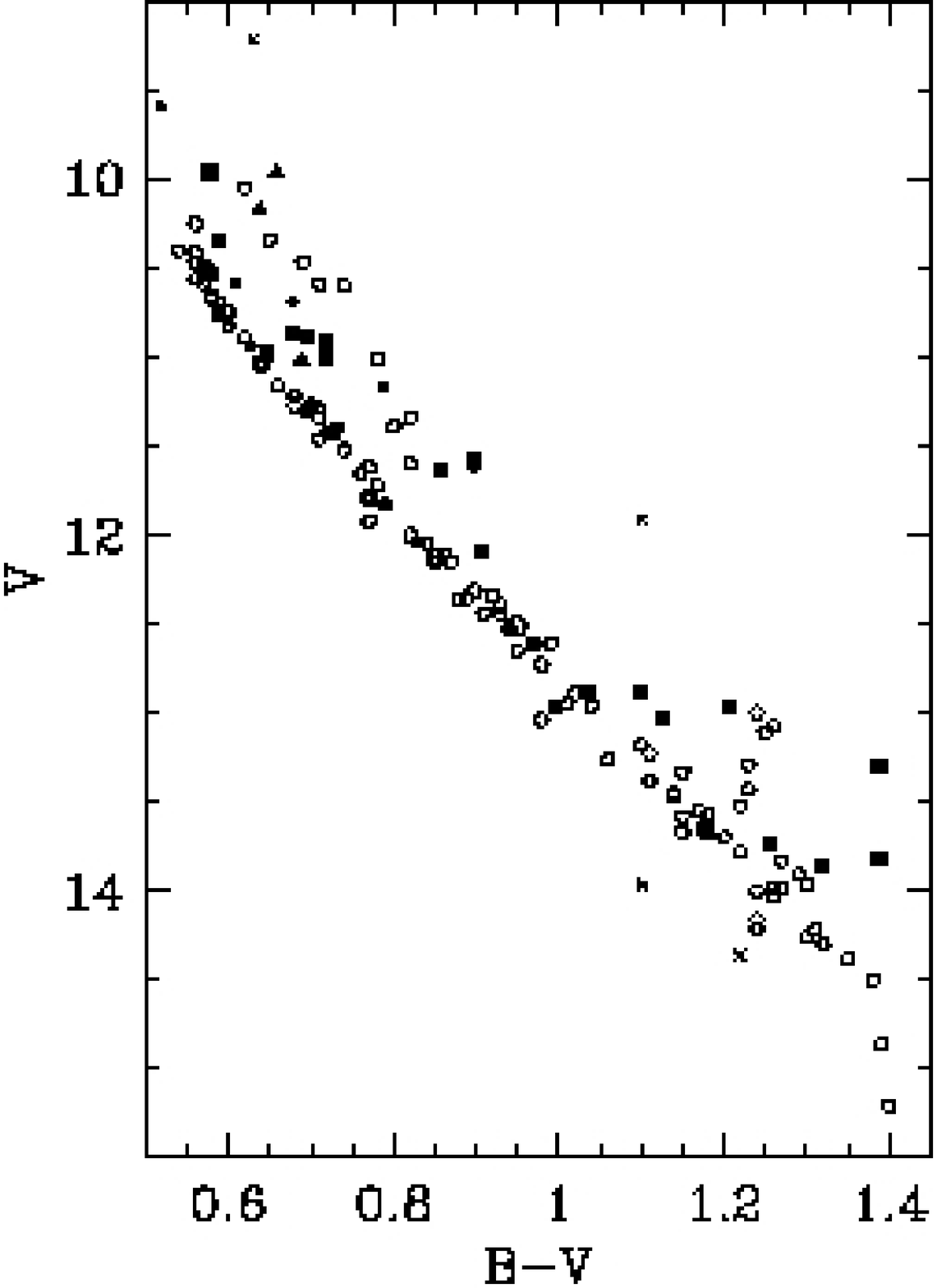}
\caption{
\label{Fig:CMD}
{\bf Left panel:} Theoretical zero-age single star and equal-mass binary main
sequences for a metallicity of $Z = 0.02$  covering a range of absolute magnitudes  $M_V$.
Also plotted for a range of primary masses, $M_1$, are binary points with
 $M_2 = q M_1$ where $q$ ranges
from $1.0 \rightarrow 0.0$ in increments of 0.1. The point at $q = 0.5$ is
an open square.
(From \citep{Hurley-1998}).
{\bf Right panel:} The $(V, B-V)$ color-magnitude diagram for low-mass Praesepe members. Symbols are as follows: filled squares: optically-resolved binary systems; filled dots: known 
spectroscopic binaries; filled triangles: triple systems; open circles: 'single' stars (although some of these are resolved in the infrared); crosses : suspected non members. 
(From \citep{Bouvier-2001})
}
\end{figure}

SyS are a specific example of a more general class of binaries having composite spectra, where the spectral features of the two components are intermingled. Spectroscopically, these systems correspond  to SB2 binaries. \citet{Ginestet-1997} and \citet{Ginestet-2002} provide lists of more than 100 systems with true composite spectra of
various spectral types and luminosity classes.

In color-color and color-magnitude diagrams, these stars occupy outlying positions and may therefore be easily identified.  In colour-magnitude diagrams of clusters for instance, binaries widen the main sequence. This effect has been quantified by \citet{Hurley-1998}, as shown on the left panel of Fig.~\ref{Fig:CMD}.
A beautiful illustration of this effect is provided by the color-magnitude diagram of
the Praesepe open cluster where the binary sequence is clearly apparent (Right panel of Fig.~\ref{Fig:CMD}).

\subsection{Detecting binaries from the partial absence of maser emission}

\begin{figure}
\includegraphics[height=.5\textheight]{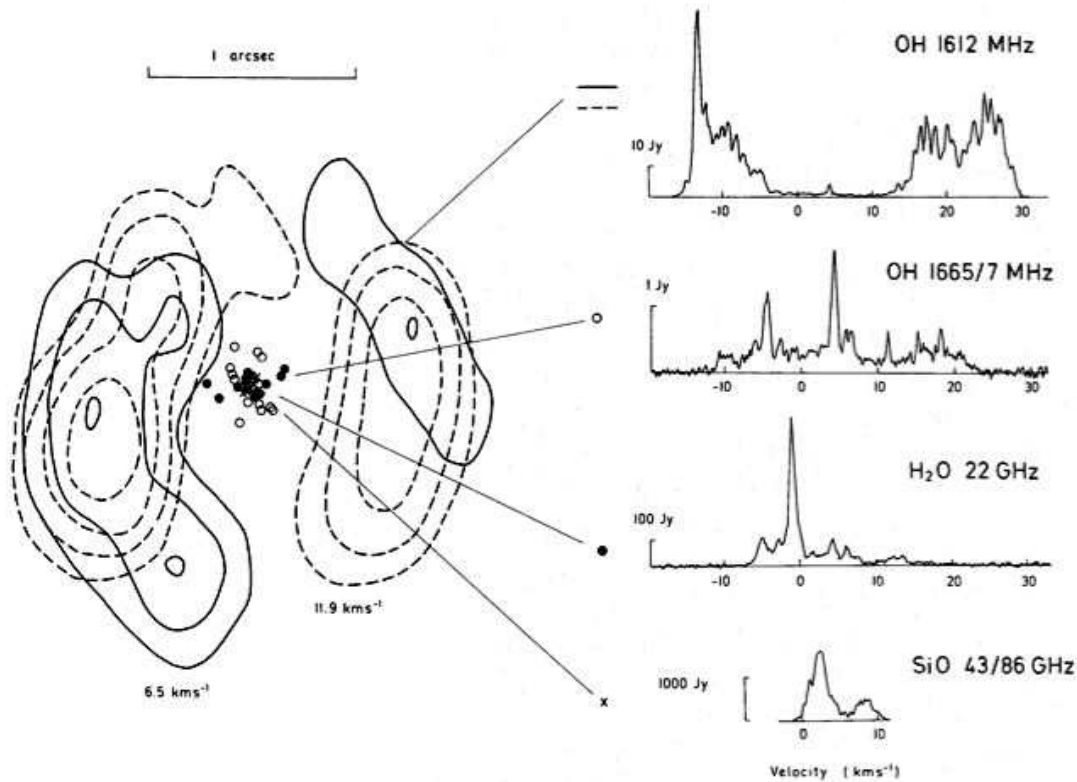}	
  			\caption{  			
  			\label{Fig:VXSgr}		
The maser emission around the supergiant star VX~Sgr. The solid and dashed lines correspond to contours of iso-intensity for the 1612 MHz OH maser line, for gas line-of-sight velocities of 6.5 and 11.9~\kms, respectively.  The stellar velocity is 5.5~\kms.
(From \citep{Chapman-1986})
}
\end{figure}

Maser emission is observed in many LPVs, and corresponds to stimulated
molecular
emission from an upper level which is more populated than the lower level (population inversion). For a review of physical processes related to astronomical masers, we refer to the textbook by \citet{Elitzur-1992}. The maser emission originates from gas in the circumstellar nebula (fed, e.g.,  by the strong stellar wind of late-type giants) and involves various molecules: SiO, H$_2$O and OH in O-rich environments, and mostly HCN in C-rich environments. Depending on the  energy required to populate the upper level, the maser emission will originate from regions closer or farther away from the central star. As discussed by \citet{Elitzur-1992} and \citet{Lewis-1989}, 
the SiO maser requires the highest excitation energies (with the involved levels lying at energies between $k \times 1730$~K and  $k \times 5500$~K, where $k$ is the Boltzmann constant) 
and thus originates at the closest distances from the star. 
The H$_2$O maser requires  intermediate excitation temperatures (the involved levels correspond to $E/k$ between 300 and 1900~K), and operates at
intermediate distances from the star, whereas the OH maser (subdivided into the so-called main OH maser at frequencies 1665 and 1667~MHz, and the satellite line 1612~MHz OH maser) involves levels
with low excitation temperatures ($\sim200$ K). Therefore, when the three O-bearing masers operate
concurrently in a circumstellar shell, they do so at different distances from the central star. A textbook example is provided by the supergiant star VX~Sgr \citep[Fig.~\ref{Fig:VXSgr};][]{Chapman-1986}.

The role of binarity
in suppressing the maser activity was suspected on theoretical
grounds \citep{HermanHabing85}, since the presence of a companion
periodically disturbs the layer where maser activity should
develop.
The systematic absence of SiO masers in symbiotic stars except in the very wide system R~Aqr \citep{Schwarz-1995,Hinkle-1989} is evidence supporting  that claim. On the other hand, very wide
systems like o~Ceti, R~Aqr and X~Oph \citep{Jorissen-03}
lack the OH maser, which involves layers several $10^5$ \Rsun\ away
from the star, in contrast to the SiO maser, which forms close to the
photosphere.
There should actually be a
critical orbital separation for every kind of maser below which the
companion sweeps through the corresponding masing layer \citep{Schwarz-1995},
thus preventing its development ($A$\,$<$\,10\,AU: no masing activity; 
10\,$\le$\,$A {\rm  (AU)}$\,$<$\,50: SiO and H$_2$O masers possible but no OH maser; $A$\,$>$\,50\,AU: all masers can operate).

A survey of IRAS sources by \citet{Lewis-1987} 
reveals that the region occupied by
OH/IR sources in the IRAS color--color diagram also contains many stars
with no OH masing activity. Furthermore, detected and undetected OH
maser sources have identical galactic latitude distributions. These
two facts find a natural explanation if binarity is the distinctive
property between detections and nondetections. 

\citet{Lewis-1989} \citep[see also][]{Stencel-1990} has presented a chronological sequence of maser states,  with increasing mass loss and thus shell thickness. The SiO maser, H$_2$O maser, the main OH maser and finally the 1612~MHz OH maser are added one by one as the shell grows in thickness. This sequence is supported by the fact that SiO and H$_2$O masers are associated with optically identified stars, while the majority of 1612~MHz OH maser sources lack an optical counterpart (OH/IR sources).
%*AF*
% Added half a sentence according tom how I understand the situation:
% the 1612 maser stay live in the outskirts all the time, even when main
% OH and H2O masers disappear, and then is dwarfed by the reborn main OH.
%%%
The masers then disappear in the reverse order as the shell continues to
thicken,
with the exception of the 1612~MHz OH maser, which remains strong.
The SiO maser disappears once steady mass loss stops. Thereafter,
the main OH maser eventually reappears and grows rapidly in strength
relative to the 1612~MHz maser in the detached fossil shell, until its
ongoing dilation causes all of the OH masers to fade away.
%%%

Anomalies in this chronological sequence 
may be another indication that the star has a companion which sweeps through the layer where the lacking maser should operate. 
Therefore, a  search for binaries among late-type giants involves 
looking  in maser catalogues \citep{Benson-1990} for O-rich stars with
SiO and OH masers but no H$_2$O maser, since this combination is never encountered in \citet{Lewis-1989} chronological sequence.
RU~Her is an example of such an anomaly; it is therefore quite significant that this star is suspected to have a composite spectrum \citep{Herbig-65} and appears in Table~\ref{Tab:VIM} as a true 'Variability-Induced Mover', which are two further indications of its binary nature.   

\begin{table}
\begin{footnotesize}
\caption{\label{Tab:masses}
A summary of the different properties that may be derived  for the various classes of binaries. The symbols $R_{A}$ and $d$ denote the radius of the eclipsed component A and the distance of the system from the Sun, respectively.
}
\begin{tabular}{lcccccc}
\hline
\mbox{}\\
Binaries &  Visual  & \multicolumn{2}{c}{Astrometric} & \multicolumn{2}{c}{Spectroscopic} & Eclipsing \\
          &                    &  AB1 & AB2      & SB1 & SB2
\medskip\\\hline
\mbox{}\medskip\\
Max $P$  $\sim$ & .. &  \multicolumn{2}{c}{100 yr} & \multicolumn{2}{c}{ 30 yr} & $\sim$ 1 yr \\ 
Min $P$ $\sim$ & 1 yr & \multicolumn{2}{c}{1 yr} & \multicolumn{2}{c}{ 1 d} &  1 h 
\medskip\\ 
$i$ ? & yes & yes & yes & no & no & yes\medskip\\
$a$ ? & $a_A + a_B$  & $a_0$ &   $a_{A,B}$ & $a_A \sin i$ & $a_{A,B} \sin i$ & $a$ \\ 
(units) & (") & (") & (") & (km) & (km) & ($R_A$) 
\medskip\\ 
Masses & $(M_A + M_B) d^{-3}$ & --- & $M_{A,B}$  & $f(M)$ & $M_{A,B} \sin^3 i$ & $(M_A + M_B) R_A^{-3}$\\ 
\\\hline 
\multicolumn{7}{l}{AB2 systems (Sirius, $\alpha$ Cen) are rare  \citep{Demarque-1986}}\\
\multicolumn{7}{l}{Ecl + SB2 (Ex:  HD 35155): individual masses may be derived}
\end{tabular}

\end{footnotesize}
\end{table}

\section{What may be known about stellar masses?}
\label{Sect:summary}

Binary stars are often considered as the golden way to derive stellar masses. 
Table~\ref{Tab:masses} shows, however, that the different binary-detection methods actually offer only partial knowledge of the masses. It is only when a given system appears as both eclipsing and SB2 that the individual masses may be derived. We refer to \citet{Andersen-1991} for a detailed review on this topic.

 \section{A list of resources on binary stars}
  \label{Sect:resources}
  \subsection{Combined data}
  
  The best entry point of any search about binary stars is probably the Besan\c{c}on database {\tt\footnotesize http://bdb.obs-besancon.fr/}, which  provides links to most of the catalogues and databases listed in this section.
  
A yearly bibliography about double stars is available at \\
{\tt\footnotesize http://ad.usno.navy.mil/wds/dsl/bib'nnnn'.html\\
 (where 'nnnn' = 2006 for instance)}.

  \subsection {IAU commissions of interest}

\begin{itemize}
\item IAU commission 8 : Astrometry\\
{\tt\footnotesize http://www.ast.cam.ac.uk/iau\_comm8/}
\item IAU commission 26 : Binary and Multiple Stars\\
{\tt\footnotesize http://ad.usno.navy.mil/wds/dsl.html\#iau}
\item IAU commission 27 : Variable Stars\\
{\tt\footnotesize http://www.konkoly.hu/IAUC27/}
\item IAU commission 30 : Radial velocities\\
{\tt\footnotesize http://www.ctio.noao.edu/science/iauc30/iauc30.html}
\item IAU commission 42 : Close Binary Stars\\
{\tt\footnotesize  http://www.konkoly.hu/IAUC42/c42info.html}
\end{itemize}

\subsection{Visual and interferometric binaries}

\begin{itemize}
\item Library of links: CHARA (integrates many sources: interferometric, speckle or occultation data, WDS, Hipparcos DMSA) \\ {\tt\footnotesize http://ad.usno.navy.mil/wds/dsl.html}\\ 
\item Instantaneous positions: 
\begin{itemize}
\item Catalogue of Interferometric Measurements of Binary Stars\\
 {\tt\footnotesize http://ad.usno.navy.mil/wds/int4.html}
\item CCDM\\ {\tt\footnotesize http://webviz.u-strasbg.fr/viz-bin/Vizier?-source=I/274}
\item Washington Double Stars (WDS)\\ {\tt\footnotesize http://ad.usno.navy.mil/wds/wds.html}
\item SIDONIE\\ {\tt\footnotesize http://sidonie.obs-nice.fr}
\end{itemize}
\item Orbits:\\
Sixth Catalogue of Orbits of Visual Binary Stars (Hartkopf \& Mason)\\
\mbox{}\indent {\tt\footnotesize http://ad.usno.navy.mil/wds/orb6.html} 
\end{itemize}

\subsection{Spectroscopic binaries}

\begin{itemize}
\item Spectral families
\begin{itemize}
\item Barium stars \citep{Lu-83}: \\
{\tt\footnotesize http://cdsads.u-strasbg.fr/full/1983ApJS...52..169L}
\item S stars \citep{Stephenson-1984}:\\ 
{\tt\footnotesize http://vizier.u-strasbg.fr/viz-bin/VizieR?-source=III/168}
\item C stars \citep{Alksnis-2001}:\\ 
{\tt\footnotesize http://vizier.u-strasbg.fr/viz-bin/VizieR?-source=III/227}
\item Metal-deficient barium and CH stars
\citep{Sleivyte-90,Bartkevicius-1996:a}:\\  
{\tt\footnotesize http://cdsads.u-strasbg.fr/abs/1990VilOB..85....3S}
\item Cataclysmic binaries and X-ray binaries
\citep{Ritter-2003}:  \\
{\tt\footnotesize http://physics.open.ac.uk/RKcat/} (update RKcat7.8, 2007)
\item Chromospherically active binaries
\citep{Strassmeier-1993}:\\ 
{\tt\footnotesize http://vizier.u-strasbg.fr/viz-bin/VizieR?-source=V/76} 
\item Symbiotic stars 
\begin{itemize} 
\item   {\tt\footnotesize http://vizier.u-strasbg.fr/viz-bin/VizieR?-source=J/A+AS/146/407} \citep{Belczynski00}
\item See as well the series of papers by Carquillat \citep{Carquillat-2008}, R.E. Griffin \citep{Griffin-2006:b}, Ginestet \citep{Ginestet-1997,Ginestet-2002}
\end{itemize} 
\end{itemize}
\item Individual radial-velocity measurements\\ 
{\tt\footnotesize http://www.casleo.gov.ar/catalogo/catalogo.htm}
\medskip\\
\item Orbits:\\
 The 9th Catalogue of Spectroscopic Binary Orbits \citep{Pourbaix-04a}\\ {\tt\footnotesize http://sb9.astro.ulb.ac.be/}
\end{itemize}

\subsection{Astrometric binaries}
\begin{itemize}
\item  Common proper-motion pairs:
\begin{itemize}
\item Luyten LDS catalogue\\
{\tt\footnotesize http://vizier.u-strasbg.fr/viz-bin/VizieR?-source=I/130}
\item Halbwachs 1986 \citep{Halbwachs-1986}\\
{\tt\footnotesize http://vizier.u-strasbg.fr/viz-bin/VizieR?-source=I/121}
\item Greaves 2004 \citep{Greaves-2004}\\
{\tt\footnotesize http://vizier.u-strasbg.fr/viz-bin/VizieR?-source=J/MNRAS/355/585}
\end{itemize}
\item $\Delta \mu$ binaries: 
\begin{itemize}
\item DMUBIN database \citep{Wielen-1999}\\
 {\tt\footnotesize http://www.ari.uni-heidelberg.de/datenbanken/dmubin}
\item Makarov \& Kaplan 2005 \citep{Makarov-Kaplan-2005} 
\item Frankowski et al. 2007 \citep{Frankowski-2007}
\end{itemize}
\item Hipparcos: DMSA/O binaries
\end{itemize}

\subsection{Photometric variables}

\begin{itemize}
\item General Catalogue of Photometric Data\\   
{\tt\footnotesize http://obswww.unige.ch/gcpd/gcpd.html}
\item Hipparcos photometry annex \\{\tt\footnotesize http://www.rssd.esa.int/SA/HIPPARCOS/apps/PlotCurve.html}
\item All Sky Automated Survey\\
{\tt\footnotesize http://archive.princeton.edu/~asas/} 
\item Northern Sky Variability Survey \\
{\tt\footnotesize http://skydot.lanl.gov/}
\item OGLE \\ {\tt\footnotesize http://bulge.princeton.edu/$\sim$ogle/}
\item MACHO\\  {\tt\footnotesize http://wwwmacho.mcmaster.ca/}
\item DENIS\\ {\tt\footnotesize http://www-denis.iap.fr}
\item Eclipsing binaries: 
\begin{itemize}
\item {\tt\footnotesize http://cdsweb.u-strasbg.fr/cgi-bin/myqcat3?eclipsing+binaries}
\item Krakow Observatory\\  {\tt\footnotesize http://www.oa.uj.edu.}
\item Contact (W UMa-type) binaries at Konkoly Observatory\\
{\tt\footnotesize http://www.konkoly.hu/staff/csizmadia/wuma.html}
%%%
\item A catalogue of eclipsing variables \citep{Malkov-2006}
\end{itemize}
\item AAVSO \\
{\tt \footnotesize http://www.aavso.org}
\item { also look for the sites of the French, British, Swiss, New Zealander variable-star-observer associations...} 
\end{itemize}

\begin{theacknowledgments}
  The authors thank D. Pourbaix for useful discussions.
\end{theacknowledgments}



\bibliographystyle{aipproc}   % if natbib is available
%%%%\bibliographystyle{aipprocl} % if natbib is missing

%\input aasjournals_habing.tex
%\bibliography{ajorisse_articles}

%
%  These Macros are taken from the AAS TeX macro package version 4.0.
%  Include this file in your LaTeX source only if you are not using
%  the AAS TeX macro package and need to resolve the macro definitions
%  in the BibTeX entries returned by the ADS abstract service.
%
%  For more information on the AASTeX macro package, please see the URL
%       http://www.ferberts.com/AAS/aastex.html
%  For more information about ADS abstract server, please see the URL
%       http://adswww.harvard.edu/ads_abstracts.html
%

% Abbreviations for journals.  The object here is to provide authors
% with convenient shorthands for the most "popular" (often-cited)
% journals; the author can use these markup tags without being concerned
% about the exact form of the journal abbreviation, or its formatting.
% It is up to the keeper of the macros to make sure the macros expand
% to the proper text.  If macro package writers agree to all use the
% same TeX command name, authors only have to remember one thing, and
% the style file will take care of editorial preferences.  This also
% applies when a single journal decides to revamp its abbreviating
% scheme, as happened with the ApJ (Abt 1991).

% Modified by HO 030509
% Included Science, and remowed \rm

\def\aj{AJ}                   % Astronomical Journal
\def\araa{ARA\&A}             % Annual Review of Astron and Astrophys
\def\apj{ApJ}                 % Astrophysical Journal
\def\apjl{ApJ}                % Astrophysical Journal, Letters
\def\apjs{ApJS}               % Astrophysical Journal, Supplement
\def\ao{Appl.~Opt.}           % Applied Optics
\def\apss{Ap\&SS}             % Astrophysics and Space Science
\def\aap{A\&A}                % Astronomy and Astrophysics
\def\aapr{A\&A~Rev.}          % Astronomy and Astrophysics Reviews
\def\aaps{A\&AS}              % Astronomy and Astrophysics, Supplement
\def\azh{AZh}                 % Astronomicheskii Zhurnal
\def\ba{Baltic~Astr.}         % Baltic Astronomy
\def\baas{BAAS}               % Bulletin of the AAS
\def\jrasc{JRASC}             % Journal of the RAS of Canada
\def\memras{MmRAS}            % Memoirs of the RAS
\def\mnras{MNRAS}             % Monthly Notices of the RAS
\def\pra{Phys.~Rev.~A}        % Physical Review A: General Physics
\def\prb{Phys.~Rev.~B}        % Physical Review B: Solid State
\def\prc{Phys.~Rev.~C}        % Physical Review C
\def\prd{Phys.~Rev.~D}        % Physical Review D
\def\pre{Phys.~Rev.~E}        % Physical Review E
\def\prl{Phys.~Rev.~Lett.}    % Physical Review Letters
\def\pasp{PASP}               % Publications of the ASP
\def\pasj{PASJ}               % Publications of the ASJ
\def\pasa{Publ.~Astron.~Soc.~Australia} % Publications of the Astronomical Society of Australia
\def\qjras{QJRAS}             % Quarterly Journal of the RAS
\def\sci{Science}             % Science
\def\skytel{S\&T}             % Sky and Telescope
\def\solphys{Sol.~Phys.}      % Solar Physics
\def\sovast{Soviet~Ast.}      % Soviet Astronomy
\def\ssr{Space~Sci.~Rev.}     % Space Science Reviews
\def\zap{ZAp}                 % Zeitschrift fuer Astrophysik
\def\nat{Nature}              % Nature
\def\iaucirc{IAU~Circ.}       % IAU Cirulars
\def\aplett{Astrophys.~Lett.} % Astrophysics Letters
\def\apspr{Astrophys.~Space~Phys.~Res.}
                % Astrophysics Space Physics Research
\def\bain{Bull.~Astron.~Inst.~Netherlands} 
                % Bulletin Astronomical Institute of the Netherlands
\def\fcp{Fund.~Cosmic~Phys.}  % Fundamental Cosmic Physics
\def\gca{Geochim.~Cosmochim.~Acta}   % Geochimica Cosmochimica Acta
\def\grl{Geophys.~Res.~Lett.} % Geophysics Research Letters
\def\jcp{J.~Chem.~Phys.}      % Journal of Chemical Physics
\def\jgr{J.~Geophys.~Res.}    % Journal of Geophysics Research
\def\jqsrt{J.~Quant.~Spec.~Radiat.~Transf.}
                % Journal of Quantitative Spectroscopy and Radiative Transfer
\def\memsai{Mem.~Soc.~Astron.~Italiana}
                % Mem. Societa Astronomica Italiana
\def\nphysa{Nucl.~Phys.~A}   % Nuclear Physics A
\def\physrep{Phys.~Rep.}   % Physics Reports
\def\physscr{Phys.~Scr.}   % Physica Scripta
\def\planss{Planet.~Space~Sci.}   % Planetary Space Science
\def\procspie{Proc.~SPIE}   % Proceedings of the SPIE
\def\sci{Science}

\let\astap=\aap
\let\apjlett=\apjl
\let\apjsupp=\apjs
\let\applopt=\ao

\IfFileExists{\jobname.bbl}{}
 {\typeout{}
  \typeout{******************************************}
  \typeout{** Please run "bibtex \jobname" to optain}
  \typeout{** the bibliography and then re-run LaTeX}
  \typeout{** twice to fix the references!}
  \typeout{******************************************}
  \typeout{}
 }

\end{document}